\documentclass[twocolumn,aps,pra,10pt,floatfix,nofootinbib]{revtex4-1}

	\usepackage[usenames,dvipsnames]{xcolor}
	\usepackage{soul}
	\usepackage{amsmath}
	\usepackage{amsfonts}
	\usepackage{amssymb}
	\usepackage[colorlinks=true,citecolor=cyan,linkcolor=magenta]{hyperref}
	\usepackage{graphicx}
	\usepackage{bbold}					
	\usepackage[makeroom]{cancel}		
	\usepackage{multirow}				
	\usepackage[normalem]{ulem}        
	\usepackage{array}
	\usepackage{mathrsfs}
	\usepackage{mathtools}

	\newcolumntype{x}[1]{>{\centering\let\newline\\\arraybackslash\hspace{0pt}}p{#1}}

	\DeclareMathOperator{\sign}{sign}  	
	\DeclareMathOperator{\diag}{diag}  	

	\DeclareMathAlphabet{\mathbbold}{U}{bbold}{m}{n}
	\def\abs#1{\left|{#1}\right|}      	
	\def\bs#1{\boldsymbol{#1}}			
	\def\imi{\mathrm{i}}				
	\def\imj{\mathrm{j}}				
	\def\imk{\mathrm{k}}				
	\def\e#1{\mathrm{e}^{#1}}				

	\def\eps{\varepsilon}					
 
	\def\mcH{\mathcal{H}}					
	\def\mcT{\mathcal{T}}					
	\def\mcP{\mathcal{P}}					
	\def\mcC{\mathcal{C}}					

	\def\intg{\mathbbold{Z}}					
	\def\ztwo{\mathbbold{Z}_2}					
	\def\triv{\mathbbold{0}}					
	\def\unit{\mathbbold{1}}					
	\def\reals{\mathbbold{R}}					
	
	\DeclareMathOperator{\im}{im}

	\newcounter{subeqn} %
	\makeatletter
	\@addtoreset{subeqn}{equation}
	\makeatother

\definecolor{TB}{rgb}{0,0,0} 

\def\TB#1{{\color{TB}#1}}
\def\TBC#1{}
\definecolor{AT}{rgb}{0.5,0.3,1}

\definecolor{MyOrange}{RGB}{255,180,0}
\definecolor{MyRed}{RGB}{204,0,0}
\definecolor{MyBlue}{RGB}{0,165.75,229.5}
\definecolor{MyIndigo}{RGB}{75,0,130}
\definecolor{MyTan}{RGB}{185,155,126}

\definecolor{MyGreen}{RGB}{0,216.75,0}

\begin{document}
\title{Non-Abelian topology of nodal-line rings in \texorpdfstring{$\mcP\mcT$}{PT}-symmetric systems}

\author{Apoorv Tiwari$^{1,2}$}
\
\author{Tom\'{a}\v{s} Bzdu\v{s}ek$^{1,2,3,4}$}\email[corresponding author: ]{tomas.bzdusek@uzh.ch}
\

\affiliation{$^{1}$Condensed Matter Theory Group, Paul Scherrer Institute, CH-5232 Villigen PSI, Switzerland}
\affiliation{$^{2}$Department of Physics, University of Zurich, Winterthurerstrasse 190, 8057 Zurich, Switzerland}
\affiliation{$^{3}$Department of Physics, McCullough Building, Stanford University, Stanford, CA 94305, USA}
\affiliation{$^{4}$Stanford Center for Topological Quantum Physics, Stanford University, Stanford, CA 94305, USA}

\date{\today}

\begin{abstract}
Nodal lines inside the momentum space of three-dimensional crystalline solids are topologically stabilized by a $\pi$-flux of Berry phase. 
Nodal-line rings in $\mcP\mcT$-symmetric systems with negligible spin-orbit coupling (here described as ``nodal class AI'') can carry an additional ``monopole charge'', which further enhances their stability.
Here, we relate two recent theoretical advancements in the study of band topology in nodal class AI. 
On \TB{the} one hand, cohomology classes of real vector bundles were used to relate the monopole charge of nodal-line rings to their linking with nodal lines formed among the occupied and among the unoccupied bands.
On the other hand, homotopy studies revealed that the generalization of the Berry-phase quantization to the case of multiple band gaps defines a non-Abelian topological charge, which governs the possible deformations of the nodal lines.
\TB{The present work has three aims.
First, we present} how to efficiently \TB{wield} 
the \TB{recently discovered} non-Abelian topological charge. 
\TB{Second, we apply} these methods to present an independent proof of the relation between the monopole charge and the linking structure, including all the fragile-topology exceptions.
\TB{Finally,} we show that the monopole charge flips sign when braided along a path with a non-trivial Berry phase. 
This facilitates a new type non-Abelian ``braiding'' of nodal-line rings inside the momentum space, that has not been previously reported. 
\TB{The work begins with a brief review of $\mcP\mcT$-symmetric band topology, and the}
geometric arguments \TB{employed in our theoretical analysis} 
are supplemented in the appendices with formal mathematical derivations. 
\end{abstract}

\maketitle

\section{Introduction}

Energy bands of electrons in crystalline solids can form degeneracies inside the crystal-momentum ($\bs{k}$) space. We call these degeneracies \emph{band-structure nodes}. They were first investigated in the context of three-dimensional (3D) metals with time-reversal symmetry ($\mcT$) by Herring~\cite{Herring:1937}, who found that such nodes generically occur along lines (at points) for systems with (without) spatial inversion symmetry ($\mcP$). This finding is compatible with an older codimension argument by von Neumann and Wigner~\cite{vonNeumann:1929}. Much later, after the concept of Berry phase~\cite{Berry:1984} was introduced to the band theory of solids~\cite{Zak:1989,Mikitik:1999,Xiao:2010}, the interplay between band-structure nodes and space-group symmetry has been examined by Michel and Zak~\cite{Michel:1999}. The topological underpinning of band-structure nodes became fully appreciated only after the discovery of topological insulators~\cite{Kane:2005,Moore:2007,Bernevig:2006b,Konig:2007,Fu:2007,Qi:2008,Hsieh:2009b,Xia:2009,Fu:2011,Soluyanov:2011,Wang:2016,Chiu:2016,Fang:2015a,Benalcazar:2017,Schindler:2018,Horava:2005,Shiozaki:2014} and superconductors~\cite{Kitaev:2009,Schnyder:2008,Ryu:2010,Fu:2008,Hasan:2010,Qi:2011}. In recent years, a plethora of band-structure nodes were investigated in 3D systems, both theoretically and experimentally, including nodal points~\cite{Wan:2011,Murakami:2007,Son:2013,Weng:2015,Xu:2015a,Lv:2015,Huang:2015b,Fang:2012,Soluyanov:2015,Deng:2016,Chang:2017c,Chang:2018,Armitage:2018,Young:2012,Yang:2014,Wu:2019,Wehling:2014,Wang:2012,Liu:2014a,Liu:2014b,Neupane:2014,Steinberg:2014,Zhao:2017,Wieder:2016,Bradlyn:2016}, 
nodal lines~\cite{Burkov:2011,Fang:2016,Yang:2017,Bzdusek:2016,Li:2017b,Bian:2015a,Bian:2015b,Chen:2015,Chen:2015b,Wu:2018b,Lian:2019,Li:2016,Xie:2015,Chan:2016,Zhao:2016b,Fang:2015,Nomura:2018,Ahn:2018,Huang:2016,Xu:2017,Kim:2015,Yu:2015,Weng:2015b,Wang:2017,Gong:2018,Yan:2017,Zhou:2018,Ezawa:2017,Bi:2017,Yang:2019,Chang:2017,Hearing:1958,Chen:2017,Heikkila:2015b,McClure:1957,Singh:2018,Schoop:2016,Feng:2017,Yu:2017,Takahashi:2017,Chang:2017b,Yi:2018,Bzdusek:2017},
and nodal surfaces~\cite{Bzdusek:2017,Liang:2016,Zhong:2016,Xiao:2017,Wu:2018,Yu:2019,Volkov:2018}. Analogous degeneracies were also studied in excitation bands of superconductors~\cite{Heikkila:2011,Kobayashi:2014,Chiu:2014,Zhao:2016,Lian:2017,Sun:2017,Agterberg:2017,Bouhon:2018b}, superfluids~\cite{xu2015structured}, photons~\cite{Lu:2013,Yan:2018,Zhou:2018b}, phonons~\cite{Po:2016,Rocklin:2016} and magnons~\cite{Li:2017}. Recently, the classification of generic band-structure nodes for all Altland-Zirnbauer symmetry classes~\cite{Altland:1997} with or without $\mcP$ was developed~\cite{Bzdusek:2017} using homotopy theory~\cite{Sun:2018,Lundell:1992,Turner:2012,Kennedy:2015}. The relation between the space-group symmetry and the location of the band-structure nodes has been investigated~\cite{Song:2018,Murakami:2017,Bouhon:2017,Bouhon:2018}, largely by considering symmetry-based indicators~\cite{Slager:2013,Kruthoff:2017,Po:2017,Bradlyn:2017,Watanabe:2017,Tang:2018}.

Nodal lines (NLs) are generically stabilized by quantization of Berry phase~\cite{Zak:1989}, which arises in various symmetry settings~\cite{Fang:2016}. NLs can form intricate compositions, including rings~\cite{Bian:2015a,Bian:2015b,Chen:2015,Xie:2015,Chan:2016,Zhao:2016b,Li:2016,Xu:2017,Fang:2015,Nomura:2018,Ahn:2018,Huang:2016}, chains~\cite{Bzdusek:2016,Wang:2017,Gong:2018}, links~\cite{Ezawa:2017,Lian:2017,Sun:2017,Yan:2017,Zhou:2018}, knots~\cite{Ezawa:2017,Bi:2017,Yang:2019}, helices~\cite{Chang:2017,Hearing:1958,Chen:2017,Sun:2017}, nexuses~\cite{McClure:1957,Heikkila:2015b,Singh:2018}, gyroscopes~\cite{Yu:2015,Kim:2015,Weng:2015b} and nets~\cite{Bzdusek:2016,Schoop:2016,Feng:2017,Yu:2017,Takahashi:2017,Chang:2017b,Yi:2018}. In this manuscript we consider in particular the case of $\mcP\mcT$-symmetric 3D systems with negligible spin-orbit coupling (SOC)~\cite{Chen:2015b,Xie:2015,Chan:2016,Zhao:2016b,Li:2016,Xu:2017,Fang:2015,Nomura:2018,Ahn:2018,Huang:2016,Kim:2015,Yu:2015,Weng:2015b,Wang:2017,Gong:2018,Lian:2019,Wu:2018b,Yan:2017,Zhou:2018,Bi:2017,Yang:2019,Ezawa:2017,Chang:2017,Hearing:1958,Chen:2017,McClure:1957,Heikkila:2015b,Singh:2018,Schoop:2016,Feng:2017,Yu:2017,Takahashi:2017,Chang:2017b,Yi:2018,Bzdusek:2017}, where the NLs are known to form all the compositions listed above. In these systems, the antiunitary $\mcP\mcT$-operator squares to $+\unit$ and commutes with the Bloch Hamiltonian at every $\bs{k}$. Motivated by the Altland-Zirnbauer classification~\cite{Altland:1997}, we call the symmetry settings with such an operator as \emph{nodal class} $\textrm{AI}$~\cite{Bzdusek:2017}. Besides electron bands in $\mcP\mcT$-symmetric 2D and 3D systems with weak SOC, nodal class AI also encompasses excitation bands of photons and of Goldstone bosons in $\mcP\mcT$-symmetric systems~\cite{Po:2016,Rocklin:2016,Li:2017}. Nodal class AI also describes 2D materials with $C_{2z}\mcT$ symmetry (both with or without SOC), where $C_{2z}$ is a $\pi$-rotation of the 2D plane~\cite{Ahn:2019,Bouhon:2019,Ahn:2018b}. 

Importantly, nodal-line rings (NL-rings) in nodal class $\textrm{AI}$~\cite{Xie:2015,Chan:2016,Zhao:2016b,Li:2016,Xu:2017,Fang:2015,Nomura:2018,Ahn:2018,Huang:2016,Kim:2015} can exhibit an \TB{additional} 
topological obstruction called the \emph{monopole charge}~\cite{Fang:2015}, which is independent of the Berry phase quantization. NL-rings with a monopole charge can be removed from the band structure only through a pairwise annihilation~\cite{Bzdusek:2017}, resembling the case of two Weyl points with opposite chirality~\cite{Wan:2011}. NL-rings with a monopole charge were proposed to exist in various systems, including the electron bands of $\textrm{ABC}$-stacked graphdiyne~\cite{Nomura:2018,Ahn:2018}, Bogoliubon bands of a superconducting phase on the hyperhoneycomb lattice~\cite{Bouhon:2018b}, and magnon bands of antiferromagnetic Cu$_3$TeO$_6$~\cite{Li:2017}.

In this work, we are especially concerned with two recent theoretical advancements in the study of band topology in nodal class $\textrm{AI}$. On the one hand, the Euler class and the second Stiefel-Whitney class of real vector bundles spanned by the occupied states were investigated~\cite{Ahn:2018,Ahn:2018b,Bouhon:2019}. This led to the discovery that the monopole charge of NL-rings (assumed to be formed near the chemical potential) is related to their linking structure with NLs formed among the occupied and among the unoccupied bands~\cite{Ahn:2018}. We emphasize that the previous statement involves several NLs formed inside \emph{various} band gaps, \emph{i.e.}~between various pairs of bands, and thus falls outside the scope of the tenfold-way classification~\cite{Schnyder:2008,Kitaev:2009, Ryu:2010}. Curiously, the Euler class was also shown to correspond to the fragile-topology obstruction present in tight-binding models of twisted bilayer graphene near the ``magic angle''~\cite{Ahn:2018b,Cao:2018,Dodaro:2018,Kang:2020}. 

On the other hand, a mathematical framework capable of the \emph{simultaneous} description of multiple NLs formed inside various band gaps has been independently sought using homotopic methods~\cite{Bzdusek:2017}. This search uncovered a non-Abelian generalization of the Berry phase quantization~\cite{Wu:2018b}, which is different from the non-Abelian Berry-Wilczek-Zee connection~\cite{Wilczek:1984}. More specifically, a pair of NLs formed inside two consecutive band gaps were shown to carry noncommuting topological charges, implying that they cannot be moved across each other~\cite{Poenaru:1977}, and suggesting the possibility of non-Abelian ``braiding'' of NLs in $\bs{k}$-space~\cite{Toulouse:1977,Mermin:1979,Alexander:2012}. Analogous behaviour was reported by Ref.~\cite{Ahn:2018b} for nodal points in 2D. Furthermore, the non-Abelian topological charge poses constraints on the admissible NL-compositions, and on the topological transitions between various such compositions. Some of these constraints were recently observed in computational studies of elemental Sc under strain~\cite{Wu:2018b} and of MgSrSi-type crystals~\cite{Lian:2019}. The non-Abelian charge has been mathematically formulated~\cite{Wu:2018b} as a spinor generalization of Wilson holonomy operators~\cite{Yu:2011,Soluyanov:2011b,Gresch:2017}. 

This manuscript has \TB{three goals}. 
First, we present an overview of the non-Abelian topological charge found by Ref.~\cite{Wu:2018b}. 
\TB{Importantly, we provide this non-Abelian charge} a simple geometric interpretation, which can be used for an efficient analysis of complicated NL-compositions using a minimal amount of equations. 
\TB{Since some of the presented methods can be understood as generalization of certain conventional techniques, we begin the manuscript with a brief but general review of band topology of $\mcP\mcT$-symmetric Bloch Hamiltonians.}
\TB{The next two goals correspond to applications of the presented methods to analyze the monopole charge of NL-rings.}
\TB{On the one hand, we present} 
an independent proof of the relation between the monopole charge of NL-rings and their linking with NLs formed inside \emph{other} band gaps~\cite{Ahn:2018}, including all the fragile-topology exceptions \TB{reported} 
by Ref.~\cite{Bzdusek:2017}. \TB{On the other hand, we reveal a previously unreported topological phenomenon, namely that} the monopole charge flips sign when \TB{moved} along a \TB{closed} path with a non-trivial Berry phase, suggesting a yet another possibility of non-Abelian braiding of NL-rings in $\bs{k}$-space. Such noncommutative behaviour is analogous to the process known to arise in uniaxial nematics, when a hedgehog defect changes into an ``antihedehog'' when it is braided around 
a \TB{disclination line} ~\cite{Mermin:1979}. The \TB{sign reversal of the monopole charge} 
can be \TB{formalized} 
using the notion of \emph{Abe homotopy}~\cite{Abe:1940,kobayashi2012abe,Sun:2019} which naturally incorporates the action of the fundamental group on the $2^{\textrm{nd}}$ (or more generally $n^{\mathrm{th}}$ homotopy group).

The manuscript is organized as follows. In Sec.~\ref{sec:topo-sum} we provide a brief \TB{review} 
of topological properties of NLs in nodal class $\textrm{AI}$. This section can be skipped by experts in the field, without impairing the continuity of our exposition. We \TB{specifically} comment here on the von Neumann-Wigner argument, the quantization of the Berry phase, the monopole charge, and on the recently revealed non-Abelian topological charge. In Sec.~\ref{sec:non-Abelian-NLs}, we present how to efficiently wield the non-Abelian topological charge, and how it leads to constraints on admissible NL compositions. This discussion simplifies, \TB{but also further expands}, the supplemental material of Ref.~\cite{Wu:2018b}. \TB{We then proceed with applications of the develoepd methods. First, in} Sec.~\ref{sec:monopole-linking-relation} we rederive the relation between the monopole charge and the linking structure using \TB{the geometric description of} the non-Abelian topological charge. Although we reproduce here the findings of Ref.~\cite{Ahn:2018}, we believe that our arguments are easier to follow for readers not knowledgeable in characteristic classes of real vector bundles. Finally, we show in Sec.~\ref{sec:NA-braiding} how the monopole charge flips sign when it is braided along a path with a non-trivial Berry phase, suggesting a non-Abelian braiding of NLs formed inside the \emph{same} band gap. We conclude in Sec.~\ref{sec:conclude} by outlining the broader scope of the possible applications of the non-Abelian topological charge in the study of band topology.

We have attempted to keep the main text accessible to a wide readership by avoiding advanced mathematical techniques. In a sharp contrast, we review in the appendices several mathematical tools which can be further employed to formalize our reasoning. This technical complement of our discussion is aimed at readers experienced with algebraic topology. Our goal is to provide these readers concrete examples of how these advanced tools can be applied in the study of band-structure nodes. This discussion, which can be read independently of the main text, is organized as follows. First, in appendix~\ref{sec:cosets-spaces} we formulate the relevant spaces of Hamiltonians as coset spaces, and we explain why such a formulation is convenient to study the homotopy invariants. In appendix~\ref{app:cosets-and-homotopy} we formulate the notion of the action of the first homotopy group $\pi_1(M)$ on the second homotopy group $\pi_2(M)$ for the case when $M$ is a coset space. We subsequently generalize this concept to general spaces $M$ using the notion of Abe homotopy in appendix~\ref{sec:crossed-module}. In the final appendices~\ref{app:pi1-on-pi2-examples} and~\ref{sec:action-for-CI} we employ a method adapted from Ref.~\cite{Mermin:1979} to explicitly derive the action of $\pi_1(M)$ on $\pi_2(M)$ for all the nodal classes supporting the monopole charge, especially for all the fragile-topology exceptions. This discussion provides an alternative derivation of the \TB{sign-reversal of a monopole charge when the NL-ring is moved along a path with a non-trivial Berry phase, } 
which we also derive \TB{with the help of} 
the non-Abelian topological charge in Sec.~\ref{sec:NA-braiding}.

\section{Nodal lines with \texorpdfstring{$\mcP\mcT$}{PT}-symmetry}\label{sec:topo-sum}

\subsection{Overview}

In this section, we \TB{briefly review} 
the basic aspects of NLs in 3D systems with $\mcP\mcT$ symmetry and with negligible SOC. Following Refs.~\cite{Bzdusek:2017,Fischer:2018}, we refer to such symmetry settings as \emph{nodal class $\textrm{AI}$}. Its special feature is the \emph{reality condition}: The weak SOC allows us to ignore the spin degree of freedom, such that the antiunitary $\mcP\mcT$ operator squares to $+\unit$. A suitable unitary rotation of the basis brings the $\mcP\mcT$ operator to complex conjugation, thus enforcing the Bloch Hamiltonian $H(\bs{k})$ to be \emph{real}. The Hamiltonian eigenstates can be locally expressed in a real gauge, although a \emph{global} real gauge may be absent, especially in the presence of band degeneracies. (The same reasoning also applies to $C_{2z}\mcT$-symmetric systems in 2D with or without SOC~\cite{Ahn:2019,Bouhon:2019,Ahn:2018b}.)  

Our discussion is structured as follows. In Sec.~\ref{subsec:codims} we review a \emph{codimension argument}, showing why generic degeneracies of nodal class AI in 3D are one-dimensional lines~\cite{vonNeumann:1929}. In Sec.~\ref{eqn:topo-classes} we review the \emph{leitmotif of topological classifications} of band structures in crystalline solids~\cite{Kitaev:2009,Ryu:2010}, with a special emphasis on the homotopy description of band-structure nodes~\cite{Bzdusek:2017,Sun:2018}. We continue in Sec.~\ref{NLs-first-homotopy} by applying the homotopy technique to explain the \emph{topological stability of NLs} in nodal class $\textrm{AI}$, which is related to the quantization of Berry phase~\cite{Zak:1989}. In Sec.~\ref{NLs-second-homotopy} we show how the homotopy approach further predicts the existence of the \emph{monopole charge of NL-rings}~\cite{Fang:2015,Bzdusek:2017}, which has recently been related to the linking structure of NLs in $\bs{k}$-space~\cite{Ahn:2018}. Finally, in Sec.~\ref{sec:non-Abelian-review} we provide a brief synopsis of the recently discovered \emph{non-Abelian topology of NLs}~\cite{Wu:2018b}, which arises when one tries to simultaneously describe NLs occurring inside multiple band gaps. We present a handful of rules that suggest how to efficiently apply the non-Abelian topological charge to analyze complicated NL compositions. These rules are properly justified in Sec.~\ref{sec:non-Abelian-NLs}, and we apply them to study the monopole charge in Sec.~\ref{sec:monopole-linking-relation} and in Sec.~\ref{sec:NA-braiding}. 

The goal of the present section is to provide a summary of the topic for non-experts, and to introduce the terminology employed in later sections. Experts in the field should be able to jump directly to Sec.~\ref{sec:non-Abelian-NLs}.

\subsection{Codimension argument}~\label{subsec:codims}

The observation that real Hamiltonians $H(\bs{k})$ form band degeneracies of codimension two can be traced back to von Neumann and Wigner~\cite{vonNeumann:1929}. Their argument is based on counting the dimension of the \emph{space of Hamiltonians with prescribed degeneracies of energy levels}. \TB{To figure out this dimension,} let us consider Hamiltonians with energy spectrum consisting of $n_\alpha$-fold degenerate levels, where $\sum_\alpha n_\alpha = N$ is the dimension of the Hilbert space. \TB{Let us further use $f$ to indicate the total number of energy levels. Any such a} Hamiltonian $H(\bs{k})$ can be diagonalized as
\begin{equation}
H(\bs{k}) = \sum_{j=1}^N | u^j_{\bs{k}}\rangle \eps^j_{\bs{k}} \langle u^j_{\bs{k}}|\qquad \textrm{with}\quad \eps^j_{\bs{k}} \leq \eps_{\bs{k}}^{j+1} \label{eqn:spect-decom-no-k}
\end{equation}
where $|u^j_{\bs{k}}\rangle$ is the eigenstate of $H(\bs{k})$ with energy $\eps^j_{\bs{k}}$. Assuming a local real gauge, the collection $\{u^j_{\bs{k}}\}_{j=1}^N \equiv \mathcal{U}_{\bs{k}} \in \mathsf{O}(N)$ is an orthogonal matrix, \TB{which constitute a space of dimension $\dim \mathsf{O}(N) = \tfrac{1}{2}N(N-1)$. Each level is further characterized by an $\mathsf{O}(n_\alpha)$ gauge degree of freedom, which reduces the space of \emph{unique} Hamiltonian matrices by $v = \sum_\alpha \tfrac{1}{2}n_\alpha(n_\alpha-1)$. Finally, one needs to specify the values of the $f$ energy levels. Counting all the degrees of freedom,} the dimension of the space of Hamiltonians \TB{with the prescribed level degeneracy equals}~\footnote{\TB{One can explicitly express this space as \begin{equation}
M^\textrm{AI}_{\{n_\alpha\}_{\alpha=1}^f} \!\!=\! \frac{\mathsf{O}(n_1+\ldots+n_f)}{\mathsf{O}(n_1)\!\times\!\ldots\!\times\!\mathsf{O}(n_f)}\!\times\! \textrm{Conf}_f(\mathbb{R}).\label{eqn:config-space}
\end{equation}
The first factor is known in the mathematical literature as the \emph{generalized real flag manifold}, and $\textrm{Conf}_f(\mathbb{R})$ is an ordered configuration of $f$ real numbers (which is \emph{not} a connected space). Explicit expressions analogous to Eq.~(\ref{eqn:config-space}) were recently found particularly useful to describe non-Hermitian band topology~\cite{Wojcik:2019,Li:2019}.}} 
\begin{equation}
d=\tfrac{1}{2}N(N-1) + f - v \label{eqn:codim-arg}
\end{equation}
\TB{as has been originally reported by Ref.~\cite{vonNeumann:1929}}.

When all the energy levels of $H(\bs{k})$ are non-degenerate, we have $f=N$ and $v=0$, leading to 
\begin{subequations}
\begin{equation}
d_{\textrm{non-deg.}} = \tfrac{1}{2}N(N+1). 
\end{equation}
On the other hand, in the presence of one double degeneracy, only $f=(N-1)$ energies need to be specified, while the internal rotations of the two degenerate states lead to $v = \dim \mathsf{O}(2) = 1$. Therefore,
\begin{equation}
d_\textrm{2-deg.} = \tfrac{1}{2}N(N+1) - 2. 
\end{equation}
The difference $\delta_\textrm{2-deg.} = d_\textrm{non-deg.} - d_\textrm{2-deg.}  = 2$, known as \emph{codimension}, implies that two parameters need to be tuned to encounter such a band degeneracy. Therefore, in 3D we anticipate band degeneracies to form $3 - \delta_\textrm{2-deg.} = 1$-dimensional manifolds inside the $\bs{k}$-space, \emph{i.e.}~NLs.
\end{subequations}

We remark that mirror symmetry can fix NLs to mirror-invariant planes~\cite{McClure:1957,Weng:2015b,Kim:2015,Yu:2015,Chen:2015b,Xie:2015,Chan:2016,Zhao:2016b,Li:2016,Singh:2018,Wang:2017,Yu:2017,Takahashi:2017,Gong:2018,Yi:2018,Wu:2018b,Lian:2019,Chang:2017b,Feng:2017,Schoop:2016}, thus allowing their explanation using mirror eigenvalues of the bands. For nodal class AI, however, breaking the mirror symmetry does \emph{not} remove such NLs from the band structure, but just releases them to roam freely throughout the whole BZ~\cite{Fang:2015,Huang:2016,Xu:2017,Li:2016,Nomura:2018,Ahn:2018,Wu:2018b,Lian:2019}. This freedom is especially apparent for the case of NL-links and NL-knots~\cite{Chang:2017,Hearing:1958,Yang:2019,Chen:2017,Sun:2017,Ezawa:2017,Bi:2017}. It is therefore preferred to explain the stability of the NLs by referring only to the $\mcP\mcT$ symmetry.

\subsection{Leitmotif of topological classifications}~\label{eqn:topo-classes}

Topological classifications of band structures are based on considering the Hamiltonian $H(\bs{k})$ in some region $\mathcal{R}$ inside the Brillouin zone (BZ). Conventionally, one partitions the $N$ bands into $n$ lower-energy (``occupied'') ones and $\ell = N-n$ higher-energy (''unoccupied'') ones~\footnote{We remark that we apply the terminology of occupied vs.~unoccupied bands even if it does not reflect the actual occupancy of the bands by electrons, \emph{e.g.}~as a result of a complicated band dispersion.} Furthermore, it is important that the region $\mathcal{R}$ exhibits an energy gap between the highest occupied (HO) and the lowest unoccupied (LU) band, \emph{i.e.}~for each $\bs{k}\in\mathcal{R}$ we require the validity of a \emph{sharp} inequality $\eps_{\bs{k}}^{n} < \eps_{\bs{k}}^{n+1}$. The \emph{tenfold way} classification of topological insulators and superconductors is based on choosing $\mathcal{R} \!=\! \textrm{BZ}$~\cite{Schnyder:2008,Kitaev:2009,Ryu:2010}, while topologically stable band degeneracies (\emph{i.e.}~\emph{nodes}) are revealed by considering $p$-spheres $\mathcal{R} \!\simeq\! \mathcal{S}^p$ inside BZ~\cite{Bzdusek:2017}. The central question of every topological classification is whether the Hamiltonian $H(\bs{k})$ on $\mathcal{R}$ can be continuously deformed into the \emph{atomic limit} (\emph{i.e.}~to a Hamiltonian $H_0$ \emph{constant} on $\mathcal{R}$), while preserving the spectral gap on $\mathcal{R}$ during the deformation. For various situations, such questions have been mathematically formalized using $K$-theory and Clifford algebras~\cite{Horava:2005,Kitaev:2009,Ryu:2010,Zhao:2013,Kobayashi:2014,Chiu:2014,Shiozaki:2014,Zhao:2016}, \TB{via eigenvalue arguments and using combinatorics of irreducible representations of little groups}~\cite{Kim:2015,Wang:2016,Bzdusek:2016,Michel:1999,Slager:2013,Po:2017,Kruthoff:2017,Bradlyn:2017,Watanabe:2017,Bouhon:2017,Bouhon:2018}, and using homotopy groups~\cite{Bzdusek:2017,Sun:2018,Wu:2018b,Fang:2015,Sun:2019,Wojcik:2019,Li:2019}.

In this work, we adopt the homotopy approach since it is well adapted for detecting band-structure nodes at generic (\emph{i.e.}~low-symmmetry) positions~\cite{Bzdusek:2017}. Loosely speaking, the \emph{$p$\textsuperscript{th} homotopy group} $\pi_p(M)$ of space $M$ lists the equivalence classes $[f]$ of continuous functions $f: S^p \to M$. The equivalence class $[f]$ includes all the functions that can be obtained using continuous deformations of $f$. Clearly, one always finds the equivalence class of the constant function on $S^p$. By \emph{nontrivial} homotopy group we mean that there is at least one function $f: S^p \to M$ that cannot be continuously deformed into (\emph{i.e.}~which is not equivalent to) a constant function. To see how to define \emph{group structure} on the set of equivalence classes $[f]\in\pi_p(M)$, see the discussion in Ref.~\cite{Hatcher:2002}, or in the supplementary materials of Refs.~\cite{Sun:2018,Wu:2018b}.

\begin{table}[b]
	\begin{tabular}{|c|cc|}
	\hline
	$\pi_1[M^{\textrm{AI}^{\phantom{\int}}}_{(n,\ell)}]$	& $n=1$						& $n\geq 2$					\\ \hline
	$\ell=1$					& ${\intg}$				& $\ztwo$ 	\\
	$\ell\geq 2$				& ${\ztwo}$		& ${\ztwo}$		\\ \hline	
	\end{tabular}\hspace{0.3cm}
	\begin{tabular}{|c|ccc|}
	\hline
	$\pi_2[M^{\textrm{AI}^{\phantom{\int}}}_{(n,\ell)}]$	& $n=1$				& $n=2$						& $n\!\geq\!3$ 				\\ \hline
	$\ell=1$					& ${\triv}$		& ${2\intg}$ 			& ${\triv}$    		\\
	$\ell=2$					& ${2\intg}$	& ${\intg\!\oplus\!\intg}$	& ${\intg}$			\\
	$\ell\geq 3$				& ${\triv}$		& ${\intg}$				& ${\ztwo}$	\\ \hline	
	\end{tabular}
	\caption{The first (left) and the second (right) homotopy groups of the space $M^{\textrm{AI}}_{(n,\ell)}$ of Hamiltonians of the form in Eq.~(\ref{eqn:projected-Ham}) with $n$ occupied and $\ell$ unoccupied bands. The first homotopy group captures the  Berry-phase quantization on closed paths to $0$ vs.~$\pi$, while the second homotopy group describes the monopole charge. The results for $n,\ell\to\infty$ are usually called the \emph{stable limit} (tables adapted from Ref.~\cite{Bzdusek:2017}). 
	}
	\label{tab:AI-homotopy}
\end{table}

When describing band-structure nodes, we take $M$ to be the \emph{space of Hamiltonians} subject to some specified constraints (such as symmetry, number of bands, and spectrum degeneracies). Functions $f$ which we want to classify correspond to compositions $H \circ \iota$, where $\iota: S^p \hookrightarrow \textrm{BZ}$ is an embedding (\emph{i.e.}~a selected realization) of $S^p$ inside $\textrm{BZ}$, while $H:\textrm{BZ}\to M$ is the Hamiltonian of the system. It is assumed that the embedding $\iota(S^p)\equiv \mathcal{S}^p$ (which we still call \emph{sphere} for simplicity) does not wind around $\textrm{BZ}$, and that $H$ has a spectral gap on $\mathcal{S}^p$. For $p=1$, we sometimes write $\mathcal{S}^1 \equiv \Gamma$ to emphasize that one-dimensional spheres (\emph{i.e.}~circles) can be interpreted as \emph{closed paths}.

Importantly, if $H(\mathcal{S}^p)$ corresponds to a non-trivial element of $\pi_p(M)$, then the sphere must enclose a node~\cite{Fang:2015,Bzdusek:2017,Sun:2018}. The reason is that if the interior $\mathcal{D}^{p+1}$ of the sphere (a contractible region of $\textrm{BZ}$ with boundary $\partial \mathcal{D}^{p+1} = \mathcal{S}^p$) did not contain a node, then $\mathcal{S}^p$ could be shrunk to a point without encountering a closing of the band gap, thus permitting a continuous deformation of $H(\mathcal{S}^p)$ into a constant. If the sphere $\mathcal{S}^p$ encloses just a single node, then we say that the node carries a \emph{topological charge} corresponding to the equivalence class of $H(\mathcal{S}^p)$ inside the group $\pi_p(M)$. 

\subsection{Topological stability of nodal lines}~\label{NLs-first-homotopy}

\TB{Following the previous work of Refs.~\cite{Fang:2015,Bzdusek:2017}, we now} apply the homotopic reasoning to nodal class $\textrm{AI}$. Since conventional topological classifications require us to preserve only the energy gap between the HO and LU bands, it is customary to \emph{spectrally flatten} the energy of the occupied (unoccupied) bands of $H(\bs{k})$ to $-1$ ($+1$). This can be achieved without closing the energy gap by continuously retracting $\eps^j_{\bs{k}}$ to $\sign[\eps^j_{\bs{k}}]$. We thus narrow our attention to Hamiltonians of the form
\begin{equation}
{\mcH}(\bs{k}) = -\sum_{j=1}^n | u^j_{\bs{k}}\rangle \langle u^j_{\bs{k}}| \, + \!\!\!\sum_{j=n+1}^{n+\ell}\!\!\! | u^j_{\bs{k}}\rangle \langle u^j_{\bs{k}}|\label{eqn:projected-Ham}
\end{equation}
where the calligraphic font indicates the spectral flattening. Taking into account the reality condition, Hamiltonians of the form in Eq.~(\ref{eqn:projected-Ham}) constitute an $(n\ell)$-dimensional space~\cite{Kitaev:2009} (called the \emph{real Grassmannian})
\begin{equation}
M^{\textrm{AI}}_{(n,\ell)} = \mathsf{O}(n+\ell)/\mathsf{O}(n)\!\times\!\mathsf{O}(\ell),\label{eqn:real-grass}
\end{equation}
where the orthogonal $\mathsf{O}(n\!+\!\ell)$ group indicates the space of all possible collections $\{|{u^j_{\bs{k}}}\rangle\}_{j=1}^{n+\ell}\equiv \mathcal{U}_{\bs{k}}$ of real-gauged eigenstates, while the factors $\mathsf{O}(n)$ and $\mathsf{O}(\ell)$ correspond to internal rotations (\TB{\emph{i.e.}~a gauge degree of freedom}) among the occupied resp.~among the unoccupied states, which leave ${\mcH}(\bs{k})$ invariant. 

The spaces in Eq.~(\ref{eqn:real-grass}) have a non-trivial first homotopy group $\pi_1[M^{\textrm{AI}}_{(n,\ell)}]$, shown in left Table~\ref{tab:AI-homotopy}, implying that robust singularities of $\mcH(\bs{k})$ can be enclosed by \emph{circles}. In 3D such singularities constitute 1D \emph{lines}. Importantly, singularities of $\mcH(\bs{k})$ in Eq.~(\ref{eqn:projected-Ham}) correspond precisely to the occurrences of a band degeneracy between the HO and LU bands, where the absence of the gap renders the spectral projection impossible~\cite{Bzdusek:2017}. Therefore, nodal class $\textrm{AI}$ in 3D supports robust nodal \emph{lines}, consistent with the codimension argument \TB{reviewed in Sec.~\ref{subsec:codims}}.

In the \emph{stable limit} ($n,\ell\to\infty$), the topological classification on circles is $\ztwo$-valued~\cite{Hatcher:2003}, and corresponds to the quantization of Berry phase on closed paths to $0$ vs.~$\pi$~\cite{Fang:2015}. The non-trivial value of the topological invariant is physically manifested by the appearance of \emph{drumhead states}~\cite{Heikkila:2011,Yu:2015} on the material surface. Mathematically, the $\ztwo$ invariant corresponds to the \emph{first Stiefel-Whitney class}~\cite{Ahn:2018,Nakahara:2003} of the real vector bundle spanned by the occupied states on the circle. It states whether the real vector bundle can (resp.~cannot) be written in a \emph{global} continuous gauge (akin to the dichotomy between an ordinary strip and a M\"{o}bius strip).

\subsection{Monopole charge of nodal-line rings}~\label{NLs-second-homotopy}

The spaces in Eq.~(\ref{eqn:real-grass}) have also a non-trivial \emph{second} homotopy group $\pi_2[M^{\textrm{AI}}_{(n,\ell)}]$, listed in right Table~\ref{tab:AI-homotopy}. This result implies robust nodal objects in 3D that can be enclosed by \emph{two}-dimensional spheres $\mathcal{S}^2$~\cite{Bzdusek:2017}. Since the codimension of generic nodes is fixed by $\mcP\mcT$ to $\delta_\textrm{2D deg.} =2$~\cite{vonNeumann:1929}, these nodal objects generically constitute NL-\emph{rings}~\cite{Zhao:2017}. Importantly, if ${\mcH}(\mathcal{S}^2)$ is non-trivial on a sphere enclosing a NL-ring, then the NL-ring cannot disappear from the inside of the $\mathcal{S}^2$ under continuous deformations of $H(\bs{k})$ which preserve the gap on the $\mathcal{S}^2$. This obstruction is called the \emph{monopole charge}~\cite{Fang:2015}. NL-rings that carry a monopole charge can be removed from the band structure only through a pairwise annihilation with a NL-ring carrying the opposite value of the charge, thus enhancing their robustness. In contrast, a NL-ring \emph{without} a monopole charge can be trivially removed on its own by shrinking. Non-trivial values of the monopole charge have been related to the appearance of anomalous Hall effect under $\mcP$-breaking perturbations~\cite{Rui:2018}, and to the appearance of axion insulator~\cite{Wilczek:1987,Qi:2008} under $\mcT$-breaking perturbations~\cite{Ahn:2018}.  Although the monopole charge in the stable limit is $\ztwo$-valued~\cite{Fang:2015,Hatcher:2003}, several interesting exceptions occur in few-band models~\cite{Bzdusek:2017}.

Mathematically, the $\ztwo$-valued stable limit of the monopole charge corresponds to the \emph{second Stiefel-Whitney class}~\cite{Nakahara:2003} of the real vector bundle spanned by the occupied bands on the $\mathcal{S}^2$~\cite{Ahn:2018}. It expresses whether the vector bundle can be consistently lifted into a \emph{spinor} bundle. Furthermore, a fragile $\intg$-valued monopole charge(s) arises in few-band models, which corresponds to the \emph{Euler class} of the vector bundle spanned by the two occupied (or the two unoccupied) bands~\cite{Ahn:2018}. With these identifications, Ref.~\cite{Ahn:2018} related the monopole charge of a NL-ring to its \emph{linking}~\cite{Gauss:1833} with NLs formed inside \emph{other} band gaps. More precisely, it was found that the \TB{stable} $\ztwo$ monopole charge of a NL-ring corresponds to the parity of the \TB{linking number of the ring with NLs} formed between the \emph{two highest occupied} bands (or equivalently, with NLs formed between the \emph{two lowest unoccuppied} bands). For further discussion of these methods, and how they are related to  the fragile-topology of twisted bilayer graphene near the magic angle, we refer the readers to \TB{the very pioneering work of} Refs.~\cite{Ahn:2018,Ahn:2018b}.

\subsection{Non-Abelian topology in multi-band models}~\label{sec:non-Abelian-review}

To handle statements about multiple NLs formed inside \emph{various} band gaps, it is necessary to generalize the ``conventional'' classification methods outlined in Sec.~\ref{eqn:topo-classes}. More specifically, if we want to preserve the information about NLs formed between all possible pairs of consecutive bands, we need to modify the spectral projection of $H(\bs{k})$ in a way that preserves the spectral gap between \emph{each} pair of bands. Such a generalized topological description was very recently developed by Ref.~\cite{Wu:2018b}, and here we summarize some of its key ingredients. A more practical summary is listed by rules (1--4) below, \TB{which we properly justify in the next Sec.~\ref{sec:non-Abelian-NLs}.}

In the presence of $n+\ell\equiv N$ bands, we \emph{spectrally normalize} Hamiltonians $H(\bs{k})$ as
\begin{equation}
\mathscr{H}(\bs{k}) = \sum_{j=1}^N | u^j_{\bs{k}}\rangle {\left(j - n - \tfrac{1}{2}\right)} \langle  u^j_{\bs{k}}|,\label{eqn:H-mulit-project}
\end{equation}
\emph{i.e.}~we set the energy of the $j^\textrm{th}$ band to $\eps_j = j - n - \tfrac{1}{2}$, where $n$ is the number of occupied bands. The script font epmhasizes that $\mathscr{H}(\bs{k})$ is obtained by different spectral projection than $\mcH(\bs{k})$ in Eq.~(\ref{eqn:projected-Ham}). We call the gap between bands with energies $j - n \pm \tfrac{1}{2} \equiv G\pm\tfrac{1}{2}$ as \emph{the $G^\textrm{th}$ band gap}~\cite{Chen:2018}. If the original Hamiltonian in Eq.~(\ref{eqn:spect-decom-no-k}) has $\eps_j = \eps_{j+1}$ for some $j$, the spectral normalization in Eq.~(\ref{eqn:H-mulit-project}) is unattainable. We say that the band gap $j - n \equiv G$ is \emph{closed}, and we call the manifold $\textrm{L}_G \subset \textrm{BZ}$ where such a condition occurs as \emph{nodal line in $G^\textrm{th}$ band gap} (or more briefly as \emph{nodal line $G$}). In this notation, NLs formed among the occupied (unoccupied) bands are characterized by $G < 0$ (by $G > 0$), while we have $G=0$ for NLs formed between the HO and LU bands. 

The space of  Hamiltonians of the form in Eq.~(\ref{eqn:H-mulit-project}) is
\begin{equation}
M^{\textrm{AI}}_{N} = \mathsf{O}(N)/\mathsf{O}(1)^{\times N}.
\label{eqn:M-AI-N}
\end{equation}
This is a straightforward generalization of Eq.~(\ref{eqn:real-grass}), \emph{i.e.}~$\mathsf{O}(N)$ still indicates the space of all ordered eigenstate combinations $\{ | u^j_{\bs{k}}\rangle \}_{j=1}^N \equiv \mathcal{U}_{\bs{k}}$. The quotient $\mathsf{O}(1)^{\times N} \cong \ztwo^{\,N}$ corresponds to flipping the orientation of some of the eigenstates, which keeps the form of $\mathscr{H}(\bs{k})$ invariant. Geometrically, $\mathsf{O}(1)^{\times N}$ is the point-symmetry group of a generic $N$-dimensional ellipsoid, generated by the reflections along the $N$ principal axes.

The work of Ref.~\cite{Wu:2018b} showed that the first two homotopy groups of the space in Eq.~(\ref{eqn:M-AI-N}) are
\begin{subequations}\label{eqn:Salingaros-def}
\begin{eqnarray}
\pi_1(M^{\textrm{AI}}_{N}) &=& \overline{\mathsf{P}}_N\qquad\textrm{(assuming $N \geq 3$)} \label{eqn:AIN-1} \\
\pi_2(M^{\textrm{AI}}_{N}) &=& \triv \label{eqn:AIN-2}
\end{eqnarray}
where $\overline{\mathsf{P}}_N$ is the \emph{$N^\textrm{th}$ Salingaros group}~\cite{Salingaros:1983}, discussed in more details in the supplementary material Ref.~\cite{Wu:2018b}. For our purposes, it suffices to state that this group is \emph{non-Abelian}, implying that the topological charge cannot be expressed as a real or a complex number, but requires matrices. For the special case $N = 3$ the charge can be expressed using the multiplicative quaternion group,
\begin{equation}
\pi_1(M_3^\textrm{AI}) = \mathsf{Q} = \{\pm 1,\pm\imi,\pm\imj,\pm\imk\},\label{eqn:quats}
\end{equation}
\end{subequations}
where ${\imi},\imj,\imk$ are three anticommuting imaginary units. In the discussion below, we indicate the composition of the group elements with ``$\cdot$'' (we call it \emph{multiplication}), and we write the identity element as $+1$. 

For general $N \geq 3$, the group $\overline{\mathsf{P}}_N$ contains a unique element $-1 \neq +1$ with the property $(-1)^2 = +1$. The final piece of knowledge important for our analysis is the correspondence between the elements of $\overline{\mathsf{P}}_N$, and the species of NLs encircled by closed paths inside BZ. It was found by Ref.~\cite{Wu:2018b} that
\begin{subequations}\label{eqn:node-charge-correspondence}
\begin{equation}
\textrm{NL in $G^\textrm{th}$ band gap}\quad\sim \quad \pm g_G\label{eqn:AI-N-conj-rel},
\end{equation}
\emph{i.e.}~NLs formed inside different band gaps are described by $N$ distinct elements $g_G$, which can be understood as the generators of $\overline{\mathsf{P}}_N$. The $\pm$ sign in Eq.~(\ref{eqn:AI-N-conj-rel}) allows us to assign each NL an \emph{orientation}. Importantly,
\begin{equation}
\forall G: \qquad g_G \cdot g_{G+1} = -g_{G+1} \cdot g_G,\label{eqn:consec-anticom}
\end{equation}
\emph{i.e.}~a pair of NLs occurring inside \emph{consecutive} band gaps carry \emph{anticommuting} topological charges, which implies orientation reversal under a pairwise exchange (see Sec.~\ref{sec:orientations}). On the other hand,
\begin{equation}
\forall G' \neq G \pm 1: \qquad g_{G'} \cdot g_G =  g_G \cdot g_{G'},\label{eqn:comut-charges}
\end{equation}
meaning that a pair of NLs formed inside the same gap or inside more distant band gaps have \emph{commuting} topological charges. Curiously,
\begin{equation}
\forall j: g_G^2 = -1, \quad\textrm{while}\quad g_G^4 = +1\label{eqn:i4-trivial}
\end{equation}
\end{subequations}
which we clarify in Sec.~\ref{sec:four-are-trivial}. Products of the form $\pm g_\alpha \cdot g_\beta \cdot \ldots \cdot g_\gamma$ correspond to paths in BZ that enclose NLs with indices $G = \alpha,\beta,\ldots ,\gamma$. The topological obstruction corresponding to $-1$ is special, because it survives in the stable limit of many bands, yet it is not revealed by the Berry phases of the individual bands~\footnote{\TB{The stable topology of a real vector bundle on a closed path is given by $\widetilde{KO}(S^1)=\ztwo$~\cite{Hatcher:2003}. Considering such an invariant for each one of the $N$ non-degenerate bands would suggest a $\ztwo^{N-1}$ topology (one $\ztwo$ substracted because the total bundle is trivial). However, such a description lacks the information about the band ordering. An \emph{ordered} set of (unoriented) vectors is known in the mathematical literature as \emph{the complete flag}~\cite{Bouhon:2020}. The topology of a flag bundle on a closed path is described by Eqs.~(\ref{eqn:node-charge-correspondence}). The element ``$-1$'' can be related to a $2\pi$-rotation of the frame spanned by the eigenstates~\cite{Bouhon:2019}.}}. For further discussion of the non-Abelian topological charge, we refer the readers to the supplemental material of Ref.~\cite{Wu:2018b}.

While the mathematical arguments leading to Eqs.~(\ref{eqn:M-AI-N}--\ref{eqn:node-charge-correspondence}) are somewhat technical, the final conclusions are remarkably simple, namely:
\begin{enumerate}
\item NLs formed inside multiple band gaps can be described simultaneously. In such a description, the charge of NLs forms a multiplicative group, of which any two elements commute or anticommute.
\item Each NL segment can be assigned an \emph{orientation}.
\item A pair of NLs formed inside \emph{consecutive} band gaps have topological charges that \emph{anticommute}. This implies that the orientation of one of these NLs is \emph{reversed} when they are pairwise exchanged. 
\item A pair of NLs formed inside more distant band gaps or inside the same band gap (\emph{i.e.}~with $\abs{G-G'}\neq 1$) have \emph{commuting} topological charges.
\end{enumerate}
With these rules, the study of NL compositions translates into a game of reversing orientations of linked lines of various colors. Especially, the rules imply
\begin{itemize}
\item[(\emph{i})] a path-dependent ability of a pair of NLs to annihiliate (\emph{i.e.}~non-trivial ``braiding'' of NLs in $\bs{k}$-space),
\item[(\emph{ii})] constraints on admissible NL compositions, and 
\item[(\emph{iii})] that a pair of NLs formed inside consecutive band gaps cannot move across each other 
\end{itemize}
which were noticed by Ref.~\cite{Wu:2018b}. In Sec.~\ref{sec:non-Abelian-NLs} we justify the rules (1--4), and we study their further implications for NL compositions. In the remaining Secs.~\ref{sec:monopole-linking-relation} and~\ref{sec:NA-braiding} we apply the derived rules to study properties of the monopole charge by invoking purely geometric arguments.

\section{Wielding the non-Abelian charge}\label{sec:non-Abelian-NLs}

\subsection{Overview}\label{sec:non-Abelian-overview}

In this section, we develop several simple rules for an efficient analysis of NL compositions through the prism of the non-Abelian topological charge defined by Eqs.~(\ref{eqn:Salingaros-def}) and~(\ref{eqn:node-charge-correspondence}). Our discussion is based on, \TB{and further expands,} the supplementary information file of Ref.~\cite{Wu:2018b}. The methods developed here are employed in Sec.~\ref{sec:monopole-linking-relation} and Sec.~\ref{sec:NA-braiding} to analyze the meaning and the braiding properties of the monopole charge, listed in the right Table~\ref{tab:AI-homotopy}. 

This section is organized as follows. In Sec.~\ref{sec:nonuniqueness} we demonstrate certain non-uniqueness of the non-Abelian topological charge of NLs. We overcome this ambiguity in Sec.~\ref{sec:orientations}, where we present a simple convention that allows us to assign each NL segment in an arbitrarily convoluted NL composition a \emph{unique} topological charge and orientation. We continue in Sec.~\ref{sec:crossings} by demonstrating that NLs carrying non-commuting charges \emph{cannot move across each other}~\cite{Poenaru:1977}. In Sec.~\ref{sec:compositions} we show that the non-Abelian topological charge implies constraints on the admissible NL compositions. Finally, in Sec.~\ref{sec:four-are-trivial} we clarify the meaning of Eq.~(\ref{eqn:i4-trivial}), \emph{i.e.}~why four NLs of the same type and orientation do not present a robust topological obstruction.

\begin{figure}[t!]
\includegraphics[width=0.35 \textwidth]{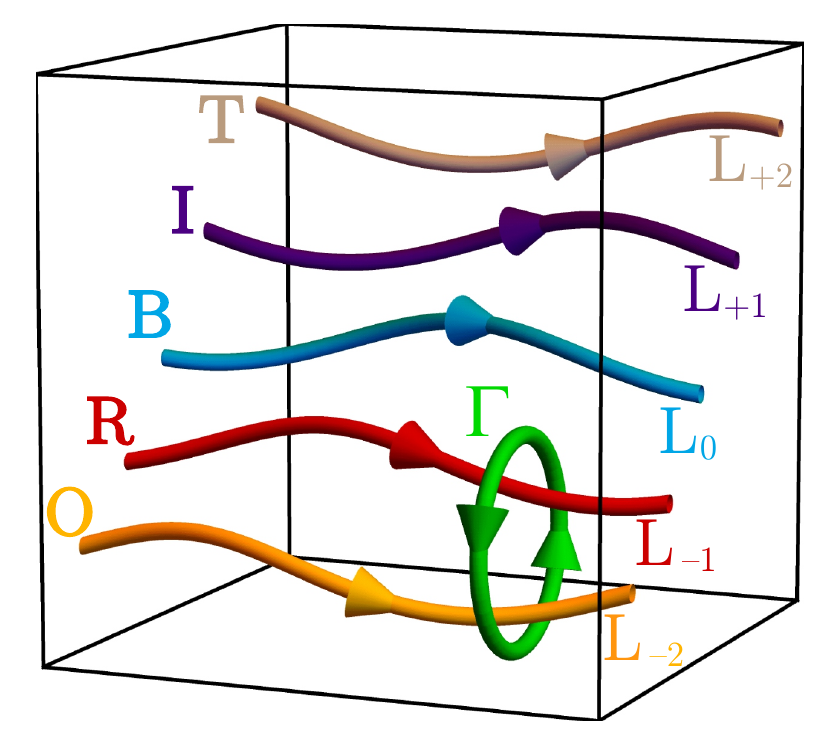}
 \caption{The black frame indicates a region of $\bs{k}$-space, and the wavy lines $\textrm{L}_G$ with $G\in\{-2,-1,0,+1,+2\}$  indicate NLs formed inside band gap $G$. In the online version, we use color scheme ``\textbf{\color{MyOrange}O}\textbf{\color{MyRed}R}\textbf{\color{MyBlue}B}\textbf{\color{MyIndigo}I}\textbf{\color{MyTan}T}'' (acronym for \textbf{\color{MyOrange}Orange}, \textbf{\color{MyRed}Red}, \textbf{\color{MyBlue}Blue}, \textbf{\color{MyIndigo}Indigo}, \textbf{\color{MyTan}Tan}) to plot NLs occurring inside the five consecutive band gaps. Closed paths ${\color{MyGreen}\Gamma}$ on which we evaluate the non-Abelian topological charge are plotted in {\color{MyGreen}green}. 
 }
\label{fig:code}
\end{figure}

Throughout the manuscript, we explicitly consider models with up to $3+3 = 6$ bands, when the $\ztwo$ stable limit of the monopole charge is reached. In that extreme case, five different species of NLs need to be considered (corresponding to $G\in\{-2,-1,0,+1,+2\}$). To make our illustrations of multi-band NL compositions more transparent, we color NLs inside the five consecutive band gaps according to a mnemonic code ``\textbf{\color{MyOrange}O}\textbf{\color{MyRed}R}\textbf{\color{MyBlue}B}\textbf{\color{MyIndigo}I}\textbf{\color{MyTan}T}'' (\textbf{\color{MyOrange}Orange}, \textbf{\color{MyRed}Red}, \textbf{\color{MyBlue}Blue}, \textbf{\color{MyIndigo}Indigo}, \textbf{\color{MyTan}Tan}) in the online version of the manuscript. Especially, NLs formed between the HO and LU bands (\emph{i.e.}~near the Fermi level, $G=0$) are always shown in {\color{MyBlue}blue}. In the print version, the species of NLs are indicated by the subscript $G$ of their labels $\textrm{L}_G$. We use shades of \textbf{\color{MyGreen}green} and symbol ${\color{MyGreen}\Gamma}$ to indicate closed paths on which we calculate the non-Abelian invariant.

\subsection{Non-uniqueness of the non-Abelian charge}\label{sec:nonuniqueness}

Non-Abelian first homotopy group implies that the charge $c$ of a NL is defined only up to conjugacy $ c \sim g \cdot c \cdot g^{-1}$ with other elements $g$ of the group~\cite{Mermin:1979}. To understand this, note that the \emph{group structure} of the homotopy group requires the possibility to \emph{compose} closed paths ${\color{MyGreen}\Gamma_i}$ on which we evaluate the topological charge. We should therefore consider only paths that begin and end at some fixed point in $\bs{k}$-space, called the \emph{base point} $\textrm{P}$. Under this constraint, a NL might be enclosed by several closed paths that cannot be continuously deformed into each other due to the presence of additional NLs. However, any such pair of paths can be related through a conjugacy with some other path based at $\textrm{P}$.

\begin{figure}[t!]
\includegraphics[width=0.29 \textwidth]{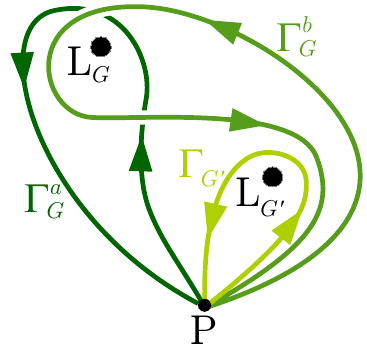}
 \caption{A point node ${\textrm{L}_G}$ inside a plane can be enclosed by paths ${\color{MyGreen}\Gamma_{G}^a}$ and ${\color{MyGreen}\Gamma_{G}^b}$, both based at point $\textrm{P}$. The two based paths cannot be continuously deformed into each other due to the presence of an additional point node ${\textrm{L}_{G'}}$. However, the two paths can be related by a conjugation with path ${\color{MyGreen}\Gamma_{G'}}$ enclosing ${\textrm{L}_{G'}}$, leading to a conjugation relation between the topological charge on ${\color{MyGreen}\Gamma_{G}^a}$ and on ${\color{MyGreen}\Gamma_{G}^b}$, cf. Eqs.~(\ref{eqn:conjuga}).}
\label{fig:conjugacy}
\end{figure}

\begin{figure*}[t!]
\includegraphics[width=0.9\textwidth]{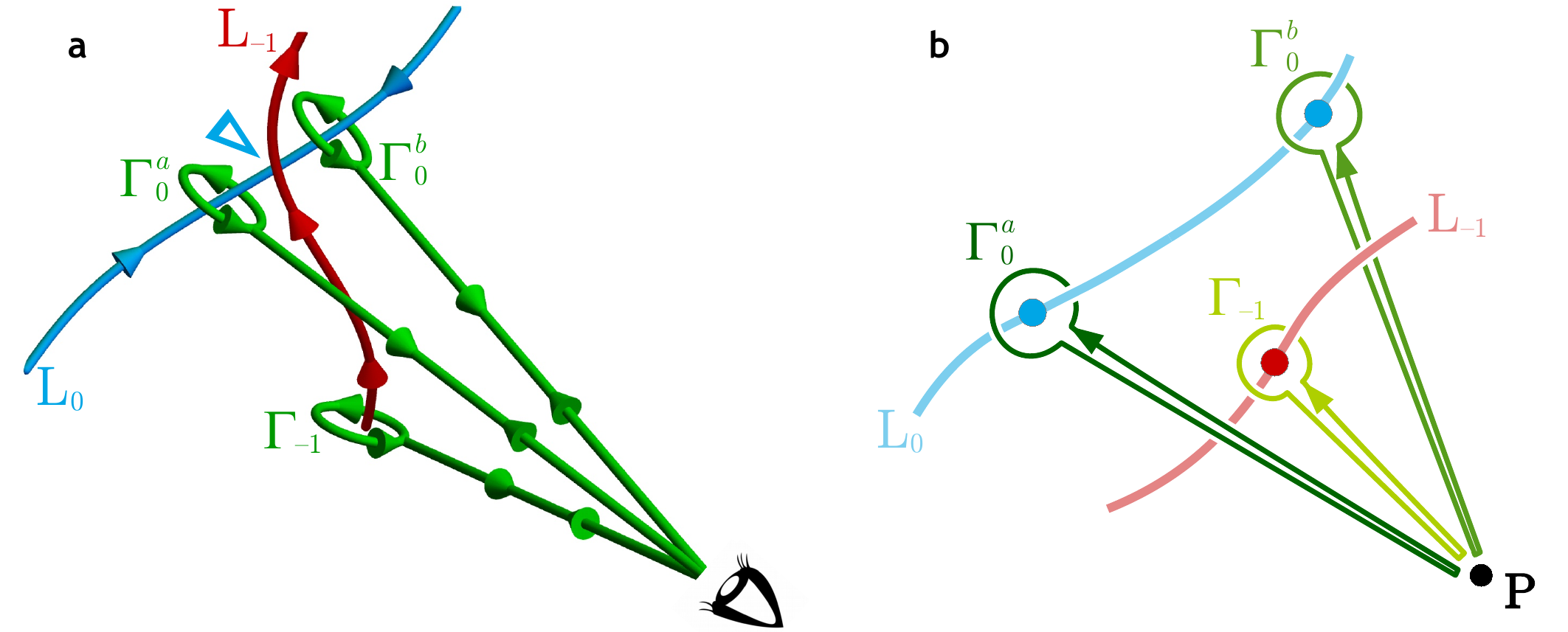}
     \caption{(a) Our convention \TB{to assign} each NL inside a 3D $\bs{k}$-space a unique orientation. We first fix a base point $\textrm{P}$ from which we observe the NL composition. To define an orientation of a chosen NL segment, we consider closed paths (green ${\color{MyGreen}\Gamma_i})$ composed of (1) a straight ray going from $\textrm{P}$ to the vicinity of the NL segment, (2) a tight loop around the NL segment, and (3) the same ray in reverse. As explained in Sec.~\ref{sec:orientations}, the NL orientation defined this way is \emph{reversed} each time we see it passing under a NL formed inside the neighboring band gap. In the illustration, the orientation of a {\color{MyBlue}blue} NL (${\color{MyBlue}\textrm{L}_0}$) is reversed when we see it passing under a {\color{MyRed}red} NL (${\color{MyRed}\textrm{L}_{-1}}$). The reversal of the NL orientation is indicated by a triangular arrowhead with the corresponding color (``${\color{MyBlue}\bs{\vartriangleright}}$''). \TB{(b) Two-dimensional projection of the situation displayed in the left panel. The colored dots along the projection of the {\color{MyBlue}blue} resp.~{\color{MyRed}red} nodal line indicate the tightly enclosed points of the nodal lines. This projection makes it clear that paths ${\color{MyGreen}\Gamma_0^a}$ and ${\color{MyGreen}\Gamma_0^b}$ are related by conjugation with path ${\color{MyGreen}\Gamma_{-1}}$, in the same way as paths ${\color{MyGreen}\Gamma_G^a}$, ${\color{MyGreen}\Gamma_G^b}$ and ${\color{MyGreen}\Gamma_{G'}}$ in Fig.~\ref{fig:conjugacy}. The centers of the tight-loop component of paths ${\color{MyGreen}\Gamma_0^a}$ and ${\color{MyGreen}\Gamma_0^b}$ here should be identified with the same point $L_{G}$ in Fig.~\ref{fig:conjugacy}.} }
\label{fig:orientations}
\end{figure*}

We illustrate this ambiguity using {point} nodes inside a 2D plane, see Fig.~\ref{fig:conjugacy}. The point node ${\textrm{L}_G}$ can be enclosed by path ${\color{MyGreen}\Gamma_{G}^a}$ as well as ${\color{MyGreen}\Gamma_{G}^b}$, both based at $\textrm{P}$. The presence of the additional point node ${\textrm{L}_{G'}}$ makes it impossible to continuously deform ${\color{MyGreen}\Gamma_{G}^a}$ into ${\color{MyGreen}\Gamma_{G}^b}$ if we keep the base point fixed. However, if we encircle ${\textrm{L}_{G'}}$ with path ${\color{MyGreen}\Gamma_{G'}}$ as shown in the figure, we find that
\begin{subequations}\label{eqn:conjuga}
\begin{equation}
{\color{MyGreen}\Gamma_{G}^b} \sim ({\color{MyGreen}\Gamma_{G'}})^{-1} \circ {\color{MyGreen}\Gamma_{G}^a} \circ {\color{MyGreen}\Gamma_{G'}}.\label{eqn:path-conjugacy}
\end{equation}
The relation ``$\sim$'' (called \emph{isotopy}) means that the composition of closed paths on the two sides of the equation can be continuously deformed into each other without encountering a band degeneracy. When composing several paths as in Eq.~(\ref{eqn:path-conjugacy}), our convention is to first move along the rightmost path in the composition, and progressively move along the paths to the left.

If we denote the representation of path ${\color{MyGreen}\Gamma}$ inside the group $\pi_1(M_N^\textrm{AI})$ as $g_{\Gamma}$, then it follows from Eq.~(\ref{eqn:path-conjugacy}) that
\begin{equation}
g_{\Gamma_G^b}^{\phantom{1}} = g_{\Gamma_{G'}}^{-1}\cdot g_{\Gamma_G^a}^{\phantom{1}}\cdot g_{\Gamma_{G'}}^{\phantom{1}},\label{eqn:group-conjugacy}
\end{equation}
\end{subequations}
which is precisely the ambiguity advertised earlier. The conjugation by all possible elements $g_{G'}$ divides the group into several \emph{conjugacy classes}. For Abelian groups, every element forms a conjugacy class of its own, therefore the path-dependence can be safely ignored. Such Abelian scenario arises for all the nodal classes in the classification of band-structure nodes developed in Ref.~\cite{Bzdusek:2017}.

However, the first homotopy group in Eq.~(\ref{eqn:AIN-1}), which describes NLs in the multi-band context, is non-Abelian. Therefore, one can uniquely assign a given NL only the whole conjugacy class of elements rather than a single group element. Owing to the (anti)commutation relations in Eq.~(\ref{eqn:node-charge-correspondence}), most conjugacy classes of this group consist of a pair of elements differing in the overall sign, \emph{i.e.}~$\{+g,-g\}$. Elements $+1$ and $-1$ form one-element conjugacy classes~\cite{Wu:2018b}. The ambiguity of the sign of the topological charge, $\pm g_G$, implies that the ability of two NL segments in $G^\textrm{th}$ band gap to annihilate depends on the path used to bring the NL segments together~\cite{Wu:2018b}.

\subsection{Defining the orientation of nodal lines}\label{sec:orientations}

We now present a simple convention that allows us to assign to each NL a unique topological charge, and thus also a unique \emph{orientation}, despite the ambiguity outlined in Sec.~\ref{sec:nonuniqueness}. First, we fix a \emph{vantage point} from which we observe a 2D projection of the NL composition. The vantage point (the observer's eye) plays the role of the base point where all the closed paths begin and end. 

We ascribe to a \emph{specific segment} of a NL a unique orientation by limiting our attention only to a special class of closed paths, illustrated in Fig.~\ref{fig:orientations}, which are composed of the following three components:
\begin{enumerate}
\item a straight ``ray'' coming from our eye to the vicinity of the NL segment, 
\item a tight loop encircling the NL segment, and 
\item the same ray in the reverse direction. 
\end{enumerate} 
When viewed from the vantage point, the motion along the straight rays is projected away. Our strategy fixes a \emph{canonical} choice of path to enclose any NL segment, thus leading to a unique topological charge and orientation of each component of the NL composition.

\begin{figure*}[t!]
\includegraphics[width=0.98 \textwidth]{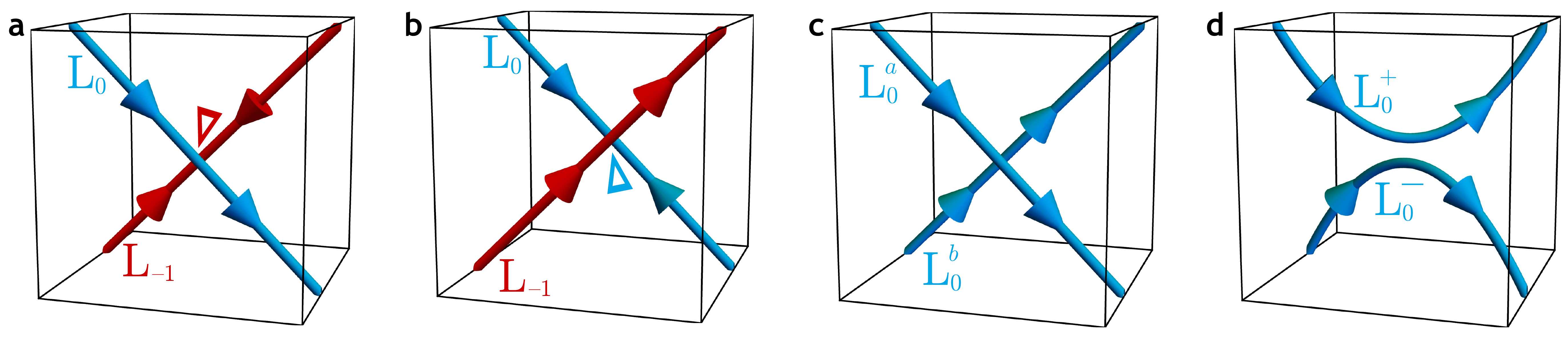}
 \caption{(a) A red NL ${\color{MyRed}\textrm{L}_{-1}}$ passing \emph{below} a blue NL ${\color{MyBlue}\textrm{L}_0}$ inside some region (the cube) of $\bs{k}$-space. The orientation of ${\color{MyBlue}\textrm{L}_0}$ is constant, while the orientation of ${\color{MyRed}\textrm{L}_{-1}}$ is reversed at the overlay (indicated by ``${\color{MyRed}\bs{\vartriangleright}}$''). (b) A red NL ${\color{MyRed}\textrm{L}_{-1}}$ passing \emph{over} a blue NL ${\color{MyBlue}\textrm{L}_0}$. In this case, the orientation of ${\color{MyRed}\textrm{L}_{-1}}$ is the same everywhere, while the orientation of ${\color{MyBlue}\textrm{L}_0}$ is flipped at the overlay (indicated by ``${\color{MyBlue}\bs{\vartriangleright}}$''). The situation (a) cannot be continuously evolved into the situation (b) because the orientations of the four outgoing NL segments do not match. \TB{(c) Analogous situation with two NLs of the \emph{same} color (here both {\color{MyBlue}blue}). Since their topological charges commute, cf.~Eq.~(\ref{eqn:comut-charges}) for $G=G'$, there is no orientation reversal at the overlay. This allows the two NLs ${\color{MyBlue}\textrm{L}_0^a}$ and ${\color{MyBlue}\textrm{L}_0^b}$ to freely move across each other. Furthermore, (d) near the critical points there is also a third phase, in which the two NLs of the same color \emph{reconnect} into ${\color{MyBlue}\textrm{L}_0^+}$ and ${\color{MyBlue}\textrm{L}_0^-}$. The three phases near the critical point are analogous to Fig.~1(b) of Ref.~\cite{Yang:2019}.}}
\label{fig:crossing}
\end{figure*}

An important and non-trivial consequence of our convention is that the orientation of a NL is reversed each time we see it passing \emph{under} a NL formed inside a neighboring band gap. To understand such orientation reversals, consider a situation, plotted in Fig.~\ref{fig:orientations}(a), where a {\color{MyRed}red} NL (${\color{MyRed}\textrm{L}_{-1}}$) passes over a {\color{MyBlue}blue} NL (${\color{MyBlue}\textrm{L}_0}$), and take two locations along ${\color{MyBlue}\textrm{L}_0}$ which are located on the opposite sides (as seen from the vantage point) of ${\color{MyRed}\textrm{L}_{-1}}$. The charge of ${\color{MyBlue}\textrm{L}_0}$ at these two locations would correspond to the homotopy classes of the Hamiltonian on paths ${\color{MyGreen}\Gamma_{0}^a}$ resp.~${\color{MyGreen}\Gamma_{0}^b}$. 

It follows from the discussion in Sec.~\ref{sec:nonuniqueness} (\TB{compare Fig.~\ref{fig:orientations}(b) to Fig.~\ref{fig:conjugacy}}) that the topological charges on the two paths are related by a conjugacy with the topological charge on path ${\color{MyGreen}\Gamma_{\textrm{-1}}}$ characterizing ${\color{MyRed}\textrm{L}_{-1}}$. Since NLs ${\color{MyRed}\textrm{L}_{-1}}$ and ${\color{MyBlue}\textrm{L}_0}$ are formed inside \emph{consecutive} band gaps (cf.~the subscripts and the color code in Fig.~\ref{fig:code}), we obtain from the anticommutation relation in Eq.~(\ref{eqn:consec-anticom}) that paths ${\color{MyGreen}\Gamma_{0}^a}$ and~${\color{MyGreen}\Gamma_{0}^b}$ carry \emph{opposite} topological charges. In other words, the orientation of ${\color{MyBlue}\textrm{L}_0}$ is \emph{reversed} when it passes under ${\color{MyRed}\textrm{L}_{-1}}$. In Fig.~\ref{fig:orientations} we indicate the orientation reversal of the {\color{MyBlue}blue} NL by a {\color{MyBlue}blue} triangular arrowhead (``${\color{MyBlue}\bs{\vartriangleright}}$''). It follows from  Eq.~(\ref{eqn:consec-anticom}) that the \TB{same} 
argument applies to any pair of NLs formed in 
consecutive band gaps.

\subsection{Moving a pair of nodal lines across each other}\label{sec:crossings}

In this section we show that a pair of NLs formed inside two consecutive band gaps (\emph{i.e.}~when the NLs carry anticommuting topological charges) cannot move across each other. On the other hand, for all \emph{other} pairs of NLs there is no such constraint. In both cases, we first present a topological argument based on the non-Abelian topological charge. Afterwards, we provide an alternative explanation using codimension arguments that generalizes the discussion from Sec.~\ref{subsec:codims}.

We begin with the topological argument, which is based on the orientation reversals discussed in the previous section. Therefore, we fix the vantage point, and we consider our canonical choice of paths (cf.~Fig.~\ref{fig:orientations}) to define the orientation of each NL segment. We argued in Sec.~\ref{sec:orientations} that the orientation of a NL is reversed each time it goes under a NL formed inside a neighboring band gap. Here, we first consider a {\color{MyRed}red} NL (${\color{MyRed}\textrm{L}_{-1}}$) passing under a {\color{MyBlue}blue} NL (${\color{MyBlue}\textrm{L}_0}$), as shown in Fig.~\ref{fig:crossing}(a). The orientation of ${\color{MyRed}\textrm{L}_{-1}}$ is reversed at the underpass, while ${\color{MyBlue}\textrm{L}_0}$ has the same orientation everywhere. We now compare this to the situation in Fig.~\ref{fig:crossing}(b), when ${\color{MyRed}\textrm{L}_{-1}}$ passes \emph{over} ${\color{MyBlue}\textrm{L}_0}$. In this scenario, ${\color{MyRed}\textrm{L}_{-1}}$ has the same orientation everywhere, while the orientation of ${\color{MyBlue}\textrm{L}_0}$ is reversed at the underpass. Clearly, the situation in Fig.~\ref{fig:crossing}(a) cannot be continuously evolved into the situation in Fig.~\ref{fig:crossing}(b), because the orientations of the four outgoing segments of ${\color{MyRed}\textrm{L}_{-1}}$ and ${\color{MyBlue}\textrm{L}_{0}}$ do not match. The argument applies to any pair of NLs formed inside consecutive band gaps. In contrast, pairs of NLs with commuting charges do not influence the orientation of each other, such that the topological obstruction described above does not arise. Therefore, NLs with commuting charges can freely move across each other.

\TB{While the argument presented in the previous section [and illustrated by Fig.~\ref{fig:crossing}(a,b)] is rigorous, we supplement it below with an alternative (although ultimately less strong) codimension argument. This is achieved by properly} adjusting the argument by von Neumann and Wigner~\cite{vonNeumann:1929}, reviewed in Sec.~\ref{subsec:codims}. \TB{Note that} if it were possible to move a pair of NLs across each other, \TB{then the phase space should be divided into two phases by a phase-transition region of codimension one. In other words, one should be able to generically move the NLs across each other by tuning a single} parameter $p$ of the Bloch Hamiltonian $H(\bs{k};p)$, \TB{and the NLs would} \emph{intersect} when $p$ is fine-tuned to one special value. 
Fine-tuning \TB{of} $p$ comes in addition to specifying the three momentum coordinates of the intersection point. Therefore, \TB{we conclude that} for the pair of NLs to be able to move across each other, the codimension to form intersecting NLs has to be ${\delta}_\textrm{c} = 4$. 

\begin{figure*}[t!]
\includegraphics[width=0.7 \textwidth]{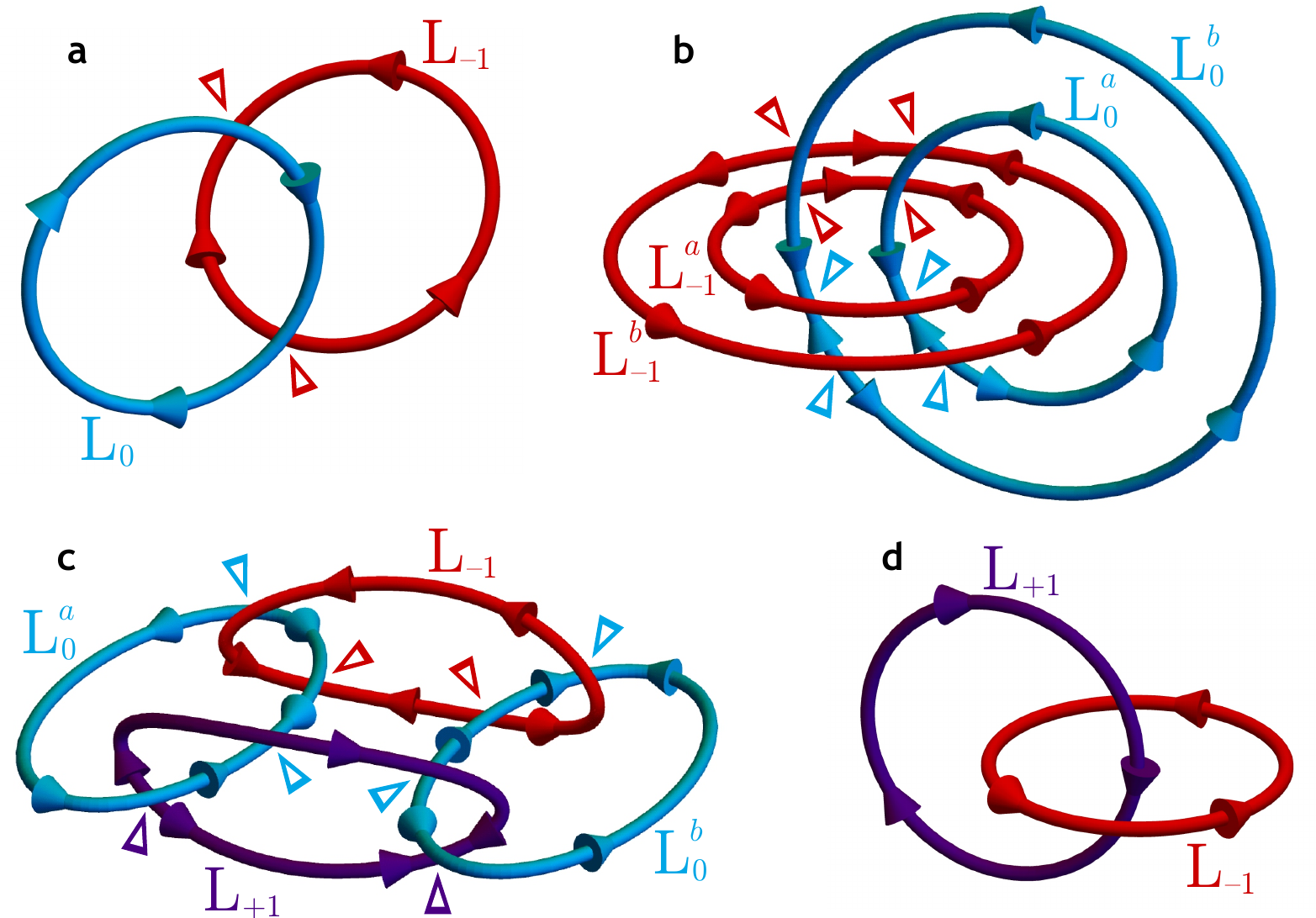}
 \caption{Four examples of NL compositions consistent with the criteria formulated in Sec.~\ref{sec:compositions}. The indexing (in the online version also the coloring) of the displayed NLs follow the scheme shown in Fig.~\ref{fig:code}. The orientation reversals of the NLs are indicated by triangular arrowheads (``${\bs{\vartriangleright}}$'') of the corresponding color. (a) A {\color{MyBlue}blue} NL-ring ${\color{MyBlue}\textrm{L}_0}$ is located \emph{above} a {\color{MyRed}red} NL-ring ${\color{MyRed}\textrm{L}_{-1}}$. Since ${\color{MyBlue}\textrm{L}_0}$ and ${\color{MyRed}\textrm{L}_{-1}}$ are formed inside consecutive band gaps, ${\color{MyRed}\textrm{L}_{-1}}$ reverses orientation each time it goes under ${\color{MyBlue}\textrm{L}_0}$. Since both NL-rings reverse orientation an even number of times ($0$ vs.~$2$), the NL composition is admissible. (b) A composition of two {\color{MyRed}red} NL-rings ${\color{MyRed}\textrm{L}_{-1}^{a,b}}$ and two {\color{MyBlue}blue} NL-rings ${\color{MyBlue}\textrm{L}_0^{a,b}}$. Each NL-ring is threaded by two NLs of the other color, leading to two orientation reversal along each NL-ring, as required by the consistency criteria. (c) A composition of one {\color{MyRed}red} NL-ring ${\color{MyRed}\textrm{L}_{-1}}$, two {\color{MyBlue}blue} NL-rings ${\color{MyBlue}\textrm{L}_0^{a,b}}$, and one {\color{MyIndigo}indigo} NL-ring ${\color{MyIndigo}\textrm{L}_{+1}}$. NL-rings ${\color{MyRed}\textrm{L}_{-1}}$ (in band gap $G=-1$) and ${\color{MyIndigo}\textrm{L}_{+1}}$ (in band gap $G=+1$) are each threaded by two {\color{MyBlue}blue} NLs ${\color{MyBlue}\textrm{L}_0^{a,b}}$ (in a neighboring band gap $G=0$), leading to two orientation reversals. On the other hand, ${\color{MyBlue}\textrm{L}_0^{a}}$ as well as ${\color{MyBlue}\textrm{L}_0^{b}}$ are each threaded by ${\color{MyRed}\textrm{L}_{-1}}$ and by ${\color{MyIndigo}\textrm{L}_{+1}}$, both inside neighboring band gaps. This implies an even number of orientation reversal of each ${\color{MyBlue}\textrm{L}_0^{a,b}}$. We find that all NL-rings in this composition reverse orientation an even number of times, implying its consistency. (d) A Hopf link formed a {\color{MyRed}red} NL-ring ${\color{MyRed}\textrm{L}_{-1}}$ and an {\color{MyIndigo}indigo} NL-ring ${\color{MyIndigo}\textrm{L}_{+1}}$. Since NLs with $G=\pm 1$ are not formed inside consecutive band gaps, they ignore the presence of each other, and no orientation reversals take place.}
\label{fig:consistent-NLCs}
\end{figure*}

We now derive the \emph{actual} codimension of such a process using Eq.~(\ref{eqn:codim-arg}). First, a pair of intersecting NLs formed inside two consecutive band gaps correspond to having a \emph{three-fold degeneracy} at the intersection point. In this case, $f= N-2$ energy levels need to be specified, while there are $v = \dim \mathsf{O}(3) = 3$ tunable parameters that do not change the Hamiltonian. The dimension of the space of Hamiltonians with a three-fold degeneracy is 
\begin{equation}
d_\textrm{3-deg.} =\tfrac{1}{2}N(N+1) - 5,
\end{equation}
from where we obtain the codimension, $\delta_\textrm{3-deg.} = d_\textrm{non-deg.} - d_\textrm{3-deg.} = 5 > \delta_\textrm{c}$. \TB{This implies that one has to tune two Hamiltonian parameters (rather than just one) to place the two nodal lines on top of each other, suggesting they cannot move across one another~\footnote{\TB{A region of codimension two may separate two phases [in our case the two nodal-line compositions in Fig.~\ref{fig:crossing}(a,b)] in a parameter space only if there is a \emph{third phase} bordering the other two phases on regions of codimension 1. However, in Fig.~\ref{fig:crossing}(a,b) there is no natural candidate for such a third phase, hence the encountered region of codimension two is \emph{not} a boundary between the two phases. This implies that the two nodal lines cannot be moved across each other.}}.}

\begin{figure*}[t!]
\includegraphics[width=0.99 \textwidth]{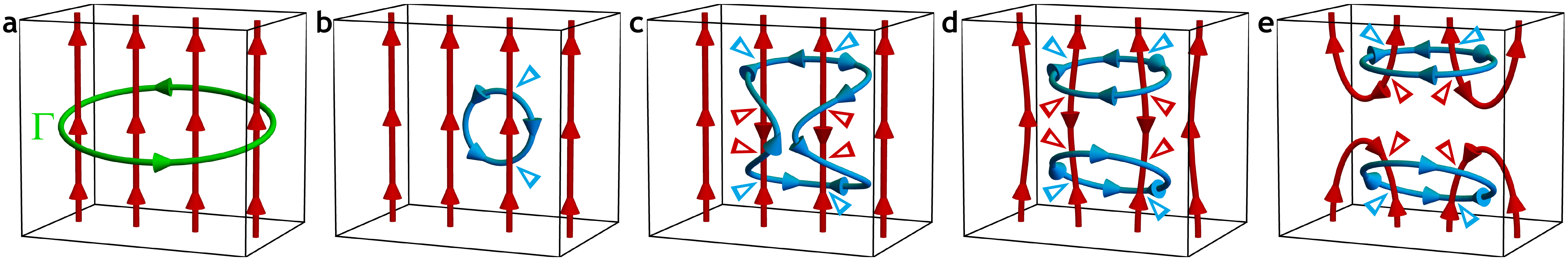}
 \caption{Manifestation of Eq.~(\ref{eqn:i4-trivial}), \emph{i.e.}~why four NLs of the same type and orientation are topologically trivial. Panels (a--e) show a process that locally annihilates segments of four parallel {\color{MyRed}red} NLs. The process involves the creation of a pair of {\color{MyBlue}blue} NL-rings located inside a neighboring band gap. For a detailed discussion of the process, see Sec.~\ref{sec:four-are-trivial}. The arrowheads (``${\bs{\vartriangleright}}$'') indicate orientation reversals of NLs of the corresponding color, as explained in Sec.~\ref{sec:orientations}.}
\label{fig:i4}
\end{figure*}

In contrast, a pair of intersecting NLs formed inside more distant band gaps correspond to forming two double degeneracies at the intersection point. In this case, one still has to specify $f = N - 2$ eigenvalues, but the number of tunable parameters that do not change the Hamiltonian is reduced to $v = \dim[\mathsf{O}(2)\times\mathsf{O}(2)] = 2$. This implies that
\begin{equation}
d_\textrm{2 $\times$ 2-deg.} =\tfrac{1}{2}N(N+1) - 4,
\end{equation}
and codimension $\delta_\textrm{2 $\times$ 2-deg.} = d_\textrm{non-deg.} - d_\textrm{2 $\times$ 2-deg.} = 4$. Since this is equal to ${\delta}_\textrm{c}$, such more distant NLs can move across each other.

We remark that it is not necessary to consider situations that exhibit several (\emph{i.e.} two or more) coinciding NLs (carrying charges $g_G^2=-1$, and higher powers of $g_G$). We describe such NLs as \emph{composite}, in contrast to the \emph{elementary} NLs with charge $g_G$. The reason that we can ignore such possibilities is that they require much fine-tuning. Therefore, if the phase diagram allows us to evolve one NL composition into another by invoking composite NLs, it is also possible to relate the two NL compositions in the phase diagram via more generic situations which involve only the elementary NLs. For this reason, in the rest of our work we safely ignore the occurrence of composite NLs.

To complete our analysis, we finally consider a pair of NLs formed inside the \emph{same} band gap. Here, the codimension analysis becomes more tricky, because apart from being located on the two sides of each other, such NLs are also allowed to reconnect, \TB{cf.~Fig.~\ref{fig:crossing}(c,d), thus} implying \emph{three} (rather than two) phases with topologically distinct NL compositions \TB{near the critical point}. Therefore, \TB{rather than} of delving into the codimension counting, we present \TB{here} an explicit two-band Hamiltonian, adapted from Ref.~\cite{Yan:2017},
\begin{equation}
\mcH(\bs{k},m) = (m_1 k_x - k_y k_z) \sigma_1 + (k_x^2 +k_y^2 - k_z^2 - m_2^2)\sigma_3,\label{eqn:two-band-example}
\end{equation}
which realizes the exchange of two NLs inside the same band gap when parameters $m_{1,2}$ are tuned. By including additional flat bands at lower and higher energies, we get an analogous model with arbitrarily many bands. Assuming first that $m_1 = m_2 \equiv m$, the model in Eq.~(\ref{eqn:two-band-example}) exhibits a pair of NLs (one at $k_x = k_z$ and $k_y = + m$, the other at $k_x = -k_z$ and $k_y = -m$) inside the same band gap, which move across each other as $m$ changes sign. If $m_1$ is not exactly equal to $m_2$, then there is an intermediate phase with a more complicated NL composition. In spite of this subtlety, the Hamiltonian in Eq.~(\ref{eqn:two-band-example}) convincingly demonstrates that a pair of NLs formed inside the same gap can move across each other.

\subsection{Admissible nodal line compositions}\label{sec:compositions}

We now show that the non-Abelian topology implies strict constraints on the admissible NL compositions. More specifically, we argue that closed NL-rings in band gap $G$ (\emph{i.e.}~with charge $\pm g_G$) can only enclose an \emph{even} number of NLs formed inside the two neighboring band gaps (\emph{i.e.}~with charges $g_{G\pm 1}$ that anticommute with $g_G$). This conclusion again follows by considering the reversals of NL orientations which we discussed in Sec.~\ref{sec:orientations}. 

Since we have adopted a convention in which the orientation of each segment of a NL is \emph{uniquely defined}, there must be an \emph{even} number of orientation reversals along every closed NL-ring. A reversal of the NL-ring orientation can occur only for the following two reasons:
\begin{enumerate}
\item There is a NL with an anticommuting charge located \emph{in front of} the considered NL-ring. Such a situation leads to an even number of orientation reversals for the NL-ring. 
\item There is a NL with anticommuting charge \emph{linked} with the considered NL ring. Such links lead to an odd number of orientation reversal for the NL-ring. 
\end{enumerate}
Therefore, for the orientation to be consistent along the whole NL-ring, it must be linked with (\emph{i.e.}~must \emph{enclose}) an even number of NLs with anticommuting charge. For example, a Hopf link formed by a pair of NLs located inside consecutive band gaps would violate such conditions, rendering it impossible. Four examples of admissible NL compositions are shown and discussed in Fig.~\ref{fig:consistent-NLCs}. 

\subsection{Four nodal lines of the same type are trivial}\label{sec:four-are-trivial}

We finally discuss the meaning of Eq.~(\ref{eqn:i4-trivial}), \emph{i.e.}~why the net charge of four NLs of the same type and orientation is trivial. We show that, indeed, segments of such four NLs can locally annihilate through a process illustrated in Fig.~\ref{fig:i4}, which involves the creation of two NL-rings inside a neighboring band gap. To see this, we consider in Fig.~\ref{fig:i4}(a) four {\color{MyRed}red} ($G=-1$) NLs pointing in the same direction, such that the green path ${\color{MyGreen}\Gamma}$ enclosing them carries the trivial topological charge $g_{-1}^4 = +1$. (The argument easily generalizes to each band gap $G$.)

By locally distorting the Hamiltonian, we create a {\color{MyBlue}blue} ($G=0$) NL-ring located \emph{behind} the {\color{MyRed}red} NLs, as shown in Fig.~\ref{fig:i4}(b). In Fig.~\ref{fig:i4}(c), we increase the size of the {\color{MyBlue}blue} NL-ring, and we fold parts of the {\color{MyBlue}blue} NL-ring such that they appear in front of the central two {\color{MyRed}red} NLs. A narrow ``neck'' of two antiparallel {\color{MyBlue}blue} NLs appears in front of the central two {\color{MyRed}red} NLs, which allows us to reconnect the {\color{MyBlue}blue} NLs as shown in Fig.~\ref{fig:i4}(d). The reconnection splits the original single {\color{MyBlue}blue} NL-ring (which did not enclose any {\color{MyRed}red} NLs) into two {\color{MyBlue}blue} NL-rings (each enclosing two {\color{MyRed}red} NLs), which can be vertically separated. Because of the non-Abelian topological charge, the orientation of the central two {\color{MyRed}red} NLs is flipped in the region between the {\color{MyBlue}blue} NL-rings, which allows us to reconnect pairs of {\color{MyRed}red} NLs as shown in Fig.~\ref{fig:i4}(e). A nodeless region appears at half the height of the indicated $\bs{k}$-space region, allowing us to shrink path ${\color{MyGreen}\Gamma}$ to a point without encountering a band degeneracy. 

With this example, we have explicitly demonstrated that four parallel NLs of the same type and orientation can locally annihilate, albeit at the cost of creating a pair of NL-rings inside a neighboring band gap. It will become apparent after the discussion in Sec.~\ref{sec:monopole-linking-relation} that these two NL-rings, which were pairwise created, carry opposite value of the monopole charge.

\section{Monopole charge from linking}\label{sec:monopole-linking-relation}

\subsection{Overview}\label{sec:linking-overview}

\TB{In this section and in the next Sec.~\ref{sec:NA-braiding}, we apply the developed geometric description of the non-Abelian topological invariant to characterize the monopole charge of NL-rings. For the purpose of the present section, recall that Ref.~\cite{Ahn:2018} used cohomology classes of vector bundles to relate the stable} $\ztwo$ monopole charge of a NL-ring to its linking structure with NLs formed inside neighboring band gaps. 
\TB{Here} we rederive the same conclusion using the non-Abelian topological charge found by Ref.~\cite{Wu:2018b}. Our analysis is purely geometrical and involves a minimal amount of equations. \TB{The presented} arguments employ the properties of NLs derived in Sec.~\ref{sec:non-Abelian-NLs}, especially that (\emph{i}) the orientation of a NL is reversed when it passes \emph{under} a NL formed inside a neighboring band gap, and that (\emph{ii}) a pair of NLs formed inside neighboring band gaps cannot move across each other. Our discussion is split into several subsections, corresponding to models with different number of occupied and of unoccupied bands, which exhibit various group structures of the monopole charge (cf.~the right Table~\ref{tab:AI-homotopy}).

\begin{figure}[b!]
\includegraphics[width=0.47\textwidth]{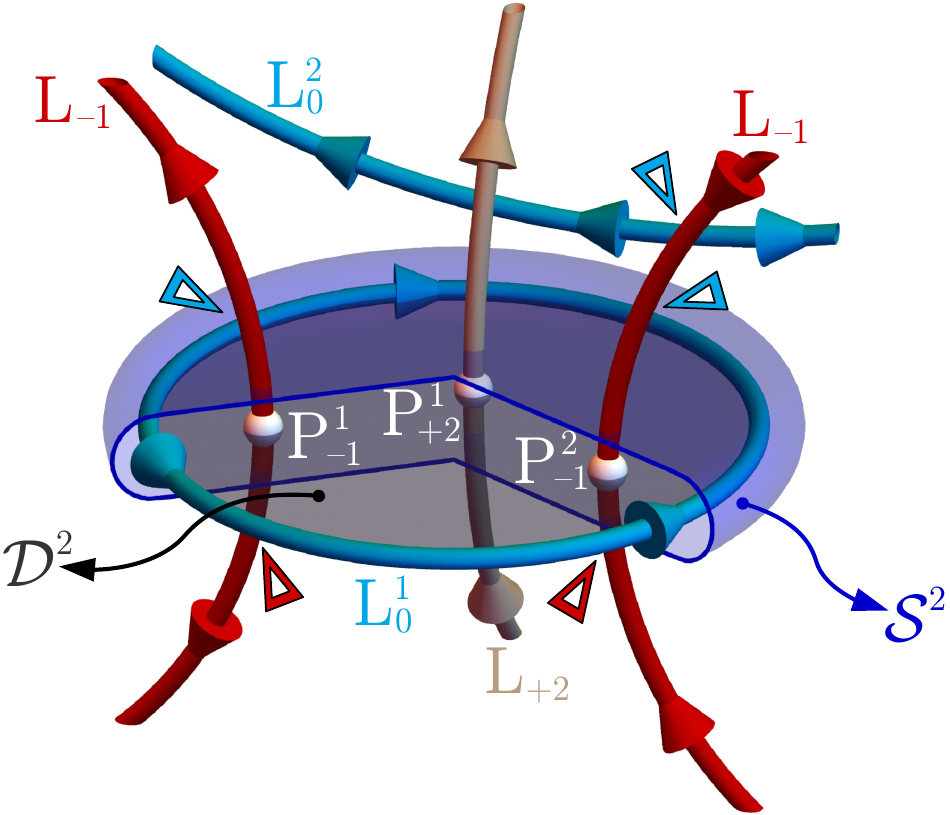}
 \caption{Summary of our notation. We consider a NL-ring ${\color{MyBlue}\textrm{L}_0^1}$ formed inside the $G=0$ band gap, which is \emph{not} linked with other NLs ${\color{MyBlue}\textrm{L}_0^{2,3,\ldots}}$ formed inside the same band gap. We find a two-dimensional disc $\mathcal{D}^2$ that is bounded by ${\color{MyBlue}\textrm{L}_0^1}$ and which is not intersected by ${\color{MyBlue}\textrm{L}_0^{2,3,\ldots}}$. Nodal lines $\textrm{L}_{G'\neq 0}$ inside other band gaps $G' \neq 0$ can be linked with ${\color{MyBlue}\textrm{L}_0^1}$, leading to intersection points $\textrm{L}_{G'} \cap \mathcal{D}^2 = \{\textrm{P}_{G'}^a\}_{a\in\mathcal{I}}$. Depending on the orientation of the NL just above the disc, we assign to each intersection point with $G' = \pm 1$ a \emph{flux} $w_{G'}^a = \pm 1$ (for the illustrated example $w_{-1}^1 = +1$, $w_{-1}^2 = -1$, and we do not define $w_{+2}^1$). To assign ${\color{MyBlue}\textrm{L}_0^1}$ a monopole charge, we inflate $\mathcal{D}^2$ into a thin ``pancake'' containing ${\color{MyBlue}\textrm{L}_0^1}$. The surface of the pancake is topologically a sphere $\mathcal{S}^2$ that preserves the $G=0$ band gap. We relate the monopole charge on $\mathcal{S}^2$ to the linking structure of ${\color{MyBlue}\textrm{L}_0^1}$ with NLs inside the neighboring band gaps. The linking structures are encoded in the fluxes $w_{\pm 1}^a$, which we use to calculate the linking numbers defined in Eqs.~(\ref{eqn:defined-charges}). For better clarity, we do not draw the pancake surface $\mathcal{S}^2$ in the following figures. The triangular arrowheads (``${\bs{\vartriangleright}}$'') indicate orientation reversals of NLs of the corresponding color.}
\label{fig:convention}
\end{figure}

\subsection{Strategy}\label{sec:linking-strategy}

To express our main finding, we need to introduce some notation, illustrated in Fig.~\ref{fig:convention}. We consider a {\color{MyBlue}blue} NL-ring formed inside the zeroth ($G=0$) band gap (\emph{i.e.}~between the HO and LU bands). We label this NL-ring as ${\color{MyBlue}\textrm{L}_0^1}$. We assume that ${\color{MyBlue}\textrm{L}_0^1}$ is \emph{not} linked with additional NLs ${\color{MyBlue}\textrm{L}_0^{2,3,\ldots}}$ formed inside the same gap. If this condition were not fulfilled, the second-homotopy charge of ${\color{MyBlue}\textrm{L}_0^1}$ would be undefined~\footnote{This is because there would be no sphere $\mathcal{S}^2$ that \TB{simultaneously} (\emph{i}) encloses ${\color{MyBlue}\textrm{L}_0^1}$, (\emph{ii}) does not enclose ${\color{MyBlue}\textrm{L}_0^{2,3,\ldots}}$, and that (\emph{iii}) preserves the band gap $G=0$}. Then we can find a two-dimensional \emph{disc} $\mathcal{D}^2$ that has the NL-ring as a boundary (\emph{i.e.}~$\partial \mathcal{D}^2 = {\color{MyBlue}\textrm{L}_0^1}$) and that preserves the $G=0$ band gap everywhere except on its boundary (\emph{i.e.}~on $\mathcal{D}^2\backslash \partial \mathcal{D}^2$). To find the monopole charge of ${\color{MyBlue}\textrm{L}_0^1}$, we inflate the disc $\mathcal{D}^2$ into a thin ``pancake'' containing ${\color{MyBlue}\textrm{L}_0^1}$. The surface of the pancake is topologically a sphere ($\mathcal{S}^2$) that does not cross ${\color{MyBlue}\textrm{L}_0^{2,3,\ldots}}$. Importantly, we \emph{allow} nodal lines $\textrm{L}_{G'\neq G}$ formed inside \emph{other} band gaps $G' \neq 0$ to be linked with ${\color{MyBlue}\textrm{L}_0^1}$.  This leads to the presence of \emph{intersection points} with the disc, $\textrm{L}_{G'} \cap \mathcal{D}^2 = \{\textrm{P}_{G'}^a\}_{a\in\mathcal{I}}$, where $\mathcal{I}$ is the appropriate index set. We assign to each intersection point $\textrm{P}^a_{G'}$ with $G' = \pm 1$ its \emph{flux} $w_{G'}^a = \pm 1$, which expresses whether the orientation (as defined in Sec.~\ref{sec:orientations}) of the intersecting NL is directed \emph{away from} ($w_i^a = +1$, \emph{i.e.}~sources) or \emph{towards} ($w_{G'}^a = -1$ \emph{i.e.}~sinks) the disc $\mathcal{D}^2$. The anticommutation of topological charge of NLs ${\color{MyBlue}\textrm{L}_0}$ with the charges of NLs ${\color{MyRed}\textrm{L}_{-1}}$ and ${\color{MyIndigo}\textrm{L}_{+1}}$, captured by Eq.~(\ref{eqn:consec-anticom}), guarantees that the ``away'' vs.~``towards'' definition leads to the same values of $w_{G'}^a$ for $G' = \pm 1$ on both sides of the disc $\mathcal{D}^2$. 

Our key finding is that the group structure of the monopole charge (listed in the right Table~\ref{tab:AI-homotopy}) on the pancake sphere is exactly mimicked by the group structure of quantitities 
\begin{subequations}\label{eqn:defined-charges}
\begin{equation}
\eta_\pm = \sum_a w_{\pm 1}^a\in\intg\label{eqn:charge-intersection}
\end{equation}
and
\begin{equation}
\nu_\pm = \tfrac{1}{2}(\eta_+ \pm \eta_-)\in \intg \label{eqn:charge-combinations}
\end{equation}
\end{subequations}
which characterize the linking of the NL-ring ${\color{MyBlue}\textrm{L}_0^1}$ with NLs formed inside the neighboring band gaps. (Below, we call these quantities \emph{linking numbers}, since they are related to the Gauss linking invariant~\cite{Gauss:1833}.) This exact correspondence allows us to \emph{interpret} the monopole charge \emph{as} the information about the linking structure.

\begin{figure}[b!]
\includegraphics[width=0.42\textwidth]{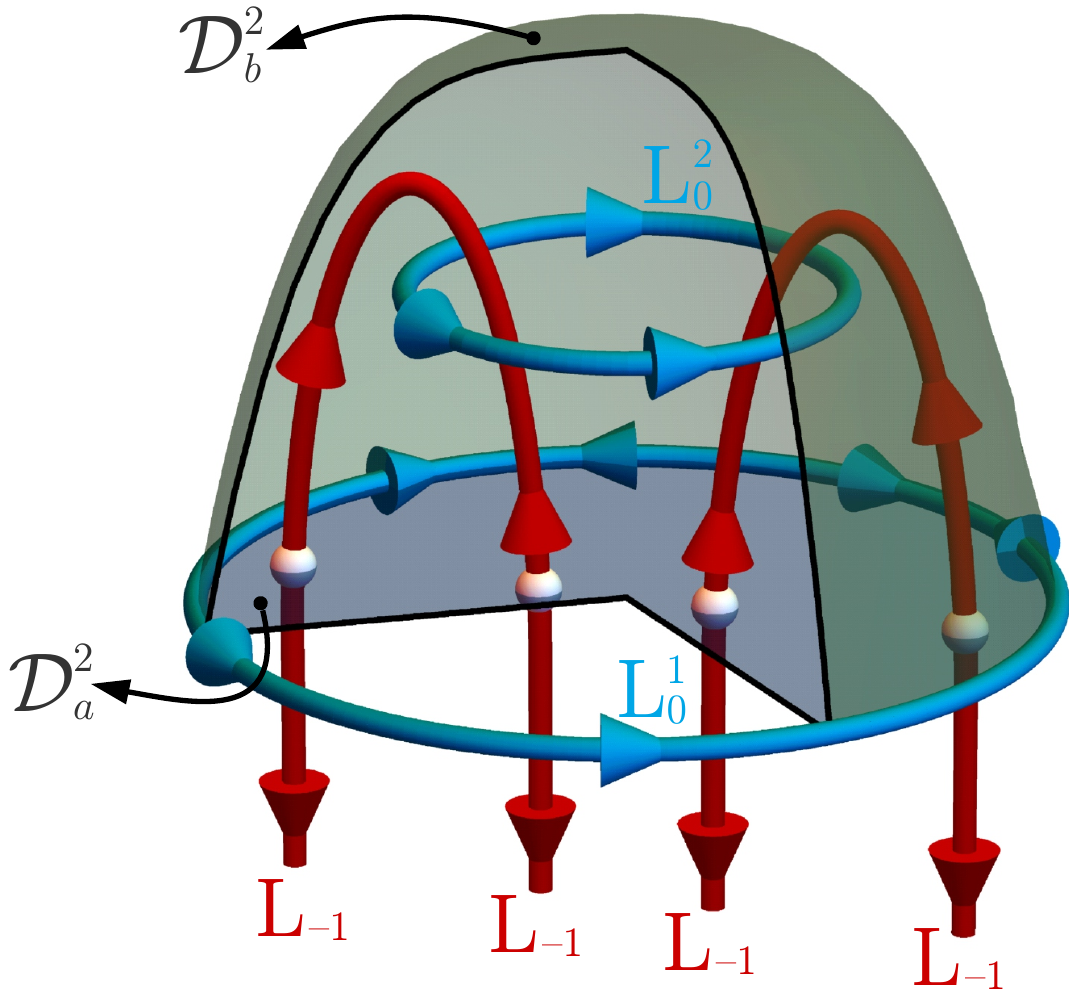}
 \caption{The monopole charge assigned to NL-ring ${\color{MyBlue}\mathrm{L}_0^1}$ may depend on the choice of the inscribed disc $\mathcal{D}^2$. The flat gray disc $\mathcal{D}^2_a$ (only a section of the disc is shown) exhibits four intersection points with ${\color{MyRed}\mathrm{L}_{-1}}$, all having the same orientation, leading to $\eta_- = +4$. In contrast, the cap-shaped green disc $\mathcal{D}^2_b$ is not intersected by ${\color{MyRed}\mathrm{L}_{-1}}$, leading to a different value $\eta_- = 0$. The ambiguity arises because an additional NL-ring ${\color{MyBlue}\mathrm{L}_0^2}$ makes it impossible to continuously deform $\mathcal{D}^2_a$ into $\mathcal{D}^2_b$ while preserving the $G=0$ spectral gap. In Sec.~\ref{sec:NA-braiding}, we relate this ambiguity to non-trivial braiding of the monopole charge around NLs of the same color.}
\label{fig:two-discs}
\end{figure}

\TB{To be more precise, the \emph{proof} of the equivalence between the monopole charges and the linking numbers (defined more precisely in the subsections below) consists of three logical steps.
\begin{enumerate}
\item \emph{Show that both quantities remain invariant under the same set of conditions.} The linking number is well-defined only if the disc is not intersected by another {\color{MyBlue}blue} node, while the element in second-homotopy group exists only if the pancake is not intersected by a {\color{MyBlue}blue} node. Since the disc and the pancake are assumed to be  infinitesimally close to each other, these are clearly equivalent conditions.
\item \emph{Show that under continuous deformations of the Hamiltonian both quantities define the same number of distinct equivalence classes.} This step is very carefully analyzed in Secs.~\ref{sec:2+1-monopole}--\ref{sec:CI-monopole} using the geometric rules reviewed in Sec.~\ref{sec:non-Abelian-NLs}.
\item \emph{Show that both quantities exhibit the same group structure under binary composition.} This follows easily, because the linking numbers (argued to be the ``manifestations’'' of the $\ztwo$ and $\intg$ monopole charges) are clearly additive under merging several NL-rings together, exactly as expected for the monopole charge.
\end{enumerate}
We find that the specific linking numbers encoding the monopole charge depends on the number of occupied and unoccupied bands, which leads us to split the discussion below into several subsections.}

\TB{Before delving into the detailed discussion further below, let us} briefly argue here why the correspondence between the monopole charge and the linking numbers in Eqs.~(\ref{eqn:defined-charges}) should appear plausible. First, we know from Sec.~\ref{sec:compositions} that each NL-ring is linked with an even number of NLs formed inside the neighboring band gaps, implying that the sum $\eta_+ + \eta_-$ is even. This already foreshadows a stable $\ztwo$ classification into cases where $\eta_\pm$ are both even vs.~both odd. [This also explains why $\nu_\pm$ are integer, cf.~Eq.~(\ref{eqn:charge-combinations}).] Since NLs formed inside band gaps $G = \pm 1$ are not able to move across the NL-ring ${\color{MyBlue}\textrm{L}_0^1}$, one may na\"{i}vely expect the quantities $\nu_\pm$ to form a $\intg\oplus\intg$ topological charge. This is indeed true for $(2+2)$-band models, when $\textrm{L}_{\pm 1}$ are the \emph{only} NLs that can be linked with the NL-ring ${\color{MyBlue}\textrm{L}_{0}^1}$. However, this simple expectation fails in models with more bands, when $\textrm{L}_{\pm 2}$ are able to freely move across the NL-ring ${\color{MyBlue}\textrm{L}_0^1}$. Although such more distant NLs do not directly enter the formulas in Eqs.~(\ref{eqn:defined-charges}), the anticommutation relations $\{g_{+1},g_{+2}\} = 0 = \{g_{-1},g_{-2}\}$ may flip the sign of some fluxes $w_{\pm 1}^a$ in Eq.~(\ref{eqn:charge-intersection}), thus leaving only the \emph{parity} of $\eta_\pm$ invariant.

One final remark is necessary \TB{before proceeding with the analysis, namely that our} definition of $\eta_\pm$ and $\nu_\pm$ in Eqs.~(\ref{eqn:defined-charges}) for a given NL-ring ${\color{MyBlue}\mathrm{L}_0^1}$ may depend on the choice of the disc $\mathcal{D}^2$ inscribed into the NL-ring. Although somewhat surprising, this is compatible with the fact that the second homotopy group, discussed in Sec.~\ref{NLs-second-homotopy}, assigns a topological invariant to a \emph{sphere} containing the NL-ring, rather than to the NL-ring itself. An example when such a non-uniqueness of the monopole charge arises is shown for a $(2+1)$-band model (with second homotopy group $\pi_2[M^\textrm{AI}_{(2,1)}] = 2\intg$) in Fig.~\ref{fig:two-discs}. In this example, we first incribe into the {\color{MyBlue}blue} NL-ring ${\color{MyBlue}\mathrm{L}_0^1}$ the flat (gray) disc $\mathcal{D}^2_a$. Since $\mathcal{D}^2_a$ is intersected by {\color{MyRed}red} nodal lines ${\color{MyRed}\mathrm{L}_{-1}}$ four times, each time with the same orientation, we find $\eta_- = +4$. On the other hand, for the cap-shaped (green) disc $\mathcal{D}^2_b$ we find no intersections with ${\color{MyRed}\mathrm{L}_{-1}}$, leading to a different result $\eta_- = 0$. Note that $\mathcal{D}^2_a$ cannot be continuously deformed into $\mathcal{D}^2_b$, because of the NL-ring ${\color{MyBlue}\textrm{L}_0^2}$ contained inside the region bounded by $\mathcal{D}^2_a\cup\mathcal{D}^2_b$. As a consequence, the pancakes associated with discs $\mathcal{D}^2_a$ and $\mathcal{D}^2_b$ cannot be continuously deformed into each other while preserving the $G=0$ band gap, thus allowing their monopole charges to differ. In Sec.~\ref{sec:NA-braiding} as well as in the appendices, we relate this ambiguity to a noncommutative braiding of NL-rings with a monopole charge around NLs of the same color. For the purpose of the current section, we simply assume the disc $\mathcal{D}^2$ to be fixed.

\subsection{Models with \texorpdfstring{$2+1$}{2+1} bands}\label{sec:2+1-monopole}

We consider models with $n=2$ occupied and $\ell=1$ unoccupied bands (analogous discussion also applies to systems with flipped values $n \leftrightarrow \ell$), when the group structure of the monopole charge (defined on spheres preserving the $G=0$ spectral gap) is 
\begin{equation}
\pi_2[M^\textrm{AI}_{(1,2)}] = \pi_2[M^\textrm{AI}_{(2,1)}] = 2\intg. 
\end{equation}
Following the color code in Fig.~\ref{fig:code}, we indicate NLs formed inside the $G=0$ band gap (\emph{i.e.}~between the HO and LU bands) with {\color{MyBlue}blue} and label them ${\color{MyBlue}\textrm{L}_0}$. NLs formed in the $G=-1$ band gap (\emph{i.e.}~formed between the two occupied bands) are indicated with {\color{MyRed}red} and labelled ${\color{MyRed}\textrm{L}_{-1}}$. 

\begin{figure}[b!]
\includegraphics[width=0.46\textwidth]{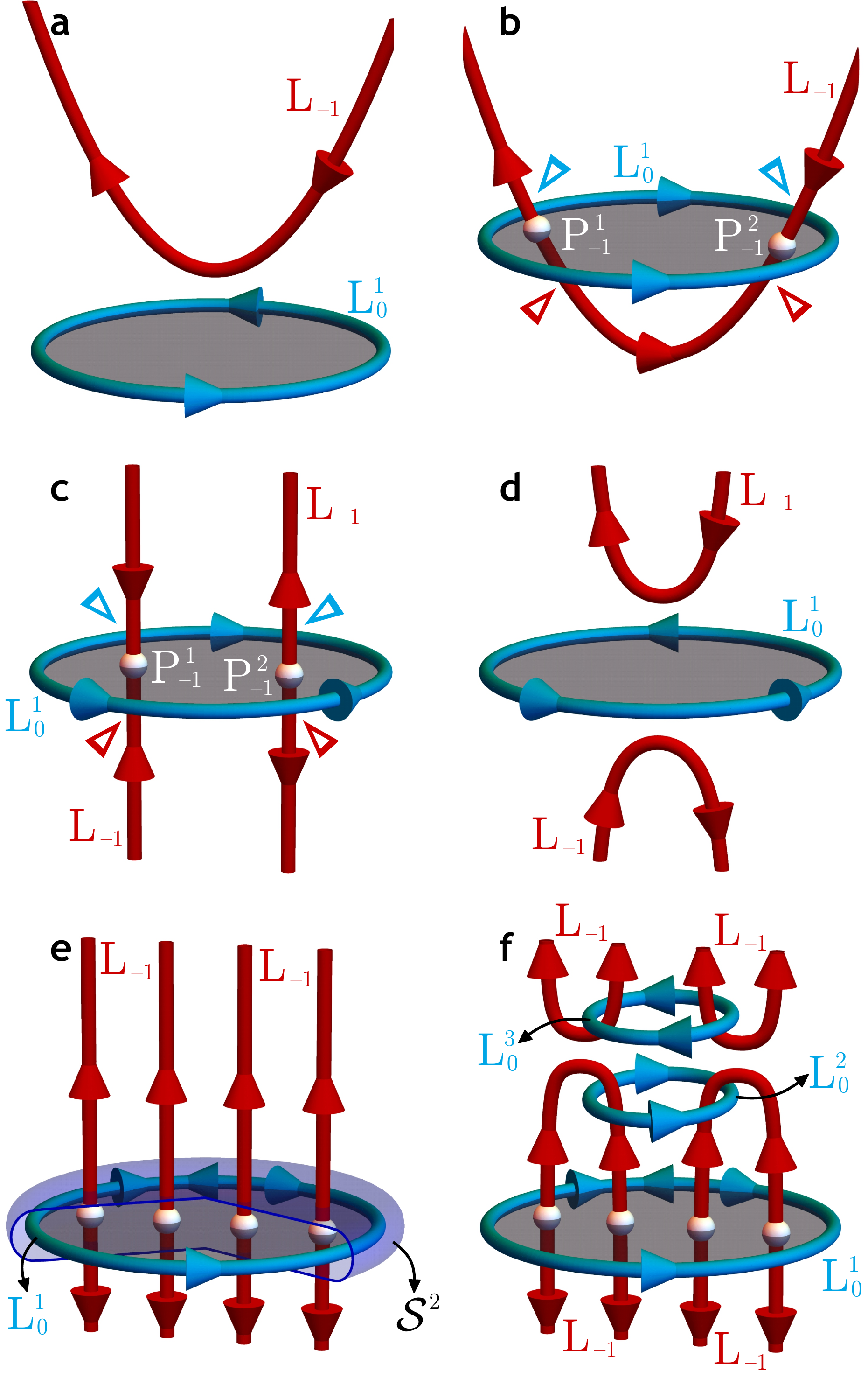}
 \caption{The possible strategies to change the linking number $\eta_-$ of a {\color{MyBlue}blue} ($G=0$) NL-ring ${\color{MyBlue}\textrm{L}_0^1}$ with {\color{MyRed}red} ($G=-1$) NLs ${\color{MyRed}\textrm{L}_{-1}}$. 
 None of these strategies is succesful. (a--b) Moving a segment of ${\color{MyRed}\textrm{L}_{-1}}$ across the disc produces two intersection points $\textrm{P}_{-1}^{1,2}$ with opposite flux. (c--d) Local reconnection of two antiparallel segments of ${\color{MyRed}\textrm{L}_{-1}}$ annihilates two intersection points $\textrm{P}_{-1}^{1,2}$ with opposite flux. (e--f) Local annihilation of four parallel segments of ${\color{MyRed}\textrm{L}_{-1}}$ is possible, following the strategy of Fig.~\ref{fig:i4}, but produces two new NL-rings ${\color{MyBlue}\textrm{L}_0^{2,3}}$ inside the $G=0$ band gap. Their presence makes it impossible to bulge the disc $\mathcal{D}_2$ into the region where the four segments of ${\color{MyRed}\textrm{L}_{-1}}$ have annihilated, because ${\color{MyBlue}\textrm{L}_0^{2}}$ would temporarily close the $G=0$ band gap on the sphere $\mathcal{S}^2$ [the sphere is shown explicitly in panel (e)] obtained by inflating the disc. In panels (b,c), the triangular arrowheads (``${\bs{\vartriangleright}}$'') indicate the orientation reversal of NLs of the corresponding color. For better clarity, such reversals are not explicitly shown in panels (e,f).}
\label{fig:2+1-bands}
\end{figure}

We apply the prescription outlined in Sec.~\ref{sec:linking-overview}, \emph{i.e.} we consider a {\color{MyBlue}blue} NL-ring ${\color{MyBlue}\textrm{L}_0^1}$ which can be linked with nodal lines ${\color{MyRed}\textrm{L}_{-1}}$. We inscribe into ${\color{MyBlue}\textrm{L}_0^1}$ a disc $\mathcal{D}^2$ that preserves the $G=0$ band gap. The disc can be inflated into a thin pancake containing ${\color{MyBlue}\textrm{L}_0^1}$. We claim that the monopole charge of ${\color{MyBlue}\textrm{L}_0^1}$ on this pancake corresponds to the linking number $\eta_- \in 2\intg$ defined in Eq.~(\ref{eqn:charge-intersection}).

Since there are no higher-energy bands, automatically $\eta_+ = 0$, and the quantities defined in Eq.~(\ref{eqn:charge-combinations}) are not relevant. The even parity of $\eta_-$ follows from the anticommutation of the charge of {\color{MyBlue}blue} ($G=0$) NLs with the charge of the {\color{MyRed}red} ($G=-1$) NLs. More precisely, the discussion in Sec.~\ref{sec:compositions} forces the {\color{MyBlue}blue} NL-ring ${\color{MyBlue}\textrm{L}_0^1}$ to be linked with an \emph{even} number of {\color{MyRed}red} NLs ${\color{MyRed}\textrm{L}_{-1}}$. This implies an even number of intersection points $\textrm{P}_{-1}^a$ of ${\color{MyRed}\textrm{L}_{-1}}$ with the disc $\mathcal{D}^2$, such that the parity of $\eta_-$ is indeed even.

We further need to show that all admissible manipulations of NLs [achieved by continuous deformations of $H(\bs{k})$] that preserve the $G=0$ band gap on $\mathcal{S}^2$ keep the value of $\eta_-\in 2\intg$ invariant. To that end, we consider the processes that might potentially change the number and the flux of the intersection points $\textrm{P}_{-1}^a$. Such changes could occur due to the following reasons:
\begin{itemize}
\item[(\emph{i})] A segment of a {\color{MyRed}red} NL ${\color{MyRed}\textrm{L}_{-1}}$ might move across the NL-ring ${\color{MyBlue}\textrm{L}_0^1}$. This would change the number of intersection points $\textrm{P}_{-1}^a$ by $\pm 1$, and also the value of $\eta_-$ by $\pm 1$. However, since the charge of {\color{MyBlue}blue} NLs anticommutes with the charge of {\color{MyRed}red} NLs, such processes are topologically forbidden (cf.~Sec.~\ref{sec:crossings}).
\item[(\emph{ii})] A segment of ${\color{MyRed}\textrm{L}_{-1}}$ could move across the disc $\mathcal{D}^2$ bounded by ${\color{MyBlue}\textrm{L}_0^1}$. Such a motion produces two intersection points $\textrm{P}_{-1}^{1,2}$ with opposite fluxes,  $w_{-1}^{1} = - w_{-1}^2$. Therefore, this process does not change the linking number $\eta_-$ in Eq.~(\ref{eqn:charge-intersection}) [cf.~Fig.~\ref{fig:2+1-bands}(a--b)].
\item[(\emph{iii})] A pair of {\color{MyRed}red} NLs that intersect the disc $\mathcal{D}^2$ at points $\textrm{P}_{-1}^{1,2}$ could reconnect, removing the two intersection points in the process. However, this is possible only if the two {\color{MyRed}red} NLs have \emph{opposite} orientation. This implies that the removed points carry opposite fluxes $w_{-1}^1 = - w_{-1}^2$, and their disappearance does not change $\eta_-$ [cf.~Fig.~\ref{fig:2+1-bands}(c--d)].
\item[(\emph{iv})] We showed in Sec.~\ref{sec:four-are-trivial} that four parallel NLs inside the same gap are, in \TB{certain} sense, trivial [cf.~Eq.~(\ref{eqn:i4-trivial})]. One could thus consider a local annihilation of four parallel {\color{MyRed}red} NL segments. However, as shown in Fig.~\ref{fig:i4}, such a process necessarily produces a pair of {\color{MyBlue}blue} NL-rings ${\color{MyBlue}\textrm{L}_0^{2,3}}$. These additional NL-rings present an obstacle standing in the way between the disc $\mathcal{D}^2$ and the region where the four {\color{MyRed}red} NL segments have annihilated [cf.~Fig.~\ref{fig:2+1-bands}(e--f)]. Therefore, this process does not allow us to change the fluxes on the disc $\mathcal{D}^2$. 
\end{itemize}
\TB{All the presented strategies to change the linking number $\eta_-$ are unsuccessful.}

\TB{Note however that} there is, in fact, a very simple process that \emph{does} change $\eta_-$: Sweeping a {\color{MyBlue}blue} NL in front of ${\color{MyBlue}\textrm{L}_0^1}$ [cf.~Fig.~\ref{fig:braiding}(a,c)]. According to Sec.~\ref{sec:orientations}, this flips the sign of all the fluxes $w_{-1}^a$, \TB{resulting in an overall sign-reversal of the linking number}. However, this process is special, and corresponds to braiding ${\color{MyBlue}\textrm{L}_0^1}$ around the NL. We show in Sec.~\ref{sec:NA-braiding} and in the appendices that this interesting exception can be treated separately. \TB{More concretely}, it corresponds to a non-trivial action induced by the first homotopy group on the second, \TB{\emph{i.e.}~this observation too is in accordance with the homotopic description of the monopole charge.}. 

Keeping in mind \TB{this caveat, which we postpone} until Sec.~\ref{sec:NA-braiding}, we find that $\eta_- \in 2\intg$ is a topological invariant characterizing ${\color{MyBlue}\textrm{L}_0^{1}}$ on the pancake sphere obtained by inflating the disc $\mathcal{D}^2$ \TB{in $(2+1)$-band models.} \TB{Following the overall strategy outlined in Sec.~\ref{sec:linking-strategy},} we have proved that the monopole charge of such models \emph{exactly corresponds} to the linking number $\eta_-$.

\begin{figure*}[t!]
\includegraphics[width=0.995\textwidth]{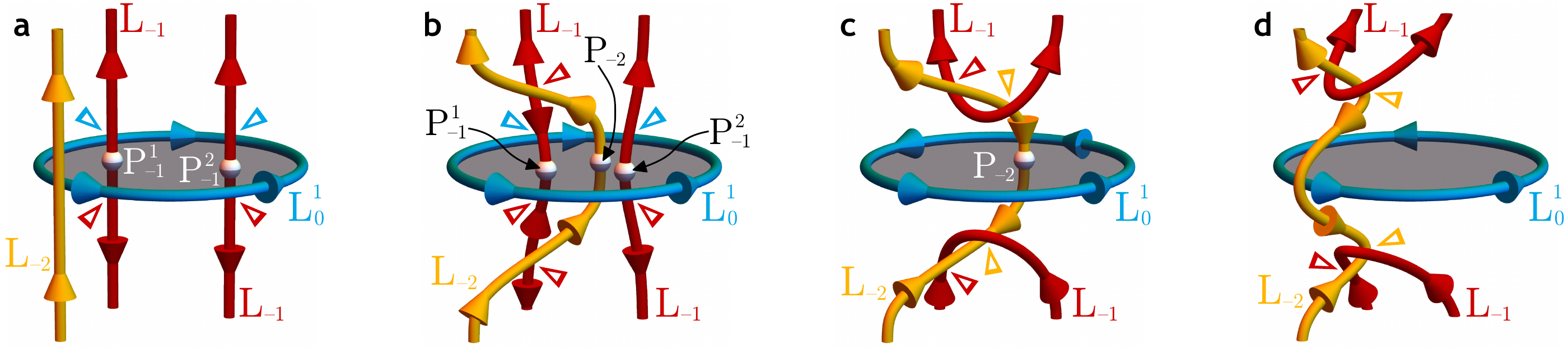}
 \caption{Changing the linking number $\eta_-$ of the {\color{MyBlue}blue} ($G=0$) NL-ring ${\color{MyBlue}\textrm{L}_0^1}$ in a $(3+1)$-band model by $\pm 2$. The change is achieved by moving an {\color{MyOrange}orange} ($G=-2$) NL ${\color{MyOrange}\textrm{L}_{-2}}$ to the inside of the NL-ring ${\color{MyBlue}\textrm{L}_0^1}$. The anticommutation of the charge $g_{-2}$ of {\color{MyOrange}orange} NLs ${\color{MyOrange}\textrm{L}_{-2}}$ with the charge $g_{-1}$ of {\color{MyRed}red} ($G=-1$) NLs ${\color{MyRed}\textrm{L}_{-1}}$ allows us to revert the flux of some of the intersection points  $\textrm{P}_{-1}^a$, thus changing the value of $\eta_-$ by multiples of $2$. As a consequence, $\eta_-$ can be made zero, and the {\color{MyBlue}blue} NL-ring ${\color{MyBlue}\textrm{L}_0^1}$ can be completely unlinked. Triangular arrowheads (``$\bs{\vartriangleright}$'') indicate the orientation reversals of NLs of the corresponding color. For a more detailed discussion of the process illustrated in panels (a--d), see Sec.~\ref{sec:3+1-monopole}.}
\label{fig:3+1-bands}
\end{figure*}

\subsection{Models with \texorpdfstring{$2+2$}{2+2} bands}\label{sec:2+2-monopole}

We now consider systems with $n=2$ occupied and $\ell=2$ unoccupied bands, when the group structure of the monopole charge is 
\begin{equation}
\pi_2[M^\textrm{AI}_{(2,2)}] = \intg\oplus\intg,
\end{equation}
This analysis easily generalizes the case of $(2+1)$-band models: Besides the {\color{MyBlue}blue} ($G=0$) NLs ${\color{MyBlue}\textrm{L}_0}$ at the Fermi level, and the {\color{MyRed}red} ($G=-1$) NLs ${\color{MyRed}\textrm{L}_{-1}}$ formed by the two occupied bands, there are now additional {\color{MyIndigo}indigo} ($G=+1$) NLs ${\color{MyIndigo}\textrm{L}_{+1}}$ formed by the two \emph{unoccupied bands}. 

We want to analyze the linking numbers of a {\color{MyBlue}blue} NL-ring ${\color{MyBlue}\textrm{L}_0^1}$ with the other two species of NLs. Since the charges $g_{+1}$ and $g_{-1}$ \emph{commute}, the {\color{MyRed}red} and the {\color{MyIndigo}indigo} NLs ignore each other's presence. This suggests that we can apply the arguments from Sec.~\ref{sec:2+1-monopole} independently to linking of ${\color{MyBlue}\textrm{L}_0^1}$ with the {\color{MyRed}red} NLs and with the {\color{MyIndigo}indigo} NLs. This is \emph{almost} true. The only caveat is that, since now \emph{both} $\eta_\pm$ could be non-zero, one gets a consistent NL composition not only when $\eta_\pm$ are both even, but also when $\eta_\pm$ are \emph{both odd}. To capture this even-even vs.~odd-odd disparity, it is more convenient to work with linking numbers $(\nu_+,\nu_-)\in\intg\oplus\intg$ defined in Eq.~(\ref{eqn:charge-combinations})~\cite{Bouhon:2020}. Arguments completely analogous to those illustrated in Fig.~\ref{fig:2+1-bands} then prove that these two integers are invariant under all admissible manipulations of the NLs which preserve the $G=0$ band gap (with the single exception -- the action of $\pi_1$ on $\pi_2$ discussed in Sec.~\ref{sec:NA-braiding}). Therefore, the monopole charge of $(2+2)$-band models \emph{corresponds} to the linking numbers $(\nu_+,\nu_-)$.

\subsection{Models with \texorpdfstring{$3+1$}{3+1} bands}\label{sec:3+1-monopole}

Let us now consider models with $n=3$ occupied and $\ell = 1$ unoccupied bands (analogous discussion also applies to models with interchanged values $n\leftrightarrow\ell$). We thus consider a {\color{MyBlue}blue} ($G=0$) NL-ring ${\color{MyBlue}\textrm{L}_0^1}$ formed between the HO and LU bands, which could be linked with {\color{MyRed}red} ($G=-1$) NLs ${\color{MyRed}\textrm{L}_{-1}}$ and with {\color{MyOrange}orange} ($G=-2$) NLs ${\color{MyOrange}\textrm{L}_{-2}}$ formed among the occupied bands. Clearly, this scenario augments the case of $(2+1)$-band models from Sec.~\ref{sec:2+1-monopole} in a \emph{different} way than the $(2+2)$-band case discussed in Sec.~\ref{sec:2+2-monopole}. The homotopy theory predicts that 
\begin{equation}
\pi_2[M^\textrm{AI}_{(3,1)}] = \pi_2[M^\textrm{AI}_{(1,3)}] = \triv,
\end{equation}
\emph{i.e.}~that there is \emph{no} monopole charge. We show that this behavior is mimicked by the absence of topological invariance of linking numbers $\eta_\pm$ defined in Eq.~(\ref{eqn:charge-intersection}). 

Let us explain this behaviour. First note that, just as for ($2+1$)-band models, we have $\eta_+ = 0$ [there is no higher-energy band to form {\color{MyIndigo}indigo} ($G=+1$) NLs], such that $\eta_-$ has to be an even integer [compatibility requires that an \emph{even} number of {\color{MyRed}red} NLs ${\color{MyRed}\textrm{L}_{-1}}$ are linked with ${\color{MyBlue}\textrm{L}_0^1}$]. What has changed since Sec.~\ref{sec:2+1-monopole} is the presence of additional and more \emph{deeply} located NLs ${\color{MyOrange}\textrm{L}_{-2}}$. Such {\color{MyOrange}orange} NLs carry topological charge $g_{-2}$ that commutes with the charge $g_0$ of {\color{MyBlue}blue} NLs, while it \emph{anti}commutes with the charge $g_{-1}$ of {\color{MyRed}red} NLs. The commmutation relation $[g_{-2},g_0]=0$ allows us to freely move ${\color{MyOrange}\textrm{L}_{-2}}$ to the inside of the NL-ring ${\color{MyBlue}\textrm{L}_0^1}$, while producing intersection points $\textrm{P}_{-2}^a$. The anticommutation relation $\{g_{-2},g_{-1}\} = 0$ allows us to flip the orientation of some {\color{MyRed}red} NLs, thus reverting the flux of some intersection points $\textrm{P}_{-1}^a$ between values $+1 \leftrightarrow -1$, which changes $\eta_-$ in Eq.~(\ref{eqn:charge-intersection}) by multiples of two. Since the linking number $\eta_-$ in $(3+1)$-band models is even, repeating the described procedure sufficiently many times can always bring the value of $\eta_-$ to zero. 

We illustrate the described process on an example shown in Fig.~\ref{fig:3+1-bands}. In Fig.~\ref{fig:3+1-bands}(a) we consider a {\color{MyBlue}blue} NL-ring ${\color{MyBlue}\textrm{L}_0^1}$ linked with two {\color{MyRed}red} NLs ${\color{MyRed}\textrm{L}_{-1}}$ of the same orientation. The NLs ${\color{MyRed}\textrm{L}_{-1}}$ intersect the disc $\mathcal{D}^2$ inscribed into ${\color{MyBlue}\textrm{L}_0^1}$ at two points $\textrm{P}_{-1}^{1,2}$, which both have the same flux $w_{-1}^1 = w_{-1}^2 = +1$, leading to $\eta_- = +2$. In Fig.~\ref{fig:3+1-bands}(b), we have moved an {\color{MyOrange}orange} NL ${\color{MyOrange}\textrm{L}_{-2}}$ across ${\color{MyBlue}\textrm{L}_0^1}$, such that it now crosses the disc $\mathcal{D}^2$ at point $\textrm{P}_{-2}$. This is allowed because the charges $g_{-2}$ and $g_0$ of these two species of NLs commute. [We remark that if the original model does not exhibit any {\color{MyOrange}orange} NL, one such NL-ring can always be produced by a local band inversion, similar to Fig.~\ref{fig:i4}(b).] Furthermore, we have deformed a segment of ${\color{MyOrange}\textrm{L}_{-2}}$ such that it passes in front of one of the {\color{MyRed}red} NLs. This flips the flux of $\textrm{P}_{-1}^1$ to $w_{-1}^1 = -1$. Since the intersection points $\textrm{P}_{-1}^{1,2}$ now carry \emph{opposite} flux, they can annihilate in front of ${\color{MyOrange}\textrm{L}_{-2}}$. This is achieved through a process of reconnecting the two {\color{MyRed}red} NLs as shown in Fig.~\ref{fig:3+1-bands}(c). The NL-ring ${\color{MyBlue}\textrm{L}_0^1}$ is now not crossed by any segments of ${\color{MyRed}\textrm{L}_{-1}}$, so clearly $\eta_- = 0$. In the last step, shown in Fig.~\ref{fig:3+1-bands}(d), ${\color{MyOrange}\textrm{L}_{-2}}$ is again moved outside of the NL-ring ${\color{MyBlue}\textrm{L}_0^1}$, such that ${\color{MyBlue}\textrm{L}_0^1}$ has become completely unlinked.

\subsection{Models with \texorpdfstring{$3+2$}{3+2} bands}\label{sec:3+2-monopole}

In this section, we consider models with $n=3$ occupied and $\ell = 2$ unoccupied bands (analogous discussion pertains to systems with flipped values $n \leftrightarrow \ell$). We consider a {\color{MyBlue}blue} ($G=0$) NL-ring ${\color{MyBlue}\textrm{L}_{0}^1}$ formed by HO and LU bands, which can be linked with {\color{MyRed}red} ($G=-1$) NLs ${\color{MyRed}\textrm{L}_{-1}}$ and {\color{MyOrange}orange} ($G=-2$) NLs ${\color{MyOrange}\textrm{L}_{-2}}$ formed among the occupied bands, as well as with {\color{MyIndigo}indigo} ($G=+1$) NLs ${\color{MyIndigo}\textrm{L}_{+1}}$ formed among the unoccupied bands. The homotopy theory predicts the monopole charge
\begin{equation}
\pi_2[M^\textrm{AI}_{(3,2)}] = \pi_2[M^\textrm{AI}_{(2,3)}] = \intg.
\end{equation}
We argue that this group structure is exactly mimicked by the linking invariant $\eta_+$. Especially, we find that the decrease of the order of the monopole charge from $\intg\oplus\intg$ in $(2+2)$-band models to $\intg$ in $(3+2)$-band models has the same origin as the decrease from $\intg$ in $(2+1)$-band models to $\triv$ in $(3+1)$-band models. Namely, the \emph{more distant} NLs with charge $g_{\pm 2}$ can be freely moved inside the NL-ring ${\color{MyBlue}\textrm{L}_{0}^1}$ with charge $g_0$, which allows us to flip the flux of some intersection points $\textrm{P}_{\pm 1}^a$, which in turn changes $\eta_\pm$ by multiples of two.

To see the argument in a bit more detail, note that $\eta_\pm$ can be \emph{arbitrary} integers. The compatibility condition discussed in Sec.~\ref{sec:compositions} only requires them to have the same parity. However, we already know from the analysis of $(3+1)$-band models in Sec.~\ref{sec:3+1-monopole} that $\eta_-$ can be changed by integer multiples of $2$, \emph{i.e.}~it can be brought to $\eta_- = 0$ (for $\eta_+$ even) or $\eta_- = +1$ (for $\eta_+$ odd). On the other hand, the absence of {\color{MyTan}tan} ($G = +2$) NLs ${\color{MyTan}\textrm{L}_{+2}}$ in $(3+2)$-band models keeps the value of the linking number $\eta_+ \in\intg$ invariant. By repeating the arguments presented in the previous sections, this is sufficient to claim that the monopole charge of $(3+2)$-band models exactly corresponds to the linking number $\eta_+$.

\subsection{Models with \texorpdfstring{$3+3$}{3+3} bands}\label{sec:3+3-monopole}
We finally consider models with $n=3$ occupied and $\ell = 3$ unoccupied bands. Assuming a {\color{MyBlue}blue} ($G=0$) NL-ring ${\color{MyBlue}\textrm{L}_{0}^1}$, the compatibility conditions from Sec.~\ref{sec:compositions} require the linking numbers $\eta_\pm\in\intg$ to obey $\eta_+ = \eta_- \;\;\textrm{(mod 2)}$. The presence of {\color{MyOrange}orange} ($G=-2$) NLs ${\color{MyOrange}\textrm{L}_{-2}}$ allows us to change $\eta_-$ by integer multiples of $2$, and the presence of {\color{MyTan}tan} ($G=+2$) NLs ${\color{MyTan}\textrm{L}_{+2}}$ implies the same ambiguity for $\eta_+$. In both cases, the procedure to achieve the change by $\pm 2$ is analogous to the process illustrated for $\eta_-$ in Fig.~\ref{fig:3+1-bands}. This freedom reduces the invariance of topological numbers $\eta_\pm$ to their $\ztwo$ \emph{parity}. This is compatible with the prediction of homotopy theory that the monopole charge is now expressed by
\begin{equation}
\pi_2[M^\textrm{AI}_{(3,3)}] =\ztwo.
\end{equation}
We remark that considering models with $n \geq 4$ or $\ell \geq 4$ is not going to augment the results derived in the individual parts of Sec.~\ref{sec:monopole-linking-relation}, since the topological charge of such very distant NLs $\textrm{L}_{\pm 3}$ \emph{commutes} both with the charge of the NL-ring ${\color{MyBlue}\textrm{L}_{0}^1}$, as well as with the charge of NLs $\textrm{L}_{\pm 1}$ which define the linking numbers $\eta_\pm$ in Eq.~(\ref{eqn:charge-intersection}). This is compatible with having reached the stable limit, observed in the right Table~\ref{tab:AI-homotopy}. We have thus proved a relation between the linking numbers $\eta_\pm$ (resp.~$\nu_\pm$) and the monopole charge of NL-rings in $\mcP\mcT$-symmetric models for arbitrary number of occupied/unoccupied bands.

\begin{figure*}[t!]
\includegraphics[width=0.895\textwidth]{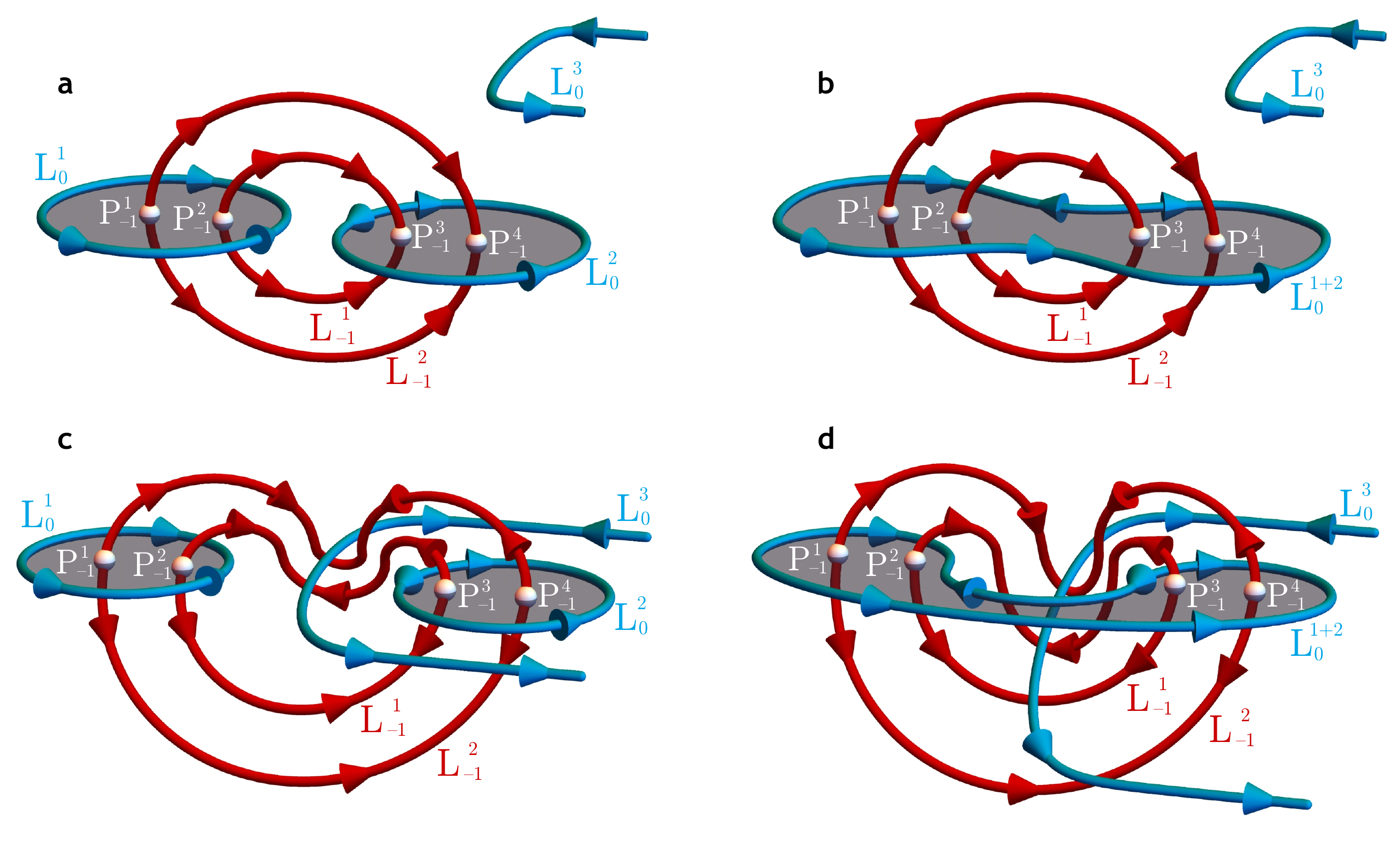}
 \caption{(a) A pair of {\color{MyBlue}blue} ($G=0$) NL-rings ${\color{MyBlue}\textrm{L}_{0}^{1,2}}$ which exhibit opposite values of the linking number $\eta_-$ with {\color{MyRed}red} ($G=-1$) NLs ${\color{MyRed}\textrm{L}_{-1}^{1,2}}$, namely $\eta_- = \pm 2$. According to the arguments presented in Sec.~\ref{sec:monopole-linking-relation}, the linking numbers reveal that ${\color{MyBlue}\textrm{L}_0^{1,2}}$ carry \emph{opposite} values of the monopole charge. (b) We merge the two NL-rings ${\color{MyBlue}\textrm{L}_{0}^{1,2}}$ along a trajectory that \emph{does not} pass under another {\color{MyBlue}blue} NL. We observe that the combined NL-ring ${\color{MyBlue}\textrm{L}_{0}^{1+2}}$ exhibits net linking number $\eta_-^\textrm{tot.} = 0$, meaning that its monopole charge is trivial.  (c) Alternatively, we bring the {\color{MyBlue}blue} NL-rings ${\color{MyBlue}\textrm{L}_{0}^{1,2}}$ together along a trajectory that encloses another {\color{MyBlue}blue} NL ${\color{MyBlue}\textrm{L}_{0}^{3}}$. In this process, the {\color{MyBlue}blue} NL ${\color{MyBlue}\textrm{L}_{0}^{3}}$ moves in front of all the intersection points inside the NL-ring ${\color{MyBlue}\textrm{L}_{0}^{2}}$ (namely $\textrm{P}_{-1}^3$ and $\textrm{P}_{-1}^4$), \TB{inducing a reversal of} 
 their fluxes (\emph{i.e.}~the numbers $w_{-1}^3$ and $w_{-1}^4$). (d) When moved together along such an alternative path, the {\color{MyBlue}blue} NL-rings ${\color{MyBlue}\textrm{L}_{0}^{1,2}}$ exhibit the \emph{same} linking number $\eta_- = +2$, such that the total linking after they are merged into ${\color{MyBlue}\textrm{L}_{0}^{1+2}}$ is $\eta_-^\textrm{tot.} = 4$, implying a non-vanishing monopole charge. The path-dependent ability of the two monopole charges to cancel resp.~to add up indicates a non-trivial action induced by the Berry phase (carried by the {\color{MyBlue}blue} NL ${\color{MyBlue}\textrm{L}_{0}^{3}}$) on the monopole charge (carried by the {\color{MyBlue}blue} NL-ring ${\color{MyBlue}\textrm{L}_{0}^{2}}$). For better clarity, in this figure we do not explicitly indicate orientation reversals of the NLs.} 
 \label{fig:braiding}
\end{figure*}

\subsection{Monopole charge in nodal class \texorpdfstring{$\textrm{CI}$}{CI}}\label{sec:CI-monopole}

Before concluding the present section, we investigate one final class of models. It has been shown by Ref.~\cite{Bzdusek:2017} that the monopole charge of NL-rings also appears in centrosymmetric and time-reversal symmetric singlet superconductors~\cite{Bouhon:2018b}. On top of $\mcP\mcT$ symmetry, such systems also exhibit the \emph{chiral symmetry} $\mcC$ that anticommutes with the Bogoliubov-de~Gennes Hamiltonian $H(\bs{k})$ at every $\bs{k}$. This implies that NLs appear symmetrically around zero energy, \emph{i.e.}~as manifolds $\textrm{L}_{G} = \textrm{L}_{-G}$, and also that $\eta_+ = \eta_-$. This produces special constraints on the situations studied in Secs.~\ref{sec:2+2-monopole} and~\ref{sec:3+3-monopole}. Here we briefly discuss the consequences. 

Let us first consider the case of $(2+2)$ bands. The linking numbers fulfill $\eta_+ = \eta_- \in \intg$. We argued in Sec.~\ref{sec:2+2-monopole} that admissible manipulations of NLs that preserve the $G=0$ spectral gap on the inflated disc cannot change $\eta_-$ nor $\eta_+$ in $(2+2)$-band models. Clearly, the \emph{more restricted} space of $\mcC$-symmetric Hamiltonians also cannot change their values, meaning that $\eta_+ = \eta_- \equiv \eta$ defines a $\intg$-valued topological invariant, which is compatible with
\begin{subequations}
\begin{equation}
\pi_2[M^\textrm{CI}_2] = \intg
\end{equation}
found by Ref.~\cite{Bzdusek:2017}. In contrast, for $n=\ell \geq 3$ the anticommuation of $\textrm{L}_{\pm 1}$ with $\textrm{L}_{\pm 2}$ implies that $\eta_+ = \eta_-$ can be \emph{simultaneously} changed by $+2$ or by $-2$. Therefore, only the $\ztwo$ \emph{parity} of $\eta_\pm$ is invariant, compatible with
\begin{equation}
\pi_2[M^\textrm{CI}_3] = \ztwo.\label{eqn:ci-stable}
\end{equation}
\end{subequations}
Similar to Sec.~\ref{sec:3+3-monopole}, adding more bands does not change the result. The group in Eq.~(\ref{eqn:ci-stable}) corresponds to the stable limit of the monopole charge in nodal class $\textrm{CI}$~\cite{Bzdusek:2017}.
\smallskip

\section{Braiding of the monopole charge}\label{sec:NA-braiding}\nopagebreak
\subsection{Overview}\nopagebreak
\TB{In this section we present another application of the geometric description of the non-Abelian band topology. Recall that} we have encountered in Sec.~\ref{sec:linking-overview} an ambiguity in the definition of the monopole charge of NL-rings, namely that it may depend on the choice of the inscribed disc $\mathcal{D}^2$ (cf.~Fig.~\ref{fig:two-discs}). Another ambiguity emerged in Sec.~~\ref{sec:2+1-monopole}, where the monopole charge of a {\color{MyBlue}blue} NL-ring ${\color{MyBlue}\textrm{L}_0^1}$ flipped sign when another {\color{MyBlue}blue} NL was swept in front of it [cf.~Fig.~\ref{fig:braiding}(a,c)]. 

Here, we argue that the encountered non-uniqueness of the monopole charge emanates from a non-trivial action induced by the Berry phase (corresponding to the first homotopy group, ``$\pi_1$'', cf.~Sec.~\ref{NLs-first-homotopy}) on the monopole charge (corresponding to the second homotopy group, ``$\pi_2$'', cf.~Sec.~\ref{NLs-second-homotopy}). This phenomenon implies that a monopole charge can be ``braided'' non-trivially around NLs, or more precisely, that the ability of two NL-rings to annihilate may depend on the trajectory used to bring them together. We emphasize that this property arises \emph{in addition} to the path-dependent ability of NLs themselves to annihilate, found by Ref.~\cite{Wu:2018b} and \TB{reviewed} 
in Sec.~\ref{sec:nonuniqueness} of this work. The noncommutative behavior of the monopole charge of NL-rings discussed here has not been previously reported. From a mathematical perspective, it is similar to the non-trivial braiding of hedgehog defects around vortex lines, known to arise in uniaxial nematics~\cite{Volovik:1977}.

Our discussion is structured as follows. In Sec.~\ref{sec:NA-braid-monopole} we explain the origin of the nontrivial action of $\pi_1$ on $\pi_2$ using the \TB{geometric description of the} non-Abelian topological charge. We find that this phenomenon arises as a consequence of the orientation reversals of NLs discussed in Sec~\ref{sec:orientations}. To strengthen the geometric arguments presented here, we provide an independent and more formal discussion of the phenomenon in the appendices. First, following Ref.~\cite{Mermin:1979}, we mathematically formulate what we mean by the action of $\pi_1$ on $\pi_2$ in appendix~\ref{app:cosets-and-homotopy}. In appendix~\ref{sec:crossed-module} we reformulate this action using the notion of \emph{Abe homotopy}~\cite{Abe:1940, kobayashi2012abe}. Finally, in appendices~\ref{app:pi1-on-pi2-examples} and~\ref{sec:action-for-CI} we explicitly compute the action of $\pi_1(M)$ on $\pi_2(M)$, where $M$ are the spaces of few-band Hamiltonians belonging to $\mcP\mcT$-symmetric nodal classes $\textrm{AI}$ and $\textrm{CI}$. The formal derivations presented in the appendices are consistent with the geometric arguments presented here.

\subsection{Explanation using the non-Abelian topology}\label{sec:NA-braid-monopole}

The argument using the non-Abelian topological charge follows in two steps. As the first step, note that when a NL-ring $\textrm{L}_G^1$ formed in $G^\textrm{th}$ band gap is moved behind another NL $\textrm{L}_G^2$ inside the \emph{same} band gap $G$, then all the NLs formed inside band gaps $G\pm 1$ which are linked with $\textrm{L}_G^1$ necessarily reverse orientations, cf.~Sec.~\ref{sec:orientations}. These orientation reversals, in turn, imply that all the fluxes $w^a_{G\pm 1}$ of the intersection points of $\textrm{L}_{G\pm 1}$ with the inscribed disc $\mathcal{D}^2$ flip sign, cf.~Sec.~\ref{sec:linking-overview}. However, we argued in Sec.~\ref{sec:monopole-linking-relation} that the monopole charge is proportional to the sum of these fluxes. We thus conclude that the monopole charge of a NL-ring flips sign when it is moved behind a NL formed inside the same band gap. Using the language of the very recent Ref.~\cite{Sun:2019}, the nodal line acts on the monopole charge as an \emph{Alice string}.

As the second step, notice that when a NL-ring $\textrm{L}_G^1$ is moved along a \emph{closed trajectory} that encircles a chosen NL, then the NL-ring $\textrm{L}_G^1$ is moved behind that NL once (or, more generally, an odd number of times). By combining the two steps, we conclude that the monopole charge of $\textrm{L}_G^1$ \emph{reverses sign} when moved along a closed trajectory that encircles a NL formed inside the same band gap. Finally, as both considered NLs are formed inside the \emph{same} band gap, we can invoke the ``conventional'' spectral flatenning from Eq.~(\ref{eqn:projected-Ham}), and state that the monopole charge [corresponding to $\pi_2(M^\textrm{AI}_{(n,\ell)})$] flips sign when braided along a closed trajectory with a non-trivial Berry phase [corresponding to $\pi_1(M^\textrm{AI}_{(n,\ell)})$].

The previous paragraph constitutes the complete proof for all values $n$, $\ell$ of occupied resp.~unoccupied bands. Nevertheless, to get a more solid understanding of the argument, let us explicitly discuss the example illustrated in Fig.~\ref{fig:braiding}, where we set $G=0$. We begin in Fig.~\ref{fig:braiding}(a) with two {\color{MyBlue}blue} ($G=0$) NL-rings ${\color{MyBlue}\textrm{L}_0^{1,2}}$, each linked with two {\color{MyRed}red} ($G=-1$) NLs ${\color{MyRed}\textrm{L}_{-1}^{1,2}}$. The orientations of the {\color{MyRed}red} NLs are such that the linking numbers are $\eta_- = +2$ and~$\eta_- = -2$ for the two {\color{MyBlue}blue} NL-rings, respectively. Therefore, if we merge the {\color{MyBlue}blue} NL-rings into a single NL-ring ${\color{MyBlue}\textrm{L}_0^{1+2}}$ along a trajectory that \emph{does not} pass under other {\color{MyBlue}blue} NLs, as shown in Fig.~\ref{fig:braiding}(b), the total linking number would be $\eta_-^\textrm{tot.} = 2+(-2) = 0$, implying a vanishing monopole charge. Indeed, it is obvious from Fig.~\ref{fig:braiding}(b) that one could remove through shrinking first the {\color{MyRed}red} NL-rings ${\color{MyRed}\textrm{L}_{-1}^{1,2}}$, and subsequently also the {\color{MyBlue}blue} NL-ring ${\color{MyBlue}\textrm{L}_0^{1+2}}$, without encountering any obstruction. 

In contrast, in Fig.~\ref{fig:braiding}(c) we move ${\color{MyBlue}\textrm{L}_0^{2}}$ towards ${\color{MyBlue}\textrm{L}_0^{1}}$ along a trajectory that passes under additional {\color{MyBlue}blue} NL ${\color{MyBlue}\textrm{L}_0^{3}}$. During this process, the fluxes $w_{-1}^{3,4}$ of intersection points $\textrm{P}_{-1}^{3,4}$ with {\color{MyRed}red} NLs reverse orientation, such that now $\eta_- = +2$ for \emph{both} {\color{MyBlue}blue} NL-rings ${\color{MyBlue}\textrm{L}_0^{1,2}}$. Therefore, merging the two NL-rings into a single {\color{MyBlue}blue} NL-ring ${\color{MyBlue}\textrm{L}_0^{1+2}}$ along this alternative path produces a total linking number $\eta_-^\textrm{tot.} = 2 + 2 = 4$, suggesting a non-vanishing monopole charge. Indeed, it is manifest from the resulting geometry, shown in Fig.~\ref{fig:braiding}(d), that there are obstructions for removing the merged {\color{MyBlue}blue} NL-ring ${\color{MyBlue}\textrm{L}_0^{1+2}}$ from the band structure. We have thus demonstrated that the ability of two NL-rings with a monopole charge to annihilate depends on the trajectory used to bring them together. The dependence of the combined monopole charge of two NL-rings on the trajectory used to bring them them together, observed in Fig.~\ref{fig:braiding}, is ultimately equivalent to the dependence of the monopole charge of a single NL-ring on the choice of the inscribed disc, observed in Fig.~\ref{fig:two-discs}.

\section{Conclusions and outlook}\label{sec:conclude}

In this manuscript, we discussed the new perspectives into the topology of NLs protected by $\mcP\mcT$-symmetry, which are offered by the non-Abelian topological charge recently discovered by Ref.~\cite{Wu:2018b}. After explaining in Sec.~\ref{sec:non-Abelian-NLs} how to efficiently wield the non-Abelian topological charge, and how to geometrically interpret it using orientation reversals of the NLs, we presented two applications of these ideas to study the topological properties of  NL-rings with a monopole charge~\cite{Fang:2015}. First, we used the non-Abelian charge in Sec.~\ref{sec:monopole-linking-relation} to provide a geometric proof of the relation between the monopole charge and the linking structure of the NLs, where we reproduced and supplemented the results of Ref.~\cite{Ahn:2018}. Second, in Sec.~\ref{sec:NA-braiding} we used this new tool to provide a simple geometric argument for a previously unreported phenomenon, when NL-rings that carry a monopole charge braid non-trivially around other NLs formed inside the same gap. The derivations that take advantage of the non-Abelian topological charge are less technical than the alternative derivations based on the ``conventional'' spectral projection of the Hamiltonian, \emph{e.g.} those presented in Refs.~\cite{Ahn:2018,Ahn:2018b} and in the appendices of this work.

We emphasize that the scope of applications of the non-Abelian topological charge is much broader than revealed by this work. Especially, the presented ideas apply generally to all the $\mcP\mcT$-symmetric models and materials with NLs studied in Refs.~\cite{Chen:2015b,Wu:2018b,Lian:2019,Li:2016,Xie:2015,Chan:2016,Zhao:2016b,Fang:2015,Nomura:2018,Ahn:2018,Huang:2016,Xu:2017,Kim:2015,Yu:2015,Weng:2015b,Wang:2017,Gong:2018,Yan:2017,Zhou:2018,Ezawa:2017,Bi:2017,Chang:2017,Hearing:1958,Chen:2017,Heikkila:2015b,McClure:1957,Singh:2018,Schoop:2016,Feng:2017,Yu:2017,Takahashi:2017,Chang:2017b,Yi:2018,Bzdusek:2017}, given that there are three or more bands. For example, the non-Abelian topological charge, together with further constraints put forward by the presence of additional crystalline symmetry~\cite{Wu:2018b}, could predict the behaviour of intricate NL compositions under applied strain, including the case of nodal chains~\cite{Wang:2017,Gong:2018}, nexuses~\cite{McClure:1957,Heikkila:2015b,Singh:2018}, gyroscopes~\cite{Yu:2015,Kim:2015,Weng:2015b} and nets~\cite{Schoop:2016,Feng:2017,Yu:2017,Takahashi:2017,Chang:2017b,Yi:2018}. It would also be interesting to check, whether constraints on the ability of band-structure nodes to annihilate, that were previously reported in different contexts~\cite{Sun:2018,Lim:2017}, could also be described using non-Abelian topological charges. Furthermore, it should be possible to generalize the presented topological classification, based on the alternative spectral normalization of the Bloch Hamiltonian in Eq.~(\ref{eqn:H-mulit-project}), to insulating systems. Some rudimentary steps in this direction were already taken by Ref.~\cite{Wu:2018b} to describe 1D \TB{insulators, 
and by Refs.~\cite{Wojcik:2019,Li:2019} to describe certain non-Hermitian topological bands.}

What is currently missing is a more tangible relation between the non-Abelian topological charge and physical observables. It is well understood that the conventional Berry phase quantization is related to bulk polarization~\cite{Vanderbilt:1993,Delpace:2011}, which explains the origin of the ``drumhead'' states on the surfaces of NL semimetals~\cite{Weng:2015}. Similarly, the work of Ref.~\cite{Wu:2018b} observed unusual edge states in 1D systems with non-trivial values of the non-Abelian charge, which could not be explained using solely the Berry phase quantization. However, the absence of a physical interpretation of the non-Abelian charge left those observations unexplained. As the first step towards understanding the surface states protected by the non-Abelian topological charges, it should prove fruitful to develop a rigorous mathematical relation between the homotopic methods applied here and the method of characteristic classes employed by Refs.~\cite{Ahn:2018,Ahn:2018b}. We leave these questions open for future studies.

\section{Acknowledgements}\label{sec:acknow}

We would like to thank J.~P.~Ang, J.-Y.~Chen, S.~A.~Kivelson, I.~Ochoa de~Alaiza, A.~A.~Soluyanov, X.-Q. Sun, P.~Teichner, G.~E.~Volovik, Q.-S.~Wu, B.-J. Yang, and S.-C.~Zhang for valuable discussions. A.~T. acknowledges funding from the European Union's Horizon 2020 Research and Innovation Programme under the Marie Sk\l odowska-Curie grant agreement No.~701647. T.~B. was supported by the Gordon and Betty Moore Foundation's EPiQS Initiative, Grant GBMF4302, and by an Ambizione grant from the Swiss National Science Foundation, Grant No.~185806. The illustrations of NL compositions in Figs.~\ref{fig:code}--\ref{fig:braiding} were produced using \texttt{Wolfram Mathematica} code, which we made publicly available online~\cite{Bzdusek:Mathematica}.

\appendix

\section{Coset spaces}\label{sec:cosets-spaces}


\subsection{Coset spaces for Lie groups}\label{sec:cosets-what-is}

Because the spaces of Hamiltonians considered in this work can be recast as \emph{coset spaces for Lie groups}, we present here some facts and results pertinent to this subject that are relevant to us. A \emph{Lie group} is essentially a smooth manifold that is also an abstract group. The (left) action $\alpha$ of a Lie group $\mathsf{G}$ on a topological space $X$ is defined as a continuous map $\alpha: \mathsf{G}\times X\to X$ such that $\alpha (g_1g_2,x)=\alpha(g_1,\alpha(g_2,x))$ for all $g_{1},g_{2}\in \mathsf{G}$ and $x\in X$, and the action of the identity element $e$ in $\mathsf{G}$ is trivial, \emph{i.e.} $\alpha(e,x)=x$. For any $g\in \mathsf{G}$, we denote by $\alpha_{g}$ a homeomorphism from $X$ to itself. The concatentation $\alpha_{h}\circ \alpha_{h^{-1}}$ is the identity homeomorphism. Therefore, one may view the action $\alpha$ as a group homomorphism between $\mathsf{G}$ and the group of automorphisms of $X$, denoted $\textrm{Aut}(X)$.  

Sets of the form $\mathsf{G}_{x}=\left\{\alpha(g,x)| \ g\in \mathsf{G}\right\}\subset X$ are called \emph{$\mathsf{G}$-orbits} of the point $x\in X$. Two $\mathsf{G}$-orbits $\mathsf{G}_{x}$,$\mathsf{G}_{x'}$ are either identical or have a null intersection. In other words the orbits constitute a partition of $X$. The set of all orbits is called the \emph{orbit space}, denoted $X/\mathsf{G}$.

Let $\mathsf{H}$ be a submanifold of $\mathsf{G}$ as well as a (sub)group. Then $\mathsf{H}$ is a \emph{Lie subgroup} of $\mathsf{G}$, denoted as $\mathsf{H}<\mathsf{G}$. There is a natural proper and free action of $\mathsf{G}$ on $\mathsf{H}$ given by the group action. The $\mathsf{H}$-orbits $\mathsf{H}_g = \{gh|h\in\mathsf{H}\}$ are called (left) cosets. The set of all such cosets is called the \emph{(left) coset space}, denoted $\mathsf{G}/\mathsf{H}$. If $\mathsf{H}$ is a \emph{normal subgroup} of $\mathsf{G}$ ($\forall h_1\in\mathsf{H}:\forall g\in \mathsf{G}:\exists h_2\in\mathsf{H}: g h_1 = h_2 g$), then the coset space has a natural group structure.

For a Lie group $\mathsf{G}$ of dimension $n$ and a Lie subgroup $\mathsf{H}< \mathsf{G}$, of dimension $k$, the coset space $\mathsf{G}/\mathsf{H}$ has a natural structure of a manifold of dimension $n-k$, such that the canonical map $\pi:\mathsf{G}\to \mathsf{G}/\mathsf{H}$ is a fiber bundle with fiber diffeomorphic to $\mathsf{H}$. In other words the sequence
\begin{align}
\triv \to \mathsf{H}\to \mathsf{G}\stackrel{\pi\;}\to \mathsf{G}/\mathsf{H}\to \triv
\label{coset_fibration}
\end{align}
is a fibration \cite{kirillov2008introduction}. This will be important for us in what follows as we will make ample use of the long exact sequence of homotopy groups for a fibration. 


\subsection{Homotopy groups of coset spaces}\label{eqn:homo-gr-cosets}

To derive the homotopy groups of a space of Hamiltonians (or of an order-parameter space) $M$, it is often possible and convenient to represent it as a {coset space} $M=\mathsf{G}/\mathsf{H}$ where $\mathsf{G}$ is a Lie group of transformations, and the \emph{isotropy subgroup} $\mathsf{H}<\mathsf{G}$ corresponds to transformations that keep a chosen Hamiltonian (or order parameter) in $M$ unchanged. In this section we review some basic relations for the homotopy groups of coset spaces. In what follows, we use $\mathsf{H}_0$ to denote the connected component of group $\mathsf{H}$ which contains the identity element. 

First we recall some properties of Lie groups. Due to a theorem by Cartan~\cite{Cartan:1936,8961}, for any Lie group 
\begin{subequations}
\begin{equation}
\pi_2(\mathsf{G}) = \triv\label{eqn:Lie-group-2nd-homo}
\end{equation}
where by $\triv$ we mean the trivial one-element group. Furthermore, any Lie group either consists of simply connected components, or it has a \emph{universal covering group} with such a property~\cite{Hatcher:2002}. This implies that we can always construct the coset space $M=\mathsf{G}/\mathsf{H}$ such that   	
\begin{equation}
\pi_1(\mathsf{G}) = \triv.\label{eqn:Lie-group-1st-homo}
\end{equation}
Finally, by taking the appropriate subgroup, we can always rewrite $M=\mathsf{G}/\mathsf{H}$ such that $\mathsf{G}$ has \emph{just one} connected component, meaning that we can also assume
\begin{equation}
\pi_0(\mathsf{G}) = \triv.\label{eqn:Lie-group-0th-homo}
\end{equation}
\end{subequations}
With this information and using the long exact sequence of homotopy groups~\cite{Hatcher:2002} for the fibration \eqref{coset_fibration},~\emph{i.e.}
\begin{align}
\cdots \to  \pi_{n}(\mathsf{H})\stackrel{i_n\;}{\to} \pi_n(\mathsf{G})\stackrel{j_n}{\to}\pi_n(\mathsf{G}/\mathsf{H}) \stackrel{\delta_{n}}{\to} \pi_{n-1}(\mathsf{H})\stackrel{i_{n-1}}{\to} \cdots
\label{LES_fibration}
\end{align}
we obtain two very useful relations. First we isolate the following two sections,
\begin{subequations}
\begin{align}
\triv &\;    \stackrel{j_2}{\to}\pi_2(\mathsf{G}/\mathsf{H}) \stackrel{\delta_{2}}{\to} \pi_{1}(\mathsf{H})\stackrel{i_{1}}{\to} \triv  \\
\triv &\;    \stackrel{j_1}{\to}\pi_1(\mathsf{G}/\mathsf{H}) \stackrel{\delta_{1}}{\to} \pi_{0}(\mathsf{H})\stackrel{i_{0}}{\to} \triv
\end{align} 
\end{subequations}
The first sequence implies that that $\text{ker}(\delta_2)=\text{im}(j_2)=\triv$,~\emph{i.e.} $\delta_2$ is injective, as well as $\text{im}(\delta_2)=\text{ker}(i_1)=\pi_1(\mathsf{H})$, which implies $\delta_2$ is surjective. Together these two statements imply the isomorphism 
\begin{equation}
\pi_2(\mathsf{G}/\mathsf{H}) = \pi_1(\mathsf{H} _0).\label{eqn:fundamental-theorem-2}
\end{equation}
Using completely parallel argumentation for $\delta_1$ instead of $\delta_2$, we arrive at the isomorphism 
\begin{equation}
\pi_1(\mathsf{G}/\mathsf{H}) = \pi_0(\mathsf{H}) = \mathsf{H}/\mathsf{H}_0.\label{eqn:fundamental-theorem}
\end{equation}
It can be checked that $\mathsf{H}_0$ is a normal subgroup of $\mathsf{H}$. It follows from the discussion in appendix~\ref{sec:cosets-what-is} that $\mathsf{H}/\mathsf{H}_0$, understood as the set of connected components of $\mathsf{H}$, has a natural group structure inherited from $\mathsf{H}$.

\section{Automorphisms induced by \texorpdfstring{$\pi_1(M)$}{first homotopy} on \texorpdfstring{$\pi_p(M)$}{higher homotopy} for \texorpdfstring{$M\simeq \mathsf{G}/\mathsf{H}$}{coset spaces}}\label{Appendix:A} \label{app:cosets-and-homotopy}

\subsection{Action of \texorpdfstring{$\pi_1(M)$}{the first homotopy group} on itself}\label{sec:non-Abelian-conjugacy}

When describing the topological charge of a band-structure node using homotopy groups, we study the homotopy class $[\mcH\circ\iota]$ of the Hamiltonian on a sphere $\mathcal{S}^p = \iota(S^p)\subset \textrm{BZ}$ that tightly encloses the node. According to the formal definition of homotopy groups~\cite{Hatcher:2002}, the sphere should be understood as $I^p/\partial I^p$, \emph{i.e.}~as a $p$-dimensional cube with quotiented boundary. We use $I = [0,1]$ to denote the closed interval on the real line with unit length. The identification of the boundary $\partial I^p$ with a single point is achieved by mapping it to a fixed point (the ``\emph{base point}'') $\mathfrak{m}\in M$. In practice, this is done by choosing a point $\mathrm{P}\in\mathrm{BZ}$ such that $\mcH(\textrm{P}) = \mathfrak{m}$, and an embedding $\iota: I^p\hookrightarrow \textrm{BZ}$ such that $\iota(\partial I^p) = \mathrm{P}$. Then the composition $\mcH\circ\iota: I^p \to M$ maps $\partial I^p$ to $\mathfrak{m}$, and can be classified using the based homotopy group $\pi_p(M,\mathfrak{m})$. The equivalence class of $[\mcH\circ \iota]$ does not change under continuous deformations of the sphere $I^p/\partial I^p\simeq S^p$ and of the Hamiltonian $\mcH$, provided that $\mcH(\mathcal{S}^p)$ is kept gapped (such that it remains inside $M$, cf.~Sec.~\ref{NLs-first-homotopy}) and that $\mcH(\textrm{P})\in M$ is kept constant. In the following, we write $f\equiv \mcH\circ\iota$ for the composed map from $S^p$ to $M$.

In the presence of multiple nodes, it may be possible to enclose the node by various spheres $\iota(S^p)$, which cannot be continuously deformed into one another when the base point $\textrm{P}$ is fixed. We already encountered this ambiguity in Sec.~\ref{sec:nonuniqueness} of the main text. Let us consider here again the case of nodes of codimension $\delta = 2$ [characterized by $\pi_1(M)$], illustrated in Fig.~\ref{fig:conjugacy} of the main text. In this example, the node $L_\textrm{G}$ can be encirlced with two based paths, $\Gamma_G^{a}$ and $\Gamma_G^b$, which cannot be deformed into each other while keeping the base point $\textrm{P}$ fixed, because of the additional node $\textrm{L}_{G'}$. However, if we encircle this additional node by a based path $\Gamma_{G'}$, then $\Gamma_G^b \sim \Gamma_{G'}\circ \Gamma_G^a \circ (\Gamma_{G'})^{-1}$ (\emph{i.e.} they are \emph{isotopic}). This implies that the corresponding elements in $\pi_1(M)$, defined by the Hamiltonian $\mcH$ along each $\Gamma$, fulfill the conjugation relation in Eq.~(\ref{eqn:group-conjugacy}) of the main text. As a consequence, the topological charge of a node is defined only up to conjugacy. Mathematically, the result in Eq.~(\ref{eqn:group-conjugacy}) can be understood as an \emph{automorphism} $\alpha_h\in \textrm{Aut}[\pi_1(M)]$ induced by an element of $h\in\pi_1(M)$, 
\begin{equation}
\forall g \in \pi_1(M): \;\; \alpha_h (g) = h g h^{-1}.
\end{equation}
We call such self-induced automorphism of $\pi_1(M)$ as \emph{the action of $\pi_1(M)$ on itself}.


\begin{figure*}[t!]
\includegraphics[width=0.73 \textwidth]{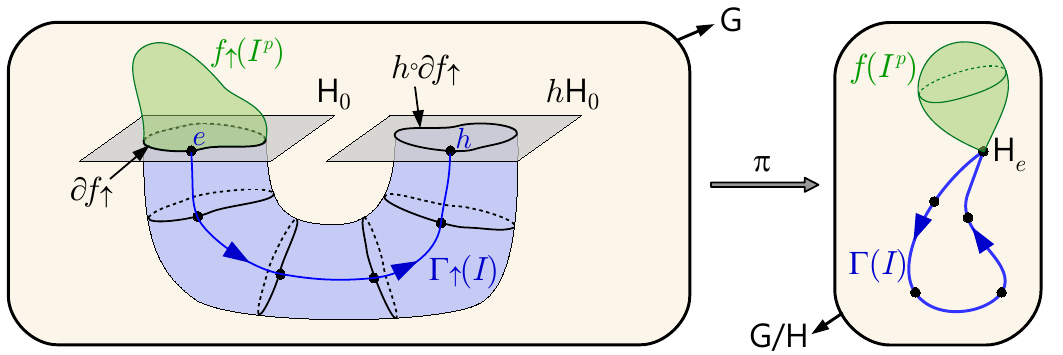}
 \caption{Schematic illustration of the considerations in appendix~\ref{subsec:action-pi1-p2}. The projection $\mathsf{G}\stackrel{\pi}{\rightarrow}\mathsf{G}/\mathsf{H}$ corresponds to the fibration in Eq.~(\ref{coset_fibration}). The closed path $\Gamma: I\to \mathsf{G}/\mathsf{H}$ shown in blue (right) is lifted to an open-ended path $\Gamma_\uparrow: I \to \mathsf{G}$ (left). The path $\Gamma_\uparrow(I)$ starts at the identity $e$ and ends at an element $h$ contained inside the connected component $h \mathsf{H}_0$ of coset $\mathsf{H}_e$. Similarly, the continuous based map $f: I^p \to \mathsf{G}/\mathsf{H}$ shown in green (right) is lifted into $f_\uparrow: I^p \to \mathsf{G}$ (left) in a way that the top face $J^{p-1} \subset \partial I^p$ of the cube is mapped to the identity $e$~\cite{Mermin:1979,Sun:2018}. The gray sheets indicate connected components $\mathsf{H}_0$ and $h \mathsf{H}_0$ of coset $\mathsf{H}_e$. The image $\textrm{im}\,f_\uparrow$ looks like a cap (green), while the image of the composition of $\Gamma_\uparrow$ with $f_\uparrow |_{\partial I^p} \equiv \partial f_\uparrow $ looks like a tube (blue). The black rings along the tube indicate several images $\Gamma_\uparrow(t)\cdot\partial f_\uparrow$ with fixed $t\in I$, and they project onto the black dots shown along $\Gamma(I)$ inside $\mathsf{G}/\mathsf{H}$. Since the green cap and the blue tube share a boundary, we can compose them. Finally, we shift this composition such that its boundary lies inside $\mathsf{H}_0$. This is achieved by a right multiplication with $h^{-1}$. We find that the composition of $f$ with $\Gamma$ induces the automorphism $\triangleright_h\in\textrm{Aut}[\pi_p(\mathsf{G}/\mathsf{H})]$, called the action of $\pi_1(\mathsf{G}/\mathsf{H})$ on $\pi_p(\mathsf{G}/\mathsf{H})$, which we describe in the text. For $p=2$, the automorphism is given explicitly by Eq.~(\ref{subequn:action-on-pi2}).}
\label{fig:theorems}
\end{figure*}

\subsection{The action of \texorpdfstring{$\pi_1(\mathsf{G}/\mathsf{H})$}{first homotopy} on \texorpdfstring{$\pi_2(\mathsf{G}/\mathsf{H})$}{the second homotopy}}\label{subsec:action-pi1-p2}

We have mathematically formalized the non-uniquness of the first-homotopy charges of nodes using the action of $\pi_1(M)$ on itself. The non-uniqueness of the second-homotopy charge, encountered in Sec.~\ref{sec:NA-braiding}, can be similarly understood as a non-trivial action of $\pi_1(M)$ on $\pi_2(M)$. Here, we describe this action for the case when $M$ is a coset space. A more general description of this action for arbitrary topological space $M$ is given in Appendix~\ref{sec:crossed-module}.

First, we consider a continuous map $f: I^p\to \mathsf{G}/\mathsf{H}$ with $p\geq 2$ that fulfills $f(\partial I^p) = \mathsf{H}_e$ (understood as a coset of $\mathsf{G}/\mathsf{H}$). In other words, we choose the base point $\mathfrak{m}\in\mathsf{G}/\mathsf{H}$ to be the coset of the identity element $e\in\mathsf{G}$. The map $f$ can be lifted into a continuous map $f_\uparrow: I^p \to \mathsf{G}$ that fulfills (1) $f_\uparrow: \partial I^p \to \mathsf{H}_e$ (understood as a subset of $\mathsf{G}$), and (2) $f_\uparrow: J^{p-1} \to e$, where $J^{p-1}$ is the top face of the cube $I^p$~\cite{Mermin:1979,Sun:2018}. We denote the restriction of $f_\uparrow$ to the boundary $\partial I^p$ as $f_\uparrow\big|_{\partial I^p} \equiv \partial f_\uparrow$, and we call it the \emph{boundary map}. The continuity of $f_\uparrow$ implies that the image $\textrm{im}\,\partial f_\uparrow $ lies inside a single connected component $\mathsf{H}_0 < \mathsf{H}_e$. For $p=2$, the isomorphism in Eq.~(\ref{eqn:fundamental-theorem-2}) implies that the equivalence class $[f] \in \pi_2(\mathsf{G}/\mathsf{H},\mathsf{H}_e)$ corresponds to the equivalence class $[\partial f_\uparrow] \in\pi_1(\mathsf{H}_0,e)$.

We further consider a closed path $\Gamma: I \to \mathsf{G}/\mathsf{H}$, which is also based at $\mathsf{H}_e$.
By the isomorphism in Eq.~(\ref{eqn:fundamental-theorem}), $\Gamma$ corresponds to some element $[\Gamma] \in \mathsf{H}/\mathsf{H}_0$.
To understand this isomorphism, note that $\Gamma$ can be lifted into an open-ended path $\Gamma_\uparrow: I \to \mathsf{G}$ that starts at $\Gamma_\uparrow(0) = e$. Since $\Gamma$ was based at $\mathsf{H}_e$, the other end-point of the lift lies at $\Gamma_\uparrow(1) = h \in \mathsf{H}$. Eq.~(\ref{eqn:fundamental-theorem}) states that the equivalence class $[\Gamma]$ corresponds to the connected component $h\mathsf{H}_0$ containing $h$.

We finally consider both maps, $f$ and $\Gamma$, simultaneously. By inflating path $\Gamma$ into a thin tube, it is possible to interpret $\Gamma$ as a based $p$-sphere (admittedly a somewhat degenerate one), allowing us to compose $\Gamma$ and $f$ to form a new $p$-sphere, denoted $\Gamma \circ f$. We ask what is the homotopy class of such a composed map. To answer this question, we construct a map
\begin{subequations}
\begin{equation}
\Gamma_\uparrow \cdot \partial f_\uparrow:\;\; I \times \partial I^p \to \mathsf{G}
\end{equation}
defined by
\begin{equation}
\Gamma_\uparrow \cdot \partial f_\uparrow: (t,\bs{y})\mapsto \Gamma_\uparrow(t)\cdot f_\uparrow(\bs{y})
\end{equation}
\end{subequations}
for all $t\in I$ and $\bs{y}\in\partial I^p$, where ``$\cdot$'' is the group composition in $\mathsf{G}$. The image $\textrm{im}\,(\Gamma_\uparrow \cdot \partial f_\uparrow)$ can be understood as a tube connecting $\im(\partial f_\uparrow)\subset \mathsf{H}_0$ to $h\cdot \im(\partial f_\uparrow) \subset \mathsf{H}_h$, shown in blue in the left part of Fig.~\ref{fig:theorems}. This tube can be continuously attached to the image $\im(f_\uparrow)$, which in Fig.~\ref{fig:theorems} we show as the green ``cap'' with boundary $\im(\partial f_\uparrow)$ inside $\mathsf{H}_0$. The composition $\textrm{im}\,(\Gamma_\uparrow \cdot \partial f_\uparrow) \cup \im (f_\uparrow)$ looks like an elongated cap, with boundary $h \cdot \partial f_\uparrow$ inside the connected component $h \mathsf{H}_0$ of $\mathsf{H}_e$.

Let us now restrict our attention to the case $p=2$. We would like to interpret the elongated cap as the lift of the composition $\Gamma \circ f$, and to classify this composition by an equivalence class of $\pi_2(\mathsf{G}/\mathsf{H},\mathsf{H}_e) = \pi_1(\mathsf{H}_0,e)$. However, since the boundary of the elongated cap lies inside $h\mathsf{H}_0$, we need to shift the cap inside $\mathsf{G}$ by $h^{-1}$. This is achieved through a right action on $\mathsf{G}$ by $h^{-1}$. Therefore, we find that the composition of $f$ with $\Gamma$ has transformed the boundary map as
\begin{subequations}\label{subequn:action-on-pi2}
\begin{equation}
\partial f_\uparrow \mapsto h (\partial f_\uparrow) h^{-1}.\label{eqn:composed-path-2-sphere}
\end{equation}
The relation in Eq.~(\ref{eqn:composed-path-2-sphere}) induces an automoprhism $\triangleright_h\in\textrm{Aut}[\pi_2(\mathsf{G}/\mathsf{H})] = \textrm{Aut}[\pi_1(\mathsf{H_0})]$, acting as
\begin{equation}
 \triangleright_h: [\partial f] \mapsto [h \cdot (\partial f )\cdot h^{-1}],\label{eqn:coset-triangle-right}
\end{equation}
\end{subequations}
which we call the \emph{action of $\pi_1(\mathsf{G}/\mathsf{H})$ on $\pi_2(\mathsf{G}/\mathsf{H})$}. The physical manifestation of this action is that the second-homotopy charge of a node depends on the choice of a based sphere $\mathcal{S}^2$ used to enclose the node.


\section{Abe homotopy and action of \texorpdfstring{$\pi_1(M)$}{the first homotopy group} on \texorpdfstring{$\pi_p(M)$}{the higher homotopy groups} for general \texorpdfstring{$M$}{M}}
\label{sec:crossed-module}
In this section we show that much of our understanding about the existence of and the interaction between the quantized Berry phase and the monopole charge can be cast into the mathematical framework of the so-called \emph{Abe homotopy groups} \cite{Abe:1940, kobayashi2012abe}. The second Abe homotopy group, denoted $\kappa_{2}(M,\mathfrak{m})$, comprises of homotopy classes of based maps from a pinched torus to $M$ and naturally encodes the action of $\pi_1(M,\mathfrak{m})$ on $\pi_2(M,\mathfrak{m})$.

Although we mostly restrict ourselves to $p=2$, the framework we describe in this section is quite general. The \emph{$p^\textrm{th}$ Abe homotopy group} $\kappa_p(M,\mathfrak{m})$ contains the based fundamental group, the based $p^\textrm{th}$ homotopy group and the action of $\pi_1(M,\mathfrak{m})$ on $\pi_p(M,\mathfrak{m})$. It can be thought of as homotopy classes of maps from $S^{p-1}\times I$ to $M$ such that $S^{p-1}\times (\partial I)$ maps to $\mathfrak{m}\in M$. This of course is the $p$-dimensional generalization of the pinched torus.

\begin{figure}[bt]
\centering
\includegraphics[scale=0.35]{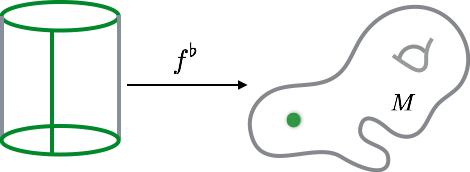}
\\
\caption{A map $f^{\flat}: S^{1} \times I \to M$ such that the green portion,~\emph{i.e.}~$S^1\times(\partial I) \;\cup\; \textrm{P} \times I$, is mapped to a base point $\mathfrak{m}\in M$.}
\label{gflat}
\end{figure}

\begin{figure}[bt]
\centering
\includegraphics[scale=0.33]{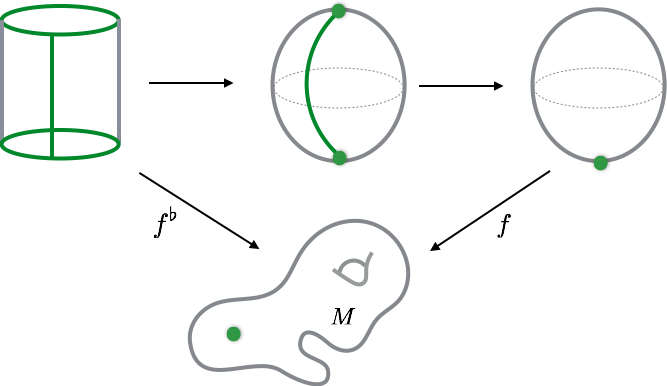}
\\
\caption{
A map $f^{\flat}$ as in Fig.~\ref{gflat} is isomorphic to a map $f\in \text{Map}[S^{2}_{\mathrm{BZ}}, M]_0 $ via the above construction. The first horizontal arrow corresponds to mapping the cylinder $S^{1}\times I$ to $S^{2}$ such that the top and bottom circles getted mapped onto chosen points in $S^{2}$. The second arrow maps $S^{2}$ to itself such that the entire green portion gets mapped to a single point.}
\label{iso}
\end{figure}

In appendix~\ref{sec:non-Abelian-conjugacy}, we discussed the construction of based maps from the Brillouin zone $\mathrm{BZ}$ to the space of Hamiltonians $M$. Before proceeding further, let us set up some notation. By $\text{Map}[{S}^{p}_{\mathrm{BZ}}, M]_{0}$ we denote the space of maps $\mathcal H \circ \iota$ that map $p$-spheres embedded in $\mathrm{BZ}$ and based at $\textrm{P}$ into $M$, such that $\textrm{P}\in S^{p}\subset \mathrm{BZ}$ is sent to $\mathfrak{m}\in M$. The physically interesting scenario corresponds to $p=1,2$ and $\pi_{1}(M)\neq \triv$ and/or $\pi_{2}(M)\neq \triv$ which corresponds to the existence of Berry phase and/or monopole charge obstructions in the corresponding models. First, since $\pi_1(M,\mathfrak{m})$ is in general a non-Abelian group, we label NLs by conjugacy classes rather than by elements in $\pi_1(M,\mathfrak{m})$. This accounts for the action of $\pi_1$ on itself as discussed previously in appendix~\ref{sec:non-Abelian-conjugacy}. In order to analyze the monopole charges, we need to study the maps $f\in \text{Map}[S^{2}_{\mathrm{BZ}},M]_0$. By $[f]$ we denote a homotopy class in $\text{Map}[S^{2}_{\mathrm{BZ}},M]_0 $. As discussed in the main text (see Sec.~\ref{sec:NA-braiding}) and in appendix~\ref{subsec:action-pi1-p2} (in the context of coset spaces), there may be the possibility of the Berry phases and monopole charges interacting. This interaction is captured by the action $\triangleright$ of $\pi_1(M)$ on $\pi_2(M)$, \emph{i.e.}
\begin{align}
\triangleright:\pi_1(M)\to \text{Aut}\left[\pi_{2}(M)\right]    
\end{align} 
In order to say something meaningful about the action of $\pi_1(M)$ on $\pi_{2}(M)$, we need to be able to define a product of a path $\Gamma:I\to M$ and some $f\in \text{Map}[S^2_{\mathrm{BZ}},M]_0$. To this end, we first define a map $f^{\flat}$ whose domain is a based cylinder $S^{1}\times I$ (see Fig.~\ref{gflat}) such that the region $S^1\times (\partial I) \;\cup\; \textrm{P} \times I$ with point $\textrm{P}\in{S}^1$
is mapped to $\mathfrak{m}\in M$.  
Such a map $f^{\flat}\sim f\in \text{Map}[S^{2}_{\mathrm{BZ}},M]_0$. This can be shown by explicitly constructing the homotopy, cf.~Fig.~\ref{iso}.

\begin{figure}[bt]
\centering
\includegraphics[scale=0.37]{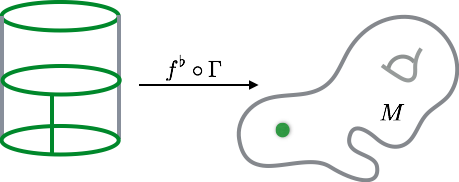}
\\
\caption{Maps
$f^{\flat}$ and $\Gamma\in \text{Map}[S^{1}_{\mathrm{BZ}},M]_0$ can be composed by stacking cylinders. }
\label{prod}
\end{figure}

\begin{figure}[bt]
\centering
\includegraphics[scale=0.37]{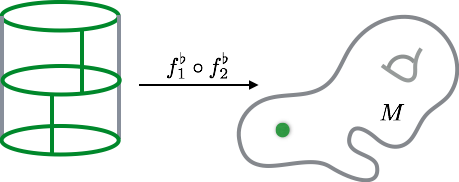}
\\
\caption{Maps 
$f^{\flat}_{1,2}\in \text{Map}[S^{2}_{\text{BZ}},M]_0$ can be composed by stacking cylinders such that all the marked (green) portions are mapped to the marked (green) point $\mathfrak{m}\in M$. }
\label{prod_pi2}
\end{figure}

Furthermore it is straightforward to define a product between a based loop $\Gamma:I\to M$ and $f^{\flat}$. This product is achieved by stacking cylinders, and by thinking of $\Gamma: I\times S^{1}\to M $ such that $\Gamma(x,\theta)=\Gamma(x,\theta')$ for all $x\in I$ and all $\theta,\theta'\in S^1$(see Fig.~\ref{prod}). Therefore a path in $\mathrm{BZ}$ is viewed as a cylinder (or a foliation of circles) such that each horizontal slice is mapped to a single point in $M$. Similarly a product structure can be defined for $\pi_{2}(M,\mathfrak{m})$ by stacking cylinders that correspond to a degenerate embedding of 2-spheres as illustrated in Fig.~\ref{prod_pi2}.  

In order to study the effect of $\pi_1(M)$ on $\pi_2(M)$ we introduce a generalization of the notion of homotopy groups wherein the notion of a base point is replaced with the notion a \emph{base space}. More precisely, let $\mu \subset M$ be a connected subspace of $M$. Then we denote the second
homotopy group based on $\mu$ as the set $\zeta_{2}(M,\mu)=\left\{ [f]\in \pi_2(M,\mathfrak{m}) \ \forall \ \mathfrak{m} \in \mu\right\}$ with the equivalence relation $\pi_2(M,\mathfrak{m}_1) \ni [f_1]\sim [f_2]\in \pi_{2}(M,\mathfrak{m}_2) $ if there exists a path $\Gamma:I\to M$ where $\Gamma(0)=\mathfrak{m}_1$ and $\Gamma(1)=\mathfrak{m}_2$ such that $[f_2] =[\Gamma]^{-1}\circ [f_1] \circ [\Gamma]$. The corresponding equivalence classes form a group which we shall denote $\pi_2(M,\mu)$. In particular if $\mu \subset M$ contains a non-trivial element of $[\gamma]\in \pi_{1}(M)$, then this implies
\begin{align}
  [f]\sim [\gamma^{-1}]\circ [f]\circ [\gamma] 
\label{action_conj}
\end{align}
where $[f]\in \pi_2(M,\mathfrak{m})$ and $\gamma\in \pi_1(M,\mathfrak{m})$.

\begin{figure}[bt]
\centering
\includegraphics[scale=0.37]{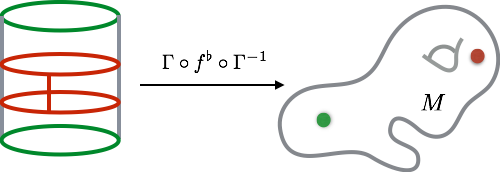}
\\
\caption{
The action of $\pi_1(M,\mathfrak{m})$ on $\pi_{2}(M,\mathfrak{m})$ is understood as a concatenation of cylinders. The bottom cylinder and the top cylinder are foliations of circles, where each foliation is mapped to the same point in $M$. The bottom and the top circles are mapped to the green point in $M$. Similarly, the middle cylinder is essentially a map from $S^{2}$ to $M$ with all the red regions mapped to the (arbitrary) red point in $M$.}
\label{pi1_action}
\end{figure}

Next we will show that the action of $\pi_{1}(M)$ on $\pi_{2}(M)$ by conjugation as in Eq.~\eqref{action_conj} can be cast into the $2^\textrm{nd}$ Abe homotopy group $\kappa_{2}(M,\mathfrak{m})$. Firstly, it is helpful to notice that the conjugation of an element $[f^{\flat}]\in \pi_{2}(M,\mathfrak{m}')$ by a path $\Gamma$ from $\mathfrak{m}$ to $\mathfrak{m}'$ can be understood as a stacking of cylinders as illustrated in Fig.~\ref{pi1_action}. The first (bottom) cylinder maps to a path from $\mathfrak{m}\in M$ (green point) to an arbitrary point $\mathfrak{m}'\in M$ (red point), the second cylinder maps to 2-sphere at $\mathfrak{m}'$ while the third cylinder maps to a path back from $\mathfrak{m}'$ to $\mathfrak{m}$. We emphasize that $\mathfrak{m}'$ is an arbitrary point, a special case of which is $\mathfrak{m}=\mathfrak{m}'$ which corresponds to Eq.~\eqref{action_conj}.    

\begin{figure}[bt]
\centering
\includegraphics[scale=0.38]{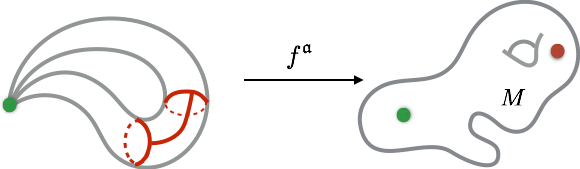}
\\
\caption{
Abe homotopy group $\kappa_{2}(M,\mathfrak{m})$ can be defined as homotopy classes of based maps from a pinched torus to $M$. As a set, $\kappa_{2}(M,\mathfrak{m})$ can be identified with $\pi_{1}(M,\mathfrak{m})\times \pi_{2}(M,\mathfrak{m})$, however the product structure on $\kappa_2(M,\mathfrak{m})$ corresponds to a semidirect product. 
}
\label{Abe}
\end{figure}

The $2^\textrm{nd}$ Abe homotopy group is defined in a similar construction as above. Consider the space of maps of the form 
\begin{align}
f^{\mathfrak {a}}:&\; S^1 \times I \to M \nonumber \\  
:&\; S^1 \times (\partial I) \mapsto \mathfrak{m}\in M
\end{align}
Since $f^{\mathfrak a}(S^{1} \times \partial I)=\mathfrak{m}$, this can be thought of as a map from a pinched torus to $M$. We denote the set of all such maps $K_2(M,\mathfrak{m})$.  The set of equivalence classes of maps of the kind $f^{\mathfrak a}$ form a based Abe homotopy groups which we shall denote $\kappa_{2}(M,\mathfrak{m})$ (Fig.~\ref{Abe}). It was shown in Ref.~\cite{kobayashi2012abe} that this group is a semi-direct product $\kappa_{2}(M,\mathfrak{m})= \pi_1(M,\mathfrak{m})\rtimes \pi_2(M,\mathfrak{m})$, where $\pi_{p}(M,\mathfrak{m})$ is the based $p^\textrm{th}$ homotopy group of $M$. 

\begin{figure}[bt]
\centering
\includegraphics[scale=0.45]{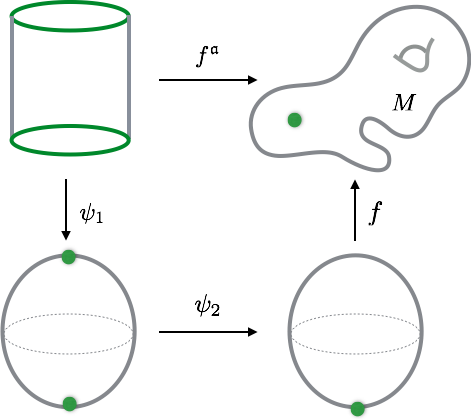}
\\
\caption{
There exists an injection $\Psi_{*}:\pi_2(M,\mathfrak{m})\!\to\! \kappa_2(M,\mathfrak{m})$ such that $\Psi_{*}(g)=g^{\mathfrak{a}}=\psi_1 \circ \psi_2 \circ g$. The map $\psi_1$ shrinks the top and the bottom circles of the cylinder to the north and the south poles of the 2-sphere, respectively, while $\psi_2$ is a smooth map that sends the north pole to an arbitrary point on the 2-sphere while keeping the south pole fixed.}
\label{inject}
\end{figure}

\begin{figure}[bt]
\centering
\includegraphics[scale=0.40]{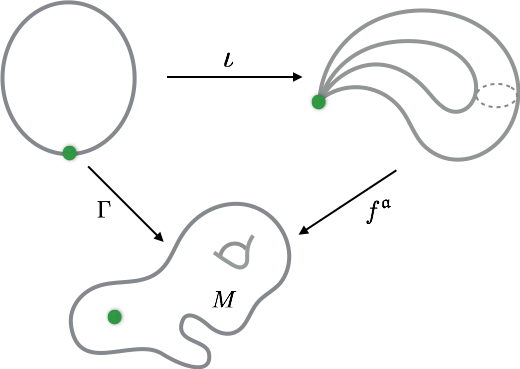}
\\
\caption{
There exists an inclusion $\iota:\pi_1(M,\mathfrak{m})\hookrightarrow \kappa_{2}(M,\mathfrak{m})$ obtained by 
``thickening'' a closed loop into a pinched torus. }
\label{inclusion}
\end{figure}

The Abe homotopy group fits in the following short exact sequence
\begin{align}
\triv\to \pi_{2}(M,\mathfrak{m})\xrightarrow{\Psi_{*}} \kappa_{2}(M,\mathfrak{m})\xrightarrow{\Phi_{*}} \pi_1(M,\mathfrak{m})\to \triv
\label{exact_abe}
\end{align}
Here, following \cite{kobayashi2012abe} we sketch a constructive proof of the exactness of the above sequence. Further details can be found in \cite{Abe:1940}. We sketch this in two steps. 
\begin{enumerate}
\item {\bf{ Construct an injection $\Psi_{*}: \pi_{2}(M,\mathfrak{m})\to \kappa_{2}(M,\mathfrak{m})$:}} Given some $f\in \text{Map}[S^{2}_{\mathrm{BZ}},M]_0$ we can obtain $f^{\mathfrak a}= f\circ  \psi_2 \circ \psi_1   \in K_{2}(M,\mathfrak{m})$ as illustrated in Fig.~\ref{inject}. This map is injective as there is an explicit construction for each $f\in \text{Map}[S^{2}_{\mathrm{BZ}},M]_0$. However, it is not a surjection as the map $\psi_2$ does not exist for all $f^{\mathfrak a}\in K_2(M,\mathfrak{m})$. In fact, the obstruction for the existence of such a map is exactly $\pi_1(M,\mathfrak{m})$. We think of $\Psi_{*}$ as a homomorphism of groups 
\begin{align}
\Psi_{*}:&\; \pi_{2}(M,\mathfrak{m})\to \kappa_{2}(M,\mathfrak{m}) \nonumber \\
:&\; [f]\mapsto [f\circ\psi_2\circ \psi_1]
\end{align}
\item {\bf{Construct a surjective map $\Phi_*: \kappa_{2}(M,\mathfrak{m}) \to \pi_{1}(M,\mathfrak{m})$:}} First, we fix a point $\textrm{P}\in S^{1}$ and construct a map $\Phi_{*}: f^{\mathfrak a}(x,t)\to \Gamma(t):= f^{\mathfrak a}(\textrm{P},t) $ where $x\in S^1$ and $t\in I$. Since $\Gamma(0)=\Gamma(1)=\mathfrak{m}$, it follows that $\Gamma\in \text{Map}[S^{1}_{\mathrm{BZ}},M]_0$. Furthermore, $\Phi_{*}([f^{\mathfrak a}])=[\Gamma]$ therefore $\Phi_{*}:\kappa_{2}(M,\mathfrak{m})\to \pi_1(M,\mathfrak{m})$.   Similarly we can construct an inclusion $\iota: \pi_1(M,\mathfrak{m})\hookrightarrow \kappa_{2}(M,\mathfrak{m})$ by simply thickening a closed loop into a pinched torus as illustrated in Fig.~\ref{inclusion}.
\end{enumerate}
Therefore, we have shown that $\text{im}(\Psi)=\text{ker}(\Phi)$. Equivalently, $\kappa_{2}(M,\mathfrak{m})$ can be expressed as $([\Gamma],[f])\in \pi_{1}(M,\mathfrak{m})\otimes \pi_2(M,\mathfrak{m})$ with the composition rule
\begin{align}
    ([\Gamma],[f])*([\Gamma'],[f'])=([\Gamma\circ \Gamma'],[\Gamma'\triangleright f \circ f'])
\end{align}
where $[\Gamma'\triangleright f]\equiv [\Gamma'] \circ [f] \circ [{\Gamma'}]^{-1}$. 
In the presence of non-trivial Berry phase, monopole charges and nodal lines should therefore be labelled by conjugacy classes of the Abe homotopy group rather than $\pi_2(M,\mathfrak{m})$ and $\pi_1(M,\mathfrak{m})$, respectively.

Before moving on to concrete computations pertaining to the \emph{Abe homotopy} groups of spaces of Hamiltonians which are of immediate interest in the present work, we make a few comments about a more general mathematical framework of which the \emph{Abe homotopy groups} are a part of. The first and the second 
homotopy groups along with some additional data can be reformulated as a categorification of the fundamental group known as the \emph{strict fundamental 2-group} or \emph{fundamental crossed module}~\cite{whitehead1941adding, whitehead1949combinatorial, baez2004higher, Martins:2008uvd, Ang:2018qur, Ang:2018rls}. More precisely this object includes (first and second) homotopy groups of the reduced CW-complex representing a space (denoted $M_{*}$), the action $\triangleright$ along with an additional piece of data which is a homomorphism from the first to the second homotopy group of $M_{*}$. Together, these four pieces of data need to satisfy some \emph{coherence relations} that ensure they work well together. Furthermore, equivalence classes of strict fundamental 2-groups are described by so-called \emph{weak 2-groups}~\cite{baez2004higher} that include $\pi_1(M,\mathfrak{m})$, $\pi_2(M,\mathfrak{m})$, $\vartriangleright$, and $[\beta]\in H^{3}_{\triangleright}(\pi_1,\pi_2)$. The last piece of data encodes the violation of associativity of $\pi_{1}$ within the weak 2-group. Recently there has been much interest in the appearance of 2-groups within the context of gauge theories \cite{Kapustin:2013uxa, Cordova:2018cvg, Benini:2018reh, Delcamp:2019fdp, Delcamp:2018wlb}. It is tempting to cast the topological interactions of Berry phases and monopole charges within the framework of homotopy 2-groups. In particular the physical interpretation of $[\beta]$ in the present context remains intriguing. We postpone this for a future work.
\section{Action of \texorpdfstring{$\pi_1$}{the Berry phase} on \texorpdfstring{$\pi_2$}{the monopole charge} in nodal class \texorpdfstring{$\textrm{AI}$}{AI}}
\label{app:pi1-on-pi2-examples}

In this appendix we explicitly compute the action of $\pi_1(M,\mathfrak{m})$ on $\pi_{2}(M,\mathfrak{m})$ for the cases where $M\equiv M^\textrm{AI}_{(n,\ell)}$ are the spaces of spectrally flattened Hamiltonians in symmetry class $\textrm{AI}$, given by Eq.~(\ref{eqn:real-grass}) of the main text. The corresponding homotopy groups are listed in Tables~\ref{tab:AI-homotopy} of the main text~\cite{Bzdusek:2017}. Case by case, we consider situations with different number of occupied and unoccupied bands, which exhibit different group structure of the monopole charge. For simplicity, we drop the base point $\mathfrak{m}$ from our notation in this appendix. Nevertheless, the reader should keep in mind that that we always consider \emph{based} homotopy groups. 

The general strategy employed to compute the action is adapted from Mermin~\cite{Mermin:1979}, and involves writing $M$ as a coset space $\mathsf{G}/\mathsf{H}$ with $\pi_{i}(\mathsf{G})=\triv$ for $i=0,1,2$. Using Eqs.~\eqref{eqn:fundamental-theorem-2} and \eqref{eqn:fundamental-theorem}, we first construct the first and the second homotopy groups of $M$. The generators of $\pi_0(\mathsf{H})$ and $\pi_1(\mathsf{H})$ correspond to the generators of $\pi_1(M)$ and $\pi_{2}(M)$, respectively. The action of $\pi_1(M)$ on $\pi_2(M)$ is extracted by looking at the action of $\pi_0(\mathsf{H})$ on $\pi_1(\mathsf{H})$ through conjugation, as outlined in appendices~\ref{subsec:action-pi1-p2} and~\ref{sec:crossed-module}. The results presented in this appendix are consistent with those obtained in Sec.~\ref{sec:NA-braiding} of the main text using the non-Abelian topological charge.

\subsection{Models with \texorpdfstring{$2+1$}{2+1} bands}\label{subsec:A21}

According to Eq.~(\ref{eqn:real-grass}) of the main text, Hamiltonians $\mcH(\bs{k})$ of nodal class $\textrm{AI}$ with two occupied and one unoccupied bands constitute the space~\cite{Bzdusek:2017} 
\begin{subequations}
 \begin{equation}
M^{\textrm{AI}}_{(1,2)} = \mathsf{O}(3)/\mathsf{O}(2)\times\mathsf{O}(1).\label{eqn:Hspace-2-1}
\end{equation}
To derive its homotopy groups, it is convenient to rewrite this space as~\cite{Mermin:1979,Wu:2018b}
\begin{equation}
M^{\textrm{AI}}_{(1,2)} = {\mathsf{SO}(3)}/{\mathsf{D}_\infty} = {\textsf{SU}(2)}/{\bar{\mathsf{D}}_\infty}\label{eqn:RP2-double-cover}
\end{equation}
\end{subequations}
To get from Eq.~(\ref{eqn:Hspace-2-1}) to Eq.~(\ref{eqn:RP2-double-cover}), we first narrowed our attention to the connected component of the group of transformations (and we performed the corresponing restriction in the isotropy subgroup). In the second step, we constructed the simply connected double cover of the group of transformations (and we performed the corresponding lift in the isotropy subgroup).

Let us discuss these two steps in more detail. First, the quotient $\mathsf{D}_\infty < \mathsf{SO}(3)$, called the \emph{infinite dihedral group}, consists of $\mathsf{O}(2)\times\mathsf{O}(1)$ matrices with positive determinat. Explicitly,
\begin{subequations}
\begin{equation}
R_0(\alpha) = \e{\alpha L_3} = \left(\begin{array}{ccc}
\cos \alpha    & -\sin\alpha  & 0     \\
\sin\alpha     & \cos\alpha   & 0     \\
0               &  0            & 1
\end{array}\right)
\end{equation}
and
\begin{equation}
R_1(\alpha) = \e{\alpha L_3}\e{\pi L_1} = \left(\begin{array}{ccc}
\cos \alpha    & \sin\alpha   & 0     \\
\sin\alpha    & -\cos\alpha  & 0     \\
0               &  0            & -1
\end{array}\right)
\end{equation}
\end{subequations}
where $\alpha \in [0,2\pi)$. The matrices $\{L_i\}_{i=1}^3$ form the basis of Lie algebra of $\mathfrak{so}(3)$, and they can be described using the Levi-Civita symbol as $(L_{i})_{jk}=-\epsilon_{ijk}$~\cite{Fecko:2006}. As a topological space, $\mathsf{D}_\infty$ looks like a disjoint union of two circles. We set the subscript $a\in\{0,1\}$ of $R_a(\alpha)$ according to the corresponding element of the zeroth homotopy group $\pi_0(\mathsf{D}_\infty) = \mathsf{D}_\infty/\mathsf{SO}(2) = \ztwo$. 

The second step in Eq.~(\ref{eqn:RP2-double-cover}) is achieved by replacing the generators 
\begin{equation}
\mathfrak{so}(3)\ni L_j \;\;\longmapsto \;\; t_j = -\tfrac{\imi}{2}\sigma_j \in \mathfrak{su}(2),
\end{equation}
where $\{\sigma_j\}_{j=1}^3$ are the Pauli matrices. This replacement preserves the structure constants of the Lie algebra, and yields the simply connected covering group $\mathsf{Spin}(3) = \mathsf{SU}(2)$ upon exponentiation~\cite{Fecko:2006}. Performing the same replacement in the isotropy subgroup leads to 
\begin{subequations}\label{eqn:AI-2-1-rot-lift}
\begin{equation}
\bar{R}_0(\alpha) = \e{\alpha t_3} = \left(\begin{array}{cc}
\e{-\imi\frac{\alpha}{2}}    & 0     \\
0     & \e{\imi\frac{\alpha}{2}}      
\end{array}\right)
\end{equation}
and
\begin{equation}
\bar{R}_1(\alpha) = \e{\alpha t_3}\e{\pi t_1} = \left(\begin{array}{cc}
0               & -\imi \e{-\imi\frac{\alpha}{2}}       \\
-\imi \e{\imi\frac{\alpha}{2}}    & 0
\end{array}\right)
\end{equation}
\end{subequations}
where $\alpha \in [0,4\pi)$. The matrices in Eqs.~(\ref{eqn:AI-2-1-rot-lift}) together constitute the double point group $\bar{\mathsf{D}}_\infty < \mathsf{SU}(2)$. As a topological space, $\bar{\mathsf{D}}_\infty$ is still homeomorphic to a disjoint union of two circles. The subscript of $\bar{R}_a(\alpha)$ is again chosen according to the corresponding element of $\pi_0(\bar{\mathsf{D}}_\infty)=\ztwo$. We remark that we use the horizontal bar, as in $\bar{R}_a(\alpha)$, to indicate double point symmetries throughout the rest of the appendices.

By applying the identities from Eqs.~(\ref{eqn:fundamental-theorem-2}) and~(\ref{eqn:fundamental-theorem}) in appendix~\ref{eqn:homo-gr-cosets}, it follows that
\begin{subequations}
\begin{eqnarray}
\pi_1(M^{\textrm{AI}}_{(1,2)}) &=& \pi_0(\bar{\mathsf{D}}_\infty) = \ztwo \label{eqn:AI-1-2-1st-homotopy} \\
\pi_2(M^{\textrm{AI}}_{(1,2)}) &=& \pi_1(\bar{\mathsf{D}}_\infty) = \intg, \label{eqn:AI-1-2-2nd-homotopy}
\end{eqnarray}
\end{subequations}
which correspond to the Berry phase and to the monopole charge, respectively. It remains to be checked whether the Berry phase induces a non-trivial action on the monopole charge. To that end, it can be checked that
\begin{equation}
\bar{R}_b(0)\bar{R}_a(\alpha)\bar{R}_b^{-1}(0) = \bar{R}_a((-1)^b\alpha),
\end{equation}
\emph{i.e.}~the conjugation with a non-trivial element $b=1$ of $\pi_1(M^\textrm{AI}_{(1,2)})$ reverses the angle of rotation, which determines the equivalence classes in $\pi_2(M^\textrm{AI}_{(1,2)})$. Comparing to Eq.~(\ref{eqn:coset-triangle-right}) of appendix~\ref{subsec:action-pi1-p2}, we say that the action induced by a non-trivial element of group~(\ref{eqn:AI-1-2-1st-homotopy}) on group group~(\ref{eqn:AI-1-2-2nd-homotopy}) is non-trivial. In physical terms, we find that the sign of the monopole charge is reversed when it is carried along a closed path with a non-trivial Berry phase. This agrees with our findings obtained in Sec.~\ref{sec:NA-braiding} using the non-Abelian topological charge.

\subsection{Models with \texorpdfstring{$2+2$}{2+2} bands}

We want to rephrase the space of Hamiltonians
\begin{align}
    M_{(2,2)}^{\text{AI}}={\mathsf{O}(4)}/{\mathsf{O}(2)\times\mathsf{O}(2)}
\end{align} as $\mathsf{G}/\mathsf{H}$, where $\mathsf{G}$ is a simply connected Lie group. Similar to the case of $2+1$ bands, we first narrow our attention to rotations with positive determinant. This lowers the group of transformations to $\mathsf{SO}(4)$. The restricted isotropy subgroup corresponds to the subset of $\mathsf{O}(2)\times\mathsf{O}(2)$ matrices with positive determinant, namely
\begin{subequations}\label{eqn:AI-2-2-isotropy-so}
\begin{eqnarray}
R_0(\alpha,\beta) &=& R^\textrm{2D}(\alpha)\oplus R^\textrm{2D}(\beta ) \\
R_1(\alpha,\beta) &=& [\sigma_3 R^\textrm{2D}(\alpha)]\oplus[\sigma_3 R^\textrm{2D}(\beta )] 
\end{eqnarray}
\end{subequations}
where
\begin{equation}
R^{2\textrm{D}}(\varphi) = \left(\begin{array}{cc}
\cos\varphi    &   -\sin\varphi           \\
\sin\varphi    &   \cos\varphi
\end{array}\right)
\end{equation}
describes rotations of a plane. From a topological perspective, the group of matrices in Eqs.~(\ref{eqn:AI-2-2-isotropy-so}) looks like a disjoint union of two tori. The subscript of $R_i(\alpha,\beta)$ corresponds to the zeroth homotopy group of this group.

We further need to find the lift of the isotropy subgroup into the double cover $\mathsf{Spin}(4) \cong \mathsf{SU}(2)\times\mathsf{SU}(2)\equiv \mathsf{G}$~\cite{Akbulut:2016}, because $\mathsf{SO}(4)$ itself is not simply connected. To achieve this, first note that the generators of $\mathsf{SO}(4)$ 
\begin{eqnarray}
L_1 = \!\left(\begin{array}{cccc}
0       &   0       &   0       &   0       \\
0       &   0       &   -1       &   0       \\
0       &   +1      &   0       &   0       \\
0       &   0       &   0       &   0       
\end{array}\right);
&\ & 
M_1 = \!\left(\begin{array}{cccc}
0       &   0       &   0       &   -1      \\
0       &   0       &   0       &   0       \\
0       &   0       &   0       &   0       \\
+1      &   0       &   0       &   0       
\end{array}\right) \nonumber  \\
L_2 = \!\left(\begin{array}{cccc}
0       &   0       &   +1       &   0       \\
0       &   0       &   0       &   0       \\
-1       &   0      &   0       &   0       \\
0       &   0       &   0       &   0       
\end{array}\right);
& \ & 
M_2= \!\left(\begin{array}{cccc}
0       &   0       &   0       &   0      \\
0       &   0       &   0       &   -1       \\
0       &   0       &   0       &   0       \\
0      &   +1       &   0       &   0       
\end{array}\right) \nonumber \\
L_3 = \!\left(\begin{array}{cccc}
0       &   -1       &   0       &   0       \\
+1       &   0       &   0       &   0       \\
0       &   0      &   0       &   0       \\
0       &   0       &   0       &   0       
\end{array}\right);
&\ & 
M_3= \!\left(\begin{array}{cccc}
0       &   0       &   0       &   0      \\
0       &   0       &   0       &   0       \\
0       &   0       &   0       &   -1       \\
0      &   0       &   +1       &   0       
\end{array}\right)  \nonumber
\end{eqnarray}
fulfill the commutation relations
\begin{equation}
  \left[L_i,L_j\right] = \epsilon_{ijk} L_k, \;
  \left[M_i,M_j\right] = \epsilon_{ijk} L_k,  \;
  \left[L_i,M_j\right] = \epsilon_{ijk}  M_k.
\end{equation}
Clearly, this is not an ideal basis to relate $\mathfrak{so}(4)$ to the direct sum $\mathfrak{su}(2)\oplus \mathfrak{su}(2)$. However, linear combinations 
\begin{equation}
X_i = \tfrac{1}{2}(L_i + M_i)\quad\textrm{and}\quad Y_i = \tfrac{1}{2}(L_i - M_i)\label{eqn:so(4)-gens}
\end{equation}
form an $\mathfrak{so}(4)$ basis with commutation relations
\begin{equation}
  \left[X_i,X_j\right] = \epsilon_{ijk} X_k, \;\;\;   \left[Y_i,Y_j\right] = \epsilon_{ijk} Y_k, \;\;\;
  \left[X_i,Y_j\right] = \triv.
\end{equation}
This coincides with the commutation relations of
\begin{equation}
\bar{X}_i = (-\tfrac{\imi}{2}\sigma_i) \oplus \triv\quad\textrm{and}\quad \bar{Y}_i = \triv\oplus (-\tfrac{\imi}{2}\sigma_i),\label{eqn:su(2)-su(2)-gens}
\end{equation}
which form a natural basis of Lie algera $\mathfrak{su}(2)\oplus\mathfrak{su}(2)$. 

Now that we know how to perform the lifts from $\mathsf{SO}(4)$ to $\mathsf{SU}(2)\times\mathsf{SU}(2)$, we express the elements of the isotropy subgroup in Eqs.~(\ref{eqn:AI-2-2-isotropy-so}) using the $\mathfrak{so}(4)$ generators as
\begin{subequations}\label{eqn:2-2-logarithms}
\begin{eqnarray}
R_0(\alpha,\beta) &=& \e{\alpha L_3}\e{\beta M_3} \nonumber \\
&=& \e{(\alpha + \beta) X_3} \e{(\alpha - \beta) Y_3} \\
R_1(\alpha,\beta) &=& \e{\pi M_2}\e{\alpha L_3}\e{\beta M_3}  \nonumber \\
&=& \e{\pi X_2}\e{(\alpha + \beta) X_3} \e{-\pi Y_2} \e{(\alpha - \beta) Y_3}.\;\;\;
\end{eqnarray}
\end{subequations}
By replacing the generators with those in Eq.~(\ref{eqn:su(2)-su(2)-gens}), we obtain the elements of the isotropy subgroup $\mathsf{H}<\mathsf{G}$,
\begin{subequations}
\begin{eqnarray}
\bar{R}_0(\alpha,\beta) &=& \e{(\alpha + \beta) \bar{X}_3} \e{(\alpha - \beta) \bar{Y}_3} \\
\bar{R}_1(\alpha,\beta) &=& \e{\pi \bar{X}_2}\e{(\alpha + \beta) \bar{X}_3} \e{-\pi \bar{Y}_2} \e{(\alpha - \beta) \bar{Y}_3},\;\;\;
\end{eqnarray}
\end{subequations}
where $\alpha,\beta \in [0, 4\pi)$, and the group of transformations $\mathsf{G} = \mathsf{SU}(2)\times\mathsf{SU}(2)$ is simply connected.

From the topological perspective, the isotropy group within $\mathsf{SU}(2)\times\mathsf{SU}(2)$ still looks like a disjoint union of two tori. The subscript of $\bar{R}_i(\alpha,\beta)$ is chosen according to the corresponding element of the zeroth homotopy group. Using the relations in Eqs.~(\ref{eqn:fundamental-theorem-2}) and~(\ref{eqn:fundamental-theorem}), we find
\begin{subequations}
\begin{eqnarray}
\pi_1(M^{\textrm{AI}}_{(2,2)}) &=& \pi_0(\mathsf{H}) = \ztwo \label{eqn:AI-2-2-1st-homotopy} \\
\pi_2(M^{\textrm{AI}}_{(2,2)}) &=& \pi_1(\mathsf{H}) = \intg\oplus\intg, \label{eqn:AI-2-2-2nd-homotopy}
\end{eqnarray}
\end{subequations}
Furthermore, an explicit calculation reveals that 
\begin{equation}
\hspace{-0.15cm}\bar{R}_b^{-1}(0,0) \bar{R}_a(\alpha,\beta) \bar{R}_b(0,0) = \bar{R}_a((-1)^b\alpha,(-1)^b\beta),
\end{equation}
implying a non-trivial action of $\pi_1$ on $\pi_2$, namely one that flips the sign of both of the $\intg$ monopole charges.

\subsection{Models with \texorpdfstring{$3+2$}{3+2} bands}

In analogy to the previous two cases, next we study models with $n=3$ occupied and $\ell=2$ unoccupied bands. The relevant space of Hamiltonians is
\begin{align}
M_{(3,2)}^{\text{AI}}={\mathsf{O}(5)}/[{\mathsf{O}(3)\times \mathsf{O}(2)}].
\end{align}
Following Mermin~\cite{Mermin:1979}, we aim to express $M^\textrm{AI}_{(3,2)}$ as $\mathsf{G}/\mathsf{H}$ with a simply connected Lie group $\mathsf{G}$. To that end, we first limit our attention to the connected component
\begin{subequations}
\begin{equation}
\widetilde{\mathsf{G}}=\mathsf{SO}(5)<\mathsf{O}(5)
\end{equation}
of the group of transformations, which narrows the isotropy subgroup to the semidirect product
\begin{equation}
\widetilde{\mathsf{H}} = (\mathsf{SO}(3)\times\mathsf{SO}(2)) \rtimes \ztwo < \mathsf{O}(3)\times\mathsf{O}(2).\label{eqn:so-3-2-stabiliz}
\end{equation}
\end{subequations}
Here, the group $\ztwo<\mathsf{SO}(5)$ is generated by a direct sum of a matrix in $\mathsf{O}(3)$ and of a matrix in $\mathsf{O}(2)$, which both have negative determinant and which both square to $+\unit$.

To further describe the isotropy subgroup, recall that $\mathsf{SO}(5)$ has $10$ generators $L_{ij}$ with $1 \leq i < j \leq 5$, which have matrix components
\begin{subequations}
\begin{equation}
(L_{ij})_{ab} = -\delta_{ia}\delta_{jb} + \delta_{ib}\delta_{ja},
\end{equation}
and which fulfill the commutation relations
\begin{equation}
[L_{ij},L_{k\ell}] = \delta_{ik} L_{j\ell} + \delta_{j\ell} L_{ik} - \delta_{i\ell}\delta_{jk} - \delta_{jk} L_{i\ell}.\label{eqn:so(5)-commutators}
\end{equation}
\end{subequations}
Let us assume that the $\mathsf{SO}(3)$ matrices act on the first three coordinates of $\reals^5$, while the $\mathsf{SO}(2)$ matrices act on the last two coordinates. Then all the elements of $\mathsf{SO}(3)\times\mathsf{SO}(2)$ can be expressed by exponentiating linear combinations of $\{L_{12},L_{23},L_{13},L_{45}\}$. Furthermore, we can set the generator of the $\ztwo$ subgroup to be $\diag{(1,1,-1,1,-1)} = \e{\pi L_{35}}$. If we define a vector $\bs{L}_{123}=(L_{12},L_{23},-L_{13})$, then all the elements of $\widetilde{\mathsf{H}}$ in Eq.~(\ref{eqn:so-3-2-stabiliz}) can be expressed as
\begin{subequations}
\begin{eqnarray}\label{eqn:rot-2-3-with-gens}
R_0(\bs{n},\alpha;\beta) &=& \e{\alpha \bs{n}\cdot \bs{L}_{123}}\e{\beta L_{45}} \\
R_1(\bs{n},\alpha;\beta) &=& \e{\pi L_{35}}\e{\alpha \bs{n}\cdot \bs{L}_{123}}\e{\beta L_{45}}
\end{eqnarray}
\end{subequations}
where $\bs{n}$ is a unit vector in $\reals^3$, and $\alpha,\beta\in[0,2\pi)$. Since the triplet $\bs{L}_{123}$ is isomoprhic to Lie algebra $\mathfrak{so}(3)$, it is manifest from Eqs.~(\ref{eqn:rot-2-3-with-gens}) that both connected components of $\widetilde{\mathsf{H}}$ are topologically equivalent to $\mathsf{SO}(3)\times \mathsf{SO}(2)$. The subscript of $R_i(\bs{n},\alpha;\beta)$ indicates the corresponding element of $\pi_0(\widetilde{\mathsf{H}})$.

In order to implement Mermin's recipe, we require $\pi_{i}(\mathsf{G})=0$ for $i=0,1,2$. Since $\pi_1(\mathsf{SO}(5))=\mathbb Z_{2}$, we need to consider the double cover of $\widetilde{\mathsf{G}}=\mathsf{SO}(5)$, \emph{i.e.} $\mathsf{Spin}(5)$, which fits into the short exact sequence 
\begin{align}
\triv \to \mathbb Z_{2}\to \mathsf{Spin}(n)\to \mathsf{SO}(n)\to \triv.
\end{align} 
It can be checked using the long exact of homotopy groups that $\mathsf{G}=\mathsf{Spin}(5)$ satisfies the required conditions. The lift $\mathsf{H}$ corresponding to $\widetilde{\mathsf{H}}$ retains two connected components. Both components are diffeomorphic to the lift of $\mathsf{SO}(3)\times\mathsf{SO}(2)$, which is topologically equivalent to $S^3\times S^1$. Therefore, 
\begin{equation}
\mathsf{H} \simeq S^0 \times S^1 \times S^3.
\end{equation}
with homotopy groups 
\begin{subequations}
\begin{align}
\pi_{0}({\mathsf{H}})=\pi_1(M^\textrm{AI}_{(3,2)})& =  \ztwo  \\
\pi_{1}({\mathsf{H}})=\pi_2(M^\textrm{AI}_{(3,2)})& =  \intg,
\end{align} 
\end{subequations}
in agreement with the results of Ref.~\cite{Bzdusek:2017} 

To extract the action of the first homotopy group on the second, we need to explicitly lift the matrices $R_{i}(\bs{n},\alpha;\beta)\in\widetilde{\mathsf{H}}\subset\mathsf{SO}(5)$ to the covering group $\mathsf{Spin}(5)$. To achieve this, we note that replacing the generators
\begin{equation}
L_{ij} \longmapsto t_{ij} = -\tfrac{1}{4}[\gamma_i,\gamma_j],
\end{equation}
where we employed Dirac matrices $\{\gamma_i\}_{i=1}^5$ obeying $\{\gamma_i,\gamma_j\}=2\delta_{ij}$, preserves the structure constants defined by Eq.~(\ref{eqn:so(5)-commutators}). Matrices $t_{ij}$ with $1\leq i < j \leq 5$ form the basis of Lie algebra $\mathfrak{spin}(5)$~\cite{Esposito:1998}. If we again organize $(t_{12},t_{23},-t_{13}) \equiv \bs{t}_{123}$, then the elements of the isotropy subrgroup $\mathsf{H}$ are obtained through exponentiation as
\begin{subequations}
\begin{eqnarray}\label{eqn:spin-2-3-with-gens}
\bar{R}_0(\bs{n},\alpha;\beta) &=& \e{\alpha \bs{n}\cdot \bs{t}_{123}}\e{\beta t_{45}} \\
\bar{R}_1(\bs{n},\alpha;\beta) &=& \e{\pi t_{35}}\e{\alpha \bs{n}\cdot \bs{t}_{123}}\e{\beta t_{45}}.
\end{eqnarray}
\end{subequations}
with $\alpha,\beta\in[0,4\pi)$. The homotopy classes of $\pi_1(M^\textrm{AI}_{(3,2)})$ are encoded by the subscript of $\bar{R}_i(\bs{n},\alpha;\beta)$, while the homotopy classes of $\pi_2(M^\textrm{AI}_{(3,2)})$ correspond to the winding of the angle $\beta$. To check the action, we choose a basis
\begin{eqnarray}
\gamma_1=\sigma_1\otimes \tau_1,\quad\; &\gamma_{2}=\sigma_1\otimes \tau_2,&\quad\; \gamma_3=\sigma_1\otimes \tau_3, \nonumber \\
\gamma_4=\sigma_2\otimes \unit \!\!&\;\;\textrm{and}\;\;& \!\!
\gamma_5 =\sigma_3\otimes \unit.
\end{eqnarray}
An explicit calculation reveals that 
\begin{eqnarray}
&\phantom{.}&\bar{R}_b(\bs{n}_1,0;0)\bar{R}_a(\bs{n}_2,0;0)\bar{R}_b^{-1}(\bs{n}_1,0;\beta) \nonumber \\
&\phantom{.}&= \bar{R}_a(\bs{n}_2,0;(-1)^b\beta).
\end{eqnarray}
Hence, the non-trivial element of $\pi_{1}(M^\textrm{AI}_{(3,2)})=\mathbb Z_2$, corresponding to Berry phase $\pi$, acts on the monopole charge $\pi_{2}(M^\textrm{AI}_{(3,2)})=\mathbb Z$ non-trivially by flipping the sign.

\subsection{Models with \texorpdfstring{$3 + 3$}{3+3} bands}

Next we consider the situation with three occupied and three unoccupied bands, when
\begin{align}
M_{(3,3)}^\textrm{AI}={\mathsf{O}(6)}/[{\mathsf{O}(3)\times \mathsf{O}(3)}]
\end{align}
Following the same strategy as in the previous subsection, we first narrow our attention to $\widetilde{\mathsf{G}}=\mathsf{SO}(6)$, which decreases the isotropy subgroup to
\begin{equation}
\widetilde{\mathsf{H}} = (\mathsf{SO}(3)\times\mathsf{SO}(3))\rtimes \ztwo \label{eqn:3-3-rot-H}
\end{equation}
in complete analogy with Eq.~(\ref{eqn:so-3-2-stabiliz}). Its homotopy groups are
\begin{equation}
\pi_{1}(\widetilde{\mathsf{H}})=\ztwo\oplus \ztwo \nonumber \qquad\textrm{and}\qquad \pi_{0}(\widetilde{\mathsf{H}})=\ztwo.
\end{equation}
However, since $\mathsf{SO}(6)$ is not simply connected, we have to perform the lift to the covering group $\mathsf{Spin}(6) = \mathsf{G}$. We further need to identify the lift $\mathsf{H}$ corresponding to $\widetilde{\mathsf{H}}$. At this step, our analysis starts to differ from the previous subsections. Especially, it is important to notice that the double group of $\mathsf{SO}(3)\times\mathsf{SO}(3)$ \emph{is not diffeomorphic to} $\mathsf{SU}(2)\times\mathsf{SU}(2)$. The reason is that $2\pi$-rotation in the first resp.~in the second $\mathsf{SO}(3)$-factor [which both correspond to $\unit\in\mathsf{SO}(6)$] should be lifted to the \emph{same} element $-\unit \in \mathsf{Spin}(6)$. In contrast, the na\"{i}ve lift wrongly produces \emph{different} matrices $(-\unit)\oplus\unit$ and $\unit\oplus(-\unit)$ in $\mathsf{SU}(2)\times\mathsf{SU}(2)$.

To see how this problem arises, we need to recall basics of singular homology theory~\cite{Hatcher:2002}. For a given manifold $M$, a \emph{$p$-cycle} $a\subset M$ is a $p$-dimensional submanifold of $M$ without boundary, and a \emph{homology class} $[a]$ is an equivalence class of $p$-cycles modulo boundaries. The \emph{torsion coefficient} is the smallest positive integer $T$ such that composing $T$ copies of $a$ is homologous to the trivial or zero cycle. Homology classes of $p$-cycles form the \emph{$p^\textrm{th}$} (singular) \emph{homology group} of $M$, which we denote $\mathbb{H}_p(M)$. Furthermore, since circles $S^1$ are $1$-cycles, there is a relation between $\pi_1(M)$ and $\mathbb{H}_1(M)$. Assuming that $M$ is path-connected, it follows from the Hurewitz theorem that $\mathbb{H}_1(M)$ is equal to the Abelianization of $\pi_1(M)$. 

Let us denote the two generators of $\pi_1(\widetilde{\mathsf{H}}) = \mathbb{H}_1(\widetilde{\mathsf{H}})$ as $[\tilde{a}]$ and $[\tilde{b}]$. We may indicate the generators explicitly by writing the group as $\ztwo^{[\tilde{a}]}\oplus \ztwo^{[\tilde{b}]}$. The space $\mathsf{SO}(3)\simeq \reals P^{3}$ has, up to homology, a single 1-cycle with torsion coefficient $T= 2$~\cite{Hatcher:2002}. Without loss of generality we may pick this cycle to be the closed path $\Gamma_{\textrm{tor.}}$ that maps $\alpha \in [0,2\pi]$ to $\e{\alpha L_1}$. We set the generators $[\tilde{a}]$ and $[\tilde{b}]$ to be the equivalence classes of an analogous path inside the first resp.~inside the second $\mathsf{SO}(3)$-factor of $\widetilde{\mathsf{H}}$ in Eq.~(\ref{eqn:3-3-rot-H}). Importantly, the lift of $\Gamma_{\text{tor.}}$ in $\mathsf{SU}(2)$ is \emph{not} a 1-cycle, but an open-ended path connecting $\unit$ to $-\unit$. However, we get a cycle by going around $\Gamma_{\text{tor.}}$ twice, although this is homologous to zero. Let us now apply analogous reasoning to the lift $\mathsf{H}$. By composing the lifts $[a],[b]$ of $[\tilde{a}],[\tilde{b}]$, we get three candidates for 1-cycles in $\mathsf{H}$, namely $2[a]$, $2[b]$ and $[a]+[b]$. The first two are homologous to zero by the same argument as before. However, $[a] + [b]$ is non-trivial, and generates a non-trivial first homology (as well as homotopy) group $\ztwo^{[a]+[b]}$.

After this detour, we can confidently state that
\begin{subequations}
\begin{eqnarray}
\pi_1({\mathsf{H}})&=&\pi_2(M^\textrm{AI}_{(3,3)}) = \ztwo \\
\pi_0(\mathsf{H})&=& \pi_1(M^\textrm{AI}_{(3,3)}) = \ztwo,
\end{eqnarray}
\end{subequations}
The presented discussion trivially generalizes to arbitrary $n,\ell \geq 3$, as anticipated in the stable limit of nodal class $\textrm{AI}$. In this case, we do not need to compute the action of $\pi_{1}$ on $\pi_{2}$, since $\ztwo$ does not admit non-trivial automorphisms (\emph{i.e.}~$\textrm{Aut}[\ztwo] = \triv$).

\section{Action of \texorpdfstring{$\pi_1$}{the winding number} on \texorpdfstring{$\pi_2$}{the monopole charge} in nodal class \texorpdfstring{\textrm{CI}}{CI}} \label{sec:action-for-CI}

We finally consider nodal class $\textrm{CI}$, which also supports a monopole charge~\cite{Bzdusek:2017,Bouhon:2018b}. In addition to $\mcP\mcT$ symmetry, systems in this symmetry class also exhibit a chiral symmetry $\mcC$ that anticommutes with the Bogoliubov-de~Gennes Hamiltonian at every $\bs{k}$. This nodal class is physically realized by $\mcP\mcT$-symmetric singlet superconductors without SOC. The chiral symmetry reduces the space of Hamiltonians to~\cite{Kitaev:2009,Bzdusek:2017b} 
\begin{equation}
M^\textrm{CI}_{(n)} = \mathsf{U}(n)/\mathsf{O}(n), 
\end{equation}
which is called the \emph{Lagrangian Grassmannian}. We are interested in their first and the second homotopy groups, as well as in the action of the first group on the second.

\subsection{General considerations}
In order to compute the homotopy groups of the Lagrangian Grassmannian, it is useful to write it in terms of a fibration~\cite{piccione2000geometry}  
\begin{align}
\triv\to\mathsf{O}(n)\to\mathsf{U}(n)\to \mathsf{U}(n)/\mathsf{O}(n)\to\triv    
\end{align}
Since the zeroth/first/second homotopy groups of both $\mathsf{O}(n)$ and $\mathsf{U}(n)$ are known, we may use the long exact sequence of a fibration \eqref{LES_fibration} to extract the homotopy groups of the quotient space. First let us focus on the following section of the long exact sequence
\begin{align}
\hspace{-0.15cm}
    \pi_{2}[\mathsf{U}(n)]\stackrel{j_2\;}{\to}
     \pi_{2}[\mathsf{U}(n)/\mathsf{O}(n)]\stackrel{\delta_2\;}{\to}
     \pi_{1}[\mathsf{O}(n)]\stackrel{i_1\;}{\to}
     \pi_{1}[\mathsf{U}(n)]
\end{align}
Since $\pi_2(\mathsf{G})=\triv$ for any Lie group \cite{Cartan:1936}, $\text{im}(j_2)=0=\text{ker}(\delta_2)$ which implies $\text{im}(\delta_2)=\pi_{2}[\mathsf{U}(n)/\mathsf{O}(n)]$.
Furthermore the map $\pi_1[\mathsf{O}(n)]\to \pi_1[\mathsf{U}(n)]$ has image $\triv$ as $\pi_1[\mathsf{U}(n)]$ corresponds to the winding number of the phase of the determinant, while the determinant of $\mathsf{O}(n)$ matrices is constant. Then using exactness, we learn that
\begin{align}
    \pi_2[\mathsf{U}(n)/\mathsf{O}(n)]=\pi_1[\mathsf{O}(n)]=
    \begin{cases}
      \mathbb Z, & \text{if}\ n=2 \\
      \mathbb Z_2, & \text{if}\ n>2.
    \end{cases}
\end{align}

Furthermore, the other unknown homotopy group fits into a short exact sequence 
\begin{align}
    \triv\to\intg \stackrel{j_1}{\to} \pi_1[\mathsf{U}(n)/\mathsf{O}(n)]\to\ztwo \to\triv
\end{align}
There are two possible group extensions of $\mathbb Z_{2}$ by $\mathbb Z$: 
\begin{itemize}
\item[(\emph{i})] either the unknown homotopy group $\pi_1[\mathsf{U}(n)/\mathsf{O}(n)]$ is $\intg\oplus\ztwo$ (in which case $\textrm{im} \, j_1 = \intg \times\{0\} < \intg\oplus\ztwo$, where $\{0\}$ is the identity of the $\ztwo$ factor)
\item[(\emph{ii})] or it is $\pi_1[\mathsf{U}(n)/\mathsf{O}(n)] = \intg$ (in which case $\im j_1 = 2\intg$ are even integers). 
\end{itemize}
In both cases, $j_1$ is \emph{not surjective}. In order to narrow in on one of the two possibilities, we take a look at the cokernel of $j_{1}$, \emph{i.e.} $\pi_1[\mathsf{U}(n)/\mathsf{O}(n)]/\text{im}(j_1)$. This comprises of a path in $\mathsf{U}(n)$ that connects a matrix with determinant $+1$ to a matrix with determinant $-1$. Within $U(n)$ this is an open-ended path, so it is not classified by homotopy group. However, upon quotienting the $\mathsf{O}(n)$ subgroup, this path becomes closed. This is the non-trivial generator of $\pi_1[\mathsf{U}(n)/\mathsf{O}(n)]$. The crucial question is what is the \emph{square} of this element. If it were trivial within $\pi_1[\mathsf{U}(n)/\mathsf{O}(n)]$, then it would follow that $\pi_1[\mathsf{U}(n)/\mathsf{O}(n)] = \intg \oplus \ztwo$. But this is not the case! Instead, taking two copies of this path produces a closed path in $\mathsf{U}(n)$ with winding number $+1$. Since $\textrm{ker}\, j_1 = \triv$, this element is non-trivial inside $\pi_1[\mathsf{U}(n)/\mathsf{O}(n)]$ too, implying that
\begin{align}
\pi_1[\mathsf{U}(n)/\mathsf{O}(n)] = \intg.  
\end{align}

\subsection{Models with \texorpdfstring{$2+2$}{2+2} bands}

To compute the action for $M^\textrm{CI}_{(2)} = \mathsf{U}(2)/\mathsf{O}(2)$, we employ a different strategy than in appendix~\ref{app:pi1-on-pi2-examples}. Recall that two matrices $u_{1,2}
\in\mathsf{U}(2)$ produce the same element of $M^\textrm{CI}_{(2)}$ (\emph{i.e.}~they are \emph{equivalent}, denoted $u_1 \sim u_2$) if they differ by an orthogonal matrix $o \in \mathsf{O}(2)$. We indicate such an equivalence by writing
\begin{equation}
\forall u\in\mathsf{U}(2):\forall o\in\mathsf{O}(2): \quad u \sim o\cdot u.\label{eqn:CI-space-1}
\end{equation}
Note also that each element $u\in \mathsf{U}(2)$ can be written as $u = \e{2\pi \imi f} s$ where $f \in \reals$ and $s \in \mathsf{SU}(2)$. We indicate this decomposition by writing $u = [f \,|\, s]$. However, such a decomposition is not unique. Performing a half-integer shift of $f$ while flipping the sign $s\mapsto -s$ encodes the same $u$, \emph{i.e.}~$[f\,|\,s] = [f+\tfrac{1}{2}\,|\,-s]$. Furthermore, since $s$ and $-s$ differ by an $\mathsf{O}(2)$ transformation, we find an equivalence
\begin{subequations}\label{eqn:so-long-and-thank-you}
\begin{equation}
\forall f \in \reals: \forall s\in\mathsf{SU}(2):\quad [f\,|\,s] \sim [f+\tfrac{1}{2}\,|\,s].\label{eqn:CI-space-2}
\end{equation}
The half-integer (rather than integer) shifts come from the $\intg_N$ quotient in $\mathsf{U}(N) = \mathsf{SU}(N) \times \mathsf{U}(1)/\intg_N$ for $N=2$.

The advantage of the encoding $u = [f\,|\,s]$ is that $\mathsf{G} = \reals \times \mathsf{SU}(2)$ is simply connected, thus fulfilling the assumptions of Eqs.~(\ref{eqn:fundamental-theorem-2}) and (\ref{eqn:fundamental-theorem}). Investigating the equivalence relations, such as the one in Eq.~(\ref{eqn:CI-space-2}), allows us to comprehend the isotropy subgroup $\mathsf{H}<\mathsf{G}$. To achieve our goal, we decompose using the sign of the determinant $\mathsf{O}(2) = \mathsf{SO}^+(2)\cup\mathsf{SO}^-(2)$, and we denote the corresponding elements as $o_+$, resp.~$o_-$. It follows from Eq.~(\ref{eqn:CI-space-1}), and by checking the determinants, that 
\begin{eqnarray}
[f\,|\,o_{+} \cdot s ] &\sim&  [f\,|\,s] \label{eqn:CI-space-3} \\
\left[f\,|\,o_{-} \cdot s \right] &\sim&  [f+\tfrac{1}{4}\,|\,s] \label{eqn:CI-space-4}
\end{eqnarray}
\end{subequations}
for all $f\in\reals$, $s\in\mathsf{SU}(2)$, and $o_\pm\in\mathsf{SO}^\pm(2)$. These relations come in addition to Eq.~(\ref{eqn:CI-space-2}). The three Eqs.~(\ref{eqn:so-long-and-thank-you}) implicitly describe the isotropy subgroup of $M^\textrm{CI}_{(2)} = \mathsf{G}/\mathsf{H}$ with a simply connected $\mathsf{G} = \reals \times \mathsf{SU}(2)$.

Note that the isotropy subgroup $\mathsf{H}$ has only two generators: (1) An infinitesimal ${o_+}\in\mathsf{SO}^+(2)$ rotation of $s$, corresponding to Eq.~(\ref{eqn:CI-space-3}), and (2) a quarter-integer shift of $f$ followed by a multiplication of $s$ by $\sigma_3 \in\mathsf{SO}^-(2)$, corresponding to Eq.~(\ref{eqn:CI-space-4}). The relation in Eq.~(\ref{eqn:CI-space-2}) is obtained by applying the second generator twice. The two generators allow us to express each element $h\in\mathsf{H}$ using two pieces of information: a quarter integer $\tfrac{n}{4}$, and an $\mathsf{O}(2)$ matrix $(\sigma_3)^p o_+(\alpha)$ with $p\in\{0,1\}$, and $\alpha \in [0,2\pi)$. Furthermore, a shift by odd multiples of $\tfrac{1}{4}$ always brings about one factor of $\sigma_3$, meaning that the power $p = p(n)$ equals the \emph{parity} of $n$. We indicate these two pieces of information concisely in the form $h = h(n,\alpha)$. This element of $\mathsf{H}$ transforms a general element of $\mathsf{G}$ as
\begin{equation}
h(n,\alpha): \quad [f\,|\,s]\mapsto [f + \tfrac{n}{4}\,|\,(\sigma_3)^{p(n)}o_+(\alpha) s] \label{eqn:CI-space-5}
\end{equation}
By applying the action of elements in $\mathsf{H}$ on $[f\,|\,s]\in \mathsf{G}$ twice [one needs to be careful about correctly commuting a power of $\sigma_3$ with $o_+(\alpha)$], we find the composition rule
\begin{equation}
h(n,\alpha)\circ h(m,\beta) = h(n+m,(-1)^{p(m)}\alpha + \beta).\label{eqn:CI-h-composition}
\end{equation}
The inverse elements are given by 
\begin{equation}
h(n,\alpha)^{-1} = h(-n,-(-1)^{p(n)}\alpha)
\end{equation}
which we use below to derive the action of $\pi_1$ on $\pi_2$.

To see what the previous discussion implies for the homotopy groups, recall that $\pi_0(\mathsf{H})$ counts the disjoint components of $\mathsf{H}$. We now know that disjoint components of $\mathsf{H}$ correspond to the quarter-integer shifts of $f$, on which one naturally defines a $\intg$ group structure, cf.~Eq.~(\ref{eqn:CI-h-composition}). Therefore,
\begin{equation}
\pi_0(\mathsf{H}) = \pi_1(M^\textrm{CI}_{(2)}) = \intg,
\end{equation}
which gets generated by $h(1,0)$. Furthermore, $\pi_1(\mathsf{H})$ counts the winding number of $\alpha$ in $\mathsf{SO}^+(2)\simeq S^1$ for any connected component with a fixed value of $n$. Therefore,
\begin{equation}
\pi_1(\mathsf{H}) = \pi_2(M^\textrm{CI}_{(2)}) = \intg,
\end{equation}
With this information, we can finally study the action of $\pi_1$ on $\pi_2$. An explicit calculation reveals that 
\begin{equation}
h(n,0) \circ h(m,\alpha) \circ h(n,0)^{-1} = h(m,(-1)^{p(n)}\alpha).
\end{equation}
Therefore, closed paths characterized by odd values of $n$ flip the sign of $\alpha$, thus reversing the sign of the second-homotopy charge. The physical manifestation is that the monopole charge of NL-rings in singlet superconductors~\cite{Bzdusek:2017} \emph{changes sign} when carried along a path with a non-trivial winding number.


\setcounter{equation}{1}
\renewcommand\theequation{\fnsymbol{equation}}
\bibliography{bib}{}
\bibliographystyle{apsrev4-1}  
\end{document}